\documentclass[a4paper,fleqn,usenatbib]{mnras}
\usepackage{natbib}
\usepackage{tabu}
\usepackage{float}
\restylefloat{table}
\usepackage[T1]{fontenc}
\usepackage{ae,aecompl}

\usepackage{graphicx}	% Including figure files
\usepackage{amsmath}	% Advanced maths commands
\usepackage{amssymb}	% Extra maths symbols
\usepackage{float}
\usepackage[flushleft]{threeparttable}
\floatstyle{plaintop}
\restylefloat{table}

\usepackage[caption=false]{subfig}
\usepackage[usenames,dvipsnames]{color}
\def\Msol{\ifmmode{\rm M}_{\mathord\odot}\else M$_{\mathord\odot}$\fi}

\title[An Improved Test of Binary Black Hole Hypothesis]{An Improved Test of the Binary Black Hole Hypothesis for Quasars with Double-peaked Broad Balmer Lines.}

\author[A. Doan et al.]{
  Anh Doan,$^{1}$\thanks{E-mail address: \texttt{dnd5@psu.edu}} Michael Eracleous,$^{1,2}$ Jessie C. Runnoe,$^{3}$\thanks{Current address: Department of Physics and Astronomy, Vanderbilt University, 6301 Stevenson Center Ln, Nashville, TN 37235, USA} Jia Liu,$^{4}$ Gavin Mathes$^{5}$
  \newauthor and Helene M. L. G. Flohic$^{6}$
\\ \\
$^{1}$Department of Astronomy \& Astrophysics, the Pennsylvania State University, 525 Davey Lab, University Park, PA 16802, USA\\
$^{2}$Institute of Gravitation and the Cosmos, the Pennsylvania State University, 104 Davey Laboratory, University Park, PA 16802, USA\\
$^{3}$Department of Astronomy, University of Michigan, 1085 S. University Ave., Ann Arbor, MI 48109, USA \\
$^{4}$Department of Astrophysical Sciences, Princeton University, Princeton, NJ 08544, USA\\
$^{5}$Department of Astronomy, New Mexico State University, Las Cruces, NM 88001, USA\\
$^{6}$Department of Physics, University of the Pacific, 3601 Pacific Avenue, Stockton, CA 95211, USA \\
}

\begin{document}
\label{firstpage}
\pagerange{\pageref{firstpage}--\pageref{lastpage}}
\maketitle

\begin{abstract}
Velocity offsets in the broad Balmer lines of quasars and their temporal variations serve as indirect evidence for bound supermassive black hole binaries (SBHBs) at sub-parsec separations. In this work, we test the SBHB hypothesis for 14 quasars with double-peaked broad  emission lines using their long-term (14--41 years) radial velocity curves. We improve on previous work by (a) using elliptical instead of circular orbits for the SBHBs, (b) adopting a statistical model for radial velocity jitter, (c) employing a Markov Chain Monte Carlo method to explore the orbital parameter space efficiently and build posterior distributions of physical parameters and (d) incorporating new observations. We determine empirically that jitter comprises approximately Gaussian distributed fluctuations about the smooth radial velocity curves that are larger than the measurement errors by factors of order a few. We initially treat jitter by enlarging the effective error bars and then verify this approach via a variety of Gaussian process models for it. We find lower mass limits for the hypothesized SBHBs in the range $10^8$--$10^{11}\;\Msol$. For seven objects the SBHB scenario appears unlikely based on goodness-of-fit tests. For two additional objects the minimum SBHB masses are unreasonably large ($>10^{10}\;\Msol$), strongly disfavoring the SBHB scenario. Using constraints on the orbital inclination angle (which requires some assumptions) makes the minimum masses of four more objects unreasonably large. We also cite physical and observational arguments against the SBHB hypothesis for nine objects. We conclude that the SBHB explanation is not the favoured explanation of double-peaked broad emission lines.

\end{abstract}

\begin{keywords}
galaxies: nuclei -- quasars: emission lines -- quasars: supermassive black holes
\end{keywords}

%%%%%%%%%%%%%%%%%%%%%%%%%%%%%%%%%%%%%%%%%%%%%%%%%%

%%%%%%%%%%%%%%%%% BODY OF PAPER %%%%%%%%%%%%%%%%%%

\section{Introduction}\label{sec:intro}

Supermassive black hole binaries (SBHBs) were first suggested by \cite{Begelman1980} as an explanation for the observed bending and apparent precession of radio jets. According to the scenario outlined by these authors, SBHBs can form from mergers of two galaxies, each with its own central black hole (BH). Since a large fraction of galaxies have central BHs, and galaxy mergers are common, it is reasonable to expect a substantial number of SBHBs in the observable universe. In the scenario of \cite{Begelman1980}, as two galaxies merge, the central BHs will sink into the core of the merger remnant due to dynamical friction. This first stage advances quickly, on a time-scale comparable to the dynamical time-scale of the host galaxies ($\sim 10^8\;$yr). The orbit then decays on a longer time-scale to separations $\sim 1\,$pc as a result of the scattering of stars. If the population of stars available for scattering can be replenished sufficiently quickly and orbital decay can continue until the separation is $\sim 10^{-3}$~pc, then gravitational wave emission will take over as the dominant mechanism of angular momentum loss and drive the binary to coalescence within a Hubble time. Early calculation, assuming that the SBHB is in a spherical star cluster with an isotropic velocity distribution \citep[e.g.,][and references therein]{Roos1981,Valtonen1996}, showed that the orbital decay of typical SBHBs is likely to slow down (or stall) at separations $\sim0.01$--1~pc. This conclusion was a consequence of the slow replenishment of stars available for scattering by the SBHB. As a result, the SBHBs would not merge within a Hubble time, which was termed the ``last parsec problem.'' Some recent works attempt to solve the last parsec problem either by proposing alternative mechanisms for orbital decay of SBHBs, such as interaction with a gaseous reservoir or close-range torques by a circumbinary disc (e.g., \citealt{Armitage2002, Escala2004, Dotti2007, Dotti2009, Hayasaki2007, Hayasaki2009, Cuadra2009, Lodato2009, Roedig2012, Roedig2014}). Other authors adopt more realistic stellar dynamical models (non-spherical, rotating galaxies; e.g., \citealt{Yu2002, Merritt2004, Khan2013, Vasiliev2015}) to show that coalescence of the binary is possible within a Hubble time.

Finding SBHBs at close separations (< 1 pc, the slowest phase of SBHB evolution, according to \citealt{Begelman1980}) will serve as an important test for both galaxy evolution models and stellar and gas dynamical models for SBHB orbital decay. During the late stages of their evolution, SBHBs are also sources of low-frequency (nHz to mHz) gravitational waves detectable by pulsar timing arrays \citep[PTAs; e.g.,][and references therein]{Manchester2013,Kramer2013,McLaughlin2013} and the upcoming Laser Interferometer Space Antenna \citep[LISA; e.g.,][and references therein]{Amaro2017,Danzmann2017}. Therefore, a great deal of effort has been directed recently towards observational searches of such objects.

Only widely separated active BHs \citep[such as NGC~6240, e.g.,][]{Komossa2003} can be resolved by direct imaging.  Recent spectroscopic surveys and follow-up imaging studies (e.g., \citealt{Comerford2009a, Comerford2009b, Liu2010a, Liu2010b, Shen2011, Smith2010, Fu2011}) have discovered a significant number of AGNs with offset or {\it narrow}, double-peaked [O~III] emission lines, indicative of dual active nuclei in merging galaxies, typically with separation of a few kiloparsec and presumably not bound to each other. CSO~0402+379 (or 4C~37.11), with a separation of approximately 7~pc \citep[studied by radio interferometry][]{Rodriguez2006, Rodriguez2009, Bansal2017}, is the best candidate for a resolved and presumably bound binary known currently.

Bound SBHBs {\it at sub-parsec separations} cannot be spatially resolved with present methods, hence such close SBHB candidates are found on the basis of indirect evidence such as regular photometric variability or emission-line radial velocity curves suggestive of orbital motion. The candidate with the longest record of photometric monitoring is OJ287, whose long-term light curve shows regular outbursts at 12-year intervals \citep[e.g,][]{Valtonen2008}. Recently, \cite{Graham2015} noted that PG~1302$-$102 could be a close SBHB, due to the periodic behaviour of the light curve with an inferred period of a few years. Many additional examples of such behaviour have been reported by \cite{Liu2015}, \cite{Graham2015b} and \cite{Charisi2016}. However, \cite{Vaughan2016} have pointed out that the typical, stochastic quasar variability could appear periodic for a few cycles \citep[see also][]{Liu2016b}.
 
Radial velocity searches for SBHBs begin with the detection of displaced emission-line peaks from the broad-line region (BLR) i.e. gas that is bound to each individual BH and follows it in its orbit. The rationale is that orbital motion of the binary will cause the emission line peaks to shift periodically, resulting in displaced single- or double-peaked profiles, depending on whether one or both of the BHs in the SBHB have a BLR associated with it \citep{Begelman1980,Komberg1968}. \cite{Gaskell1984} noted that the two quasars 3C~227 and Mrk~668 had broad H$\beta$ emission lines shifted by 2,000--3,000$\;\mathrm{km\; s^{-1}}$, and suggested that they could be examples of SBHBs where only one BLR is observable. This interpretation fits in well with the scenario of \cite{Begelman1980}, where the BHs in the SBHBs spend most of their lifetimes at separations of 0.1--0.01~pc and periods $\sim 10^2\;$yr, which would lead to broad line velocity displacement on the order of $10^3\;\mathrm{km\; s^{-1}}$. But testing this hypothesis further is made difficult by the long orbital periods. Therefore, work to date in this direction has focused on using the radial velocity curves of the displaced peaks to obtain a {\it lower limit} on the total mass of the SBHB and using that limit to evaluate the plausibility of the SBHB hypothesis. The monitoring campaigns to date have targeted quasars with single and double displaced broad lines using similar techniques but time series of different durations for the two different families.

Quasars with double-peaked broad Balmer lines  (see examples in Figure~\ref{fig:sample}; hereafter, double-peaked emitters) have been known and studied for some time \citep[e.g.,][and references therein]{Eracleous1994,Eracleous2003,Strateva2003}. The SBHB hypothesis is not the only hypothesis for their unusual line profiles. Other proposed explanations include emission from an accretion disc, a bipolar outflow, and an anisotropically-illuminated BLR \citep[see summary and critique in][and references therein]{Eracleous2009}. \cite{Halpern1988} tested the SBHB hypothesis using a 5-year radial velocity curve of Arp~102B but did not detect any change in the radial velocity. \cite{Gaskell1996} noted that the failure to detect such a change could be explained by large BH masses, and proposed that 3C~390.3 is also a SBHB in which both BLRs are visible. The observed  drift of the displaced peaks in the spectrum of 3C~390.3 between 1968 and 1988 was consistent with this hypothesis, and a sinusoidal model fit to the data suggested a period of 300~yr and a total mass of $\sim 7.7 \times 10^9 \; \Msol $. However, the sinusoidal trend did not continue past 1990, and \cite{Eracleous1997} rejected the SBHB interpretation of 3C~390.3, together with two other double-peaked emitters, 3C~332 and Arp~102B. The main argument was that the lower limits on the total masses, based on the best-fitting orbital parameters, were in excess of $ 10^{10} \; \Msol $, and such large SBHB masses are difficult to reconcile with other observations. \cite{Liu2016} followed up with an examination of the SBHB hypothesis for 13 double-peaked emitters, many of which had 10--20~yr of observational data. They adopted a circular orbit model to fit the observed radial velocity curves, and derived lower limits on the total SBHB masses using the resulting best-fitting orbital parameters. The lower limits effectively ruled out the SBHB hypothesis for half of the targets, since the minimum masses are even greater than the most massive BHs measured so far, approximately $ 2 \times 10^{10} \; \Msol $ \citep{McConnell2012, Thomas2016}. \cite{Inayoshi2016} also argue that BHs are prevented from growing above a few$\,\times 10^{10} \; \Msol $ by small-scale accretion physics.

\begin{table}
\begin{minipage}{0.9\linewidth}
\renewcommand{\thefootnote}{\alph{footnote}}
\caption{List of Objects and Sources of Data}
\label{tab:targets}
\begin{center}
\begin{tabular}{lccl}
\hline
\hline
       &                     & Data Span &  \\
Object & $z$\,\footnotemark[1] & (years)   & References\,\footnotemark[2] \\
\hline
3C~59           & 0.1096 & 22 & 1, 2, 3 \\
PKS~0235+023    & 0.2072 & 24 & 1, 2 \\
NGC~1097	& 0.0042 & 21 & 4, 5\\
1E 0450$-$1817  & 0.0616 & 23 & 1, 3\\
Pictor~A        & 0.0350 & 15 & 1, 3, 6\\
CBS~74          & 0.0919 & 14 & 1, 3, 7 \\
PKS~0921$-$213  & 0.0531 & 16 & 1, 3 \\
PKS~1020$-$103  & 0.1965 & 12 & 3\\
3C~227          & 0.0860 & 35 & 1, 2, 7, 8, 9 \\
Mrk~668         & 0.0768 & 24 & 2, 8, 10\\
3C~332          & 0.1510 & 25 & 1, 2, 7 \\
Arp~102B        & 0.0244 & 32 & 1, 8 \\
PKS~1739+184    & 0.1859 & 23 & 1, 3 \\
3C~390.3        & 0.0555 & 41 & 2, 8, 11\\
\hline
\end{tabular}
\end{center}
\footnotetext[1]{Redshifts from \citet{Eracleous2004}, except for NGC~1097, whose redshift was measured from the spectra of \cite{Storchi-Bergmann1993}.}
\footnotetext[2]{{\it References to sources of data:} (1) New observations presented in this paper, (2) \cite{Gezari2007}, (3)\cite{Lewis2010}, (4) \cite{Storchi-Bergmann2003} , (5) \cite{Schimoia2015}, (6) \cite{Gaskell1996}, (7) SDSS DR2, (8) \cite{Liu2016}, (9) \cite{Osterbrock1976}, (10) \cite{Shapovalova2013}, (11) \cite{Popovic2014}}
\end{minipage}
\end{table}

Recently, large systematic searches for spectroscopic signatures of SBHBs have been made possible thanks to large sky surveys such as the Sloan Digital Sky Survey (SDSS). The data from these surveys have yielded $\sim 100$ candidates with single-peaked Balmer lines that have large velocity displacements \citep{Tsalmantza2011, Eracleous2012, Decarli2013, Shen2013, Liu2014, Runnoe2015, Runnoe2017, Guo2019}. In the same spirit as the work on double-peaked emitters, these authors have attempted to assess the plausibility of the candidates by obtaining lower limits on the hypothesised SBHB masses. However, the available radial velocity curves are short and sparsely sampled, leading so far to mass limits that are not restrictive. 

The analysis in previous works assumed circular orbits, motivated by the assessment of \cite{Begelman1980} that frictional drag will erase any eccentricity in the orbit. However, more recent simulations have shown that the eccentricity can be very sensitive to the initial conditions of the binary, as well as the properties of the surrounding stellar cluster \citep{Khan2012}, and there are a variety of orbital solutions that are highly eccentric at 0.1--0.01~pc separations. The eccentricity of the SBHB can also increase dramatically in counter-rotating systems (e.g., \citealt{Amaro2010, Sesana2011, Wang2014, Bockelmann2015}). Simulations also show that the SBHB can acquire a substantial eccentricity over time via interactions with inhomogeneities (i.e., dense clumps) in a circumbinary disc \citep[e.g.,][]{Roedig2011, Fiacconi2013, delValle2015}. Therefore, the goal of this work is to re-examine the binary hypothesis assuming more general, eccentric orbits by employing a more sophisticated fitting method than in our previous work. The introduction of additional model parameters (eccentricity, $e$, and argument of periapsis, $\omega$) will significantly enlarge the parameter space, and the wide variety of possible behaviours of the radial velocity curves could provide better fits for many of the targets. More importantly, using eccentric orbits may result in lower limits on the orbital period, hence the total mass of the hypothesized SBHBs and change the conclusions of previous studies. We utilise Markov Chain Monte Carlo (MCMC) methods to derive the entire probability distributions of all model parameters, instead of a single best-fitting solution (see Section~\ref{sec:simulation}). Thus, we are able to derive rigorous lower limits on the total masses from all possible solutions. This is an improvement compared to the lower limits derived by previous works that considered only deviations from the best-fitting models. Another new feature of our improved analysis is that it takes into account radial velocity jitter. Our targets include the 13 double-peaked emitters in \cite{Liu2016} plus another object, NGC~1097, which has similar Balmer line profiles and radial velocity curves spanning two decades. Half of the targets also have additional observations since \cite{Liu2016}, which serve to better constrain the models.

In addition to addressing the SBHB hypothesis, this work are also relevant to the broader question of the origin of double-peaked broad emission lines observed in some quasars. Since a number of alternative explanations have been considered \citep[see][and Section~\ref{sec:discussion} of this paper]{Eracleous2009}, testing and, potentially, disfavouring any one of the scenarios represents significant progress.

In Section~\ref{sec:observations} of this paper, we describe the data we use, including previously published data and new observations. In Section~\ref{sec:jitter}, we study radial velocity jitter and quantify its magnitude empirically. In Section~\ref{sec:fitting}, we describe our eccentric orbit model, our tests of the MCMC code, and our approach for incorporating jitter in the analysis. We then present our findings in Section~\ref{sec:results} and discuss the results in Section~\ref{sec:discussion}.

\begin{table*}

\begin{minipage}{0.9\linewidth}
\renewcommand{\thefootnote}{\alph{footnote}}
\caption{New Observations and Radial Velocity Measurements}\label{tab:observations}
\begin{center}
\begin{tabular}{lllcc}
\hline
\hline
       & Observation & Telescope and &  $u_b$ (Blue) & $u_r$ (Red)\\
Object & Date (UT)   & Instrument\,\footnotemark[2] & ($10^3$ km s$^{-1}$)   & ($10^3$ km s$^{-1}$)  \\
\hline
3C~59          & 2011/01/01 & HET+LRS & $-1.2 \pm 0.3$ & $4.8 \pm 0.2$ \\
               & 2011/08/28 & HET+LRS & \dots & $4.8 \pm 0.1$ \\
               & 2011/10/25 & HET+LRS & \dots & $4.7 \pm 0.1$ \\
               & 2012/08/11 & HET+LRS &  $-0.7 \pm 0.3$ & $4.8 \pm 0.5$\\
               & 2014/08/29 & ARC 3.5m+DIS & $-0.6 \pm 0.2$ & $4.7 \pm 0.4$ \\
 & & \\
PKS~0235+023   & 2011/01/27 & HET+LRS & $-4.6 \pm 0.1 $ & \dots \\     
               & 2011/08/27 & HET+LRS & $-4.7 \pm 0.1 $ & \dots \\     
               & 2011/10/26 & HET+LRS & $-4.7 \pm 0.1 $ & \dots \\     
               & 2014/08/29 & ARC 3.5m+DIS & $-4.0 \pm 0.1 $ & \dots \\
 & & \\
1E 0450$-$1817 & 2017/12/13 & MDM 2.4m+OSMOS &  $-5.4 \pm 0.8 $ &  $4.2 \pm 0.6 $ \\
& & \\
Pictor~A       & 2016/10/30 & Magellan Baade+MagE & $-5.0 \pm 2 $ & $4.0 \pm 2$ \\  
 & & \\
CBS~74         & 2011/01/01 & HET+LRS &  $-1.2 \pm 0.2 $ & \dots \\
               & 2011/10/25 & HET+LRS & \dots & $2.5 \pm 0.2$ \\
 & & \\
PKS~0921$-$213\,\footnotemark[1] & 2009/11/20 & CTIO 1.5m+CSPEC & $-3.1 \pm 0.1 $ & $3.0 \pm 0.2$ \\
               & 2010/03/18 & CTIO 1.5m+CSPEC & $-3.2 \pm 0.1 $ & $3.1 \pm 0.1$ \\
               & 2011/02/12 & CTIO 1.5m+CSPEC & $-2.7 \pm 0.1 $ & $3.0 \pm 0.1$ \\
 & & \\
3C~227         & 2010/11/08 & HET+LRS &  $-2.6 \pm 0.3 $ &\dots \\
               & 2011/11/06 & HET+LRS &  $-2.9 \pm 0.1 $ &\dots \\
               & 2012/03/22 & HET+LRS &  $-2.4 \pm 0.1 $ &\dots \\
               & 2013/05/08 & HET+LRS \\
 & & \\
Arp~102B       & 2011/02/05 & HET+LRS      &  $-5.6 \pm 0.1 $   &\dots  \\
	       & 2011/10/15 & HET+LRS      &  $-5.4 \pm 0.1 $ &\dots  \\
               & 2012/04/20 & HET+LRS      &  $-5.6 \pm 0.1 $ &\dots  \\
               & 2013/05/07 & HET+LRS      &  $-5.5 \pm 0.1 $ &\dots  \\
               & 2014/05/27 & APO 3.5m+DIS &  $-5.8 \pm 0.1 $   &\dots  \\         
 & & \\
3C~332         & 2011/01/30 & HET+LRS &  $-5.1 \pm 0.1 $ & $4.3 \pm 0.1$     \\     
               & 2011/05/21 & HET+LRS &  $-5.3 \pm 0.1 $ & $4.11 \pm 0.1$  \\     
               & 2012/04/15 & HET+LRS &  $-5.7 \pm 0.1 $ & $3.86 \pm 0.1$ \\     
               & 2014/05/05 & ARC 3.5m+DIS &  $-4.4 \pm 0.1 $ & \dots \\
 & & \\
PKS~1739+184   & 2011/05/23 & HET+LRS       &  $-1.9 \pm 0.5 $ & $3.3 \pm 0.1$  \\
               & 2012/09/18 & HET+LRS       &  \dots & $2.9 \pm 0.2$            \\
               & 2013/05/08 & HET+LRS       &  $-2.2\pm 0.5 $ & $3.1 \pm 0.1$ \\
               & 2014/05/27 & APO 3.5m+DIS  &  $-2.6 \pm 0.1 $ & $2.3 \pm 0.2$  \\
\hline
\end{tabular}
\end{center}
\footnotetext[1]{The observations of PKS~0921$-$213 were part of two
  intensive monitoring campaigns carried out during
  2009/11/07--2010/05/29 and 2011/01/01--2011/06/27. As we explain in
  Section~\ref{sec:observations}, we report here selected measurements
  from data taken in those two campaigns.}
\footnotetext[2]{See Section~\ref{sec:observations} for details of the telescopes and instruments.}
\end{minipage}
\end{table*}

\section{Previously Published Data and New Observations}\label{sec:observations}

We compiled radial velocity measurements for 14 AGNs from \citet{Eracleous1997}, \citet{Gezari2007}, \citet{Lewis2010}, \citet{Storchi-Bergmann2003}, \citet{Schimoia2012, Schimoia2015}, and \citet{Liu2016}. We supplemented the data collected from the literature with new measurements from spectra obtained from 2009 to 2017, as we describe in detail below. The objects included in this study are listed in Table~\ref{tab:targets} along with their redshifts, the span of the radial velocity measurements, and the sources of the data. Typically, the length of the monitoring period for each object is 20--30~yr.

\begin{figure*}
  \centerline{
    \includegraphics[width=.45\linewidth]{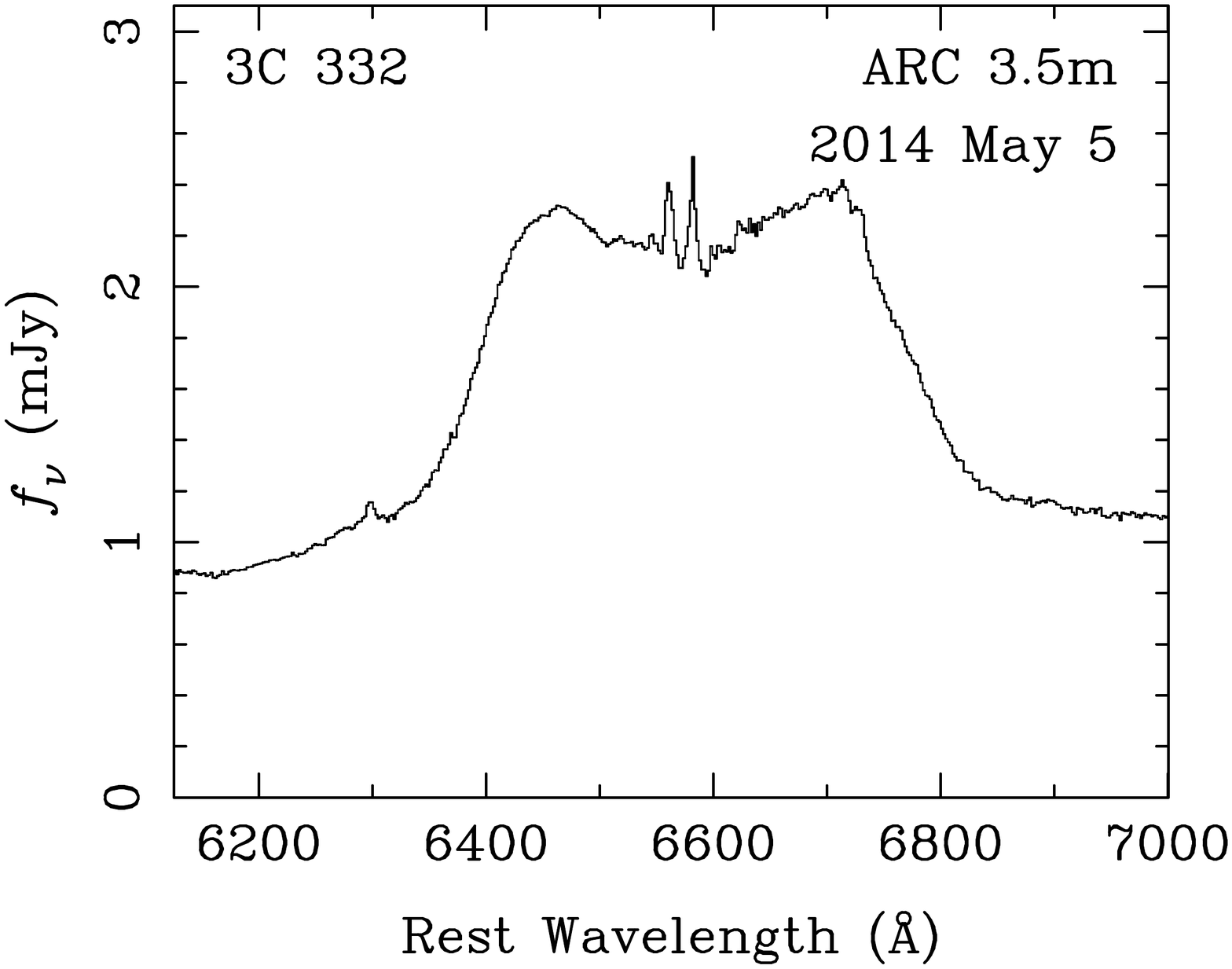}
    \includegraphics[width=.45\linewidth]{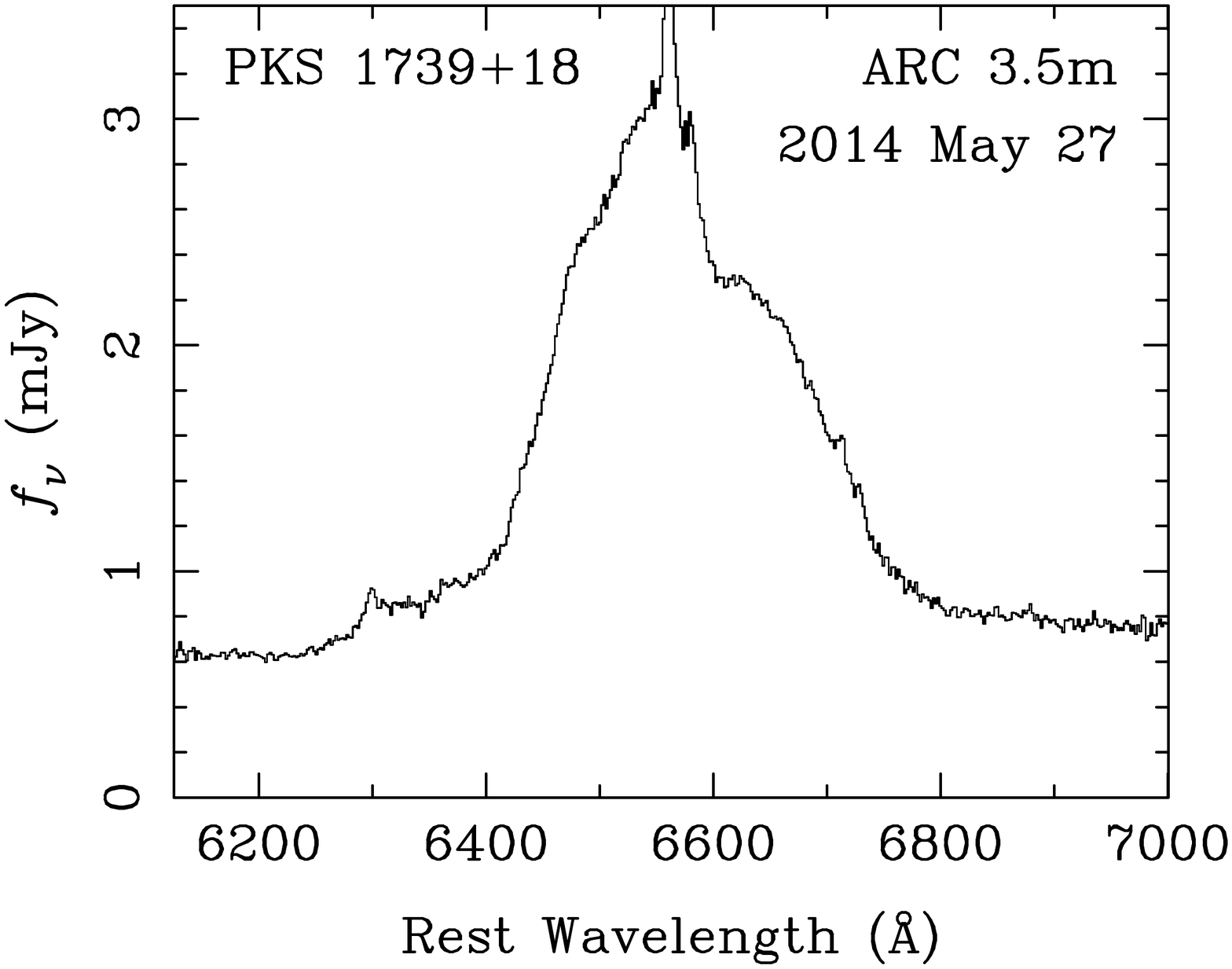}
  }
  \centerline{
    \includegraphics[width=.45\linewidth]{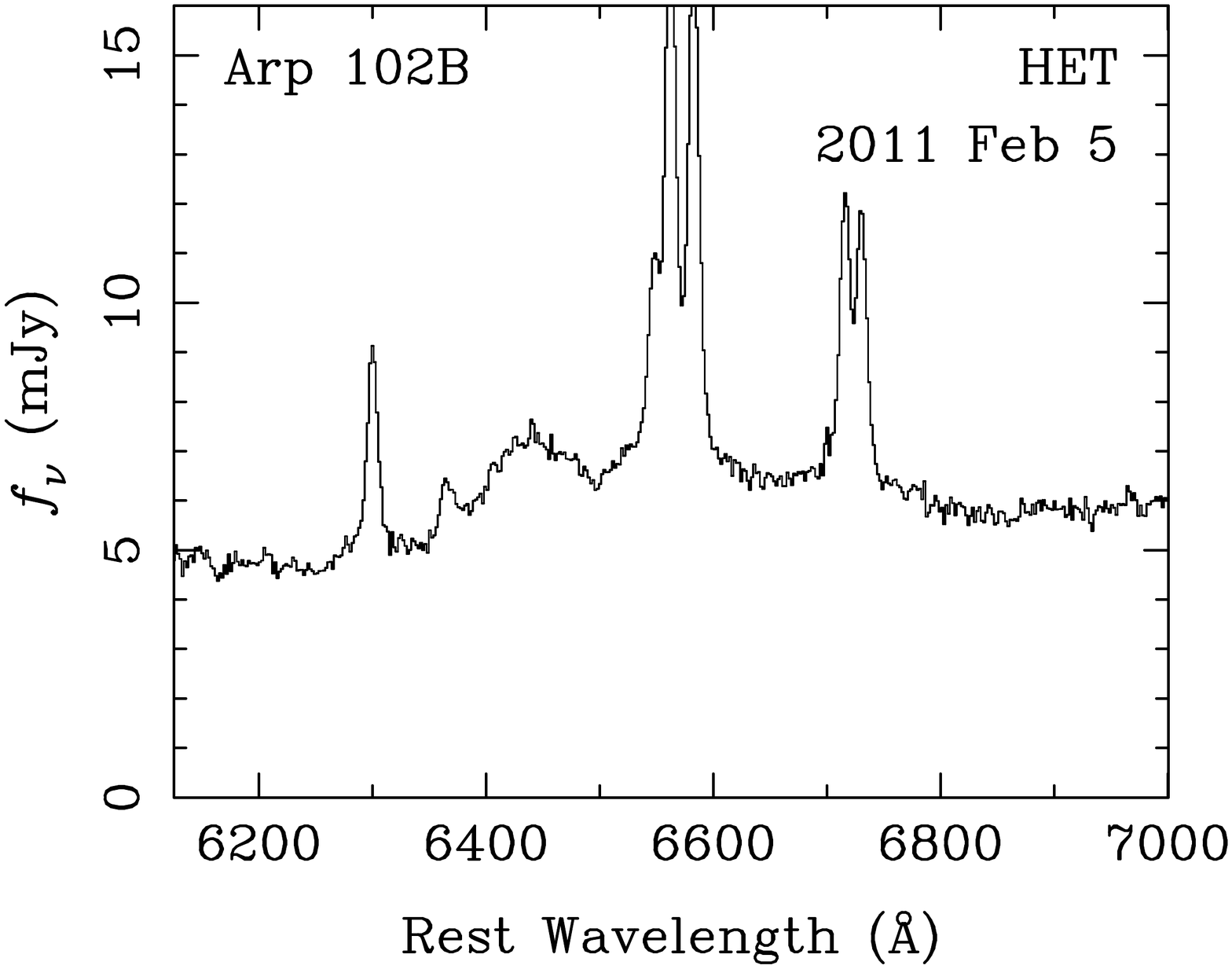}
    \includegraphics[width=.45\linewidth]{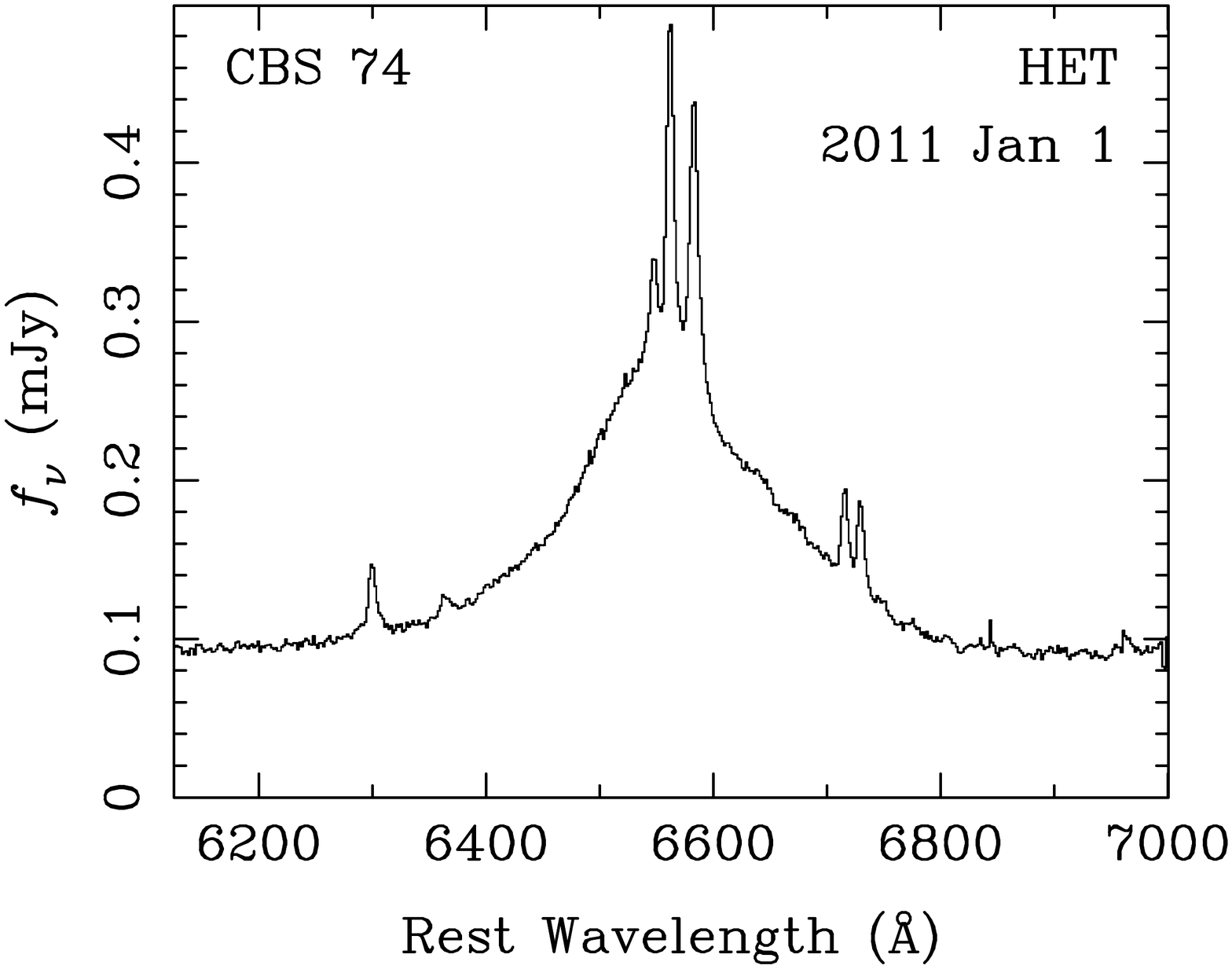}
  }
\caption{Sample spectra taken with the HET and the ARC 3.5m illustrating a variety of line profiles of objects in our sample. The log of new observations is given in Table~\ref{tab:observations} and the instruments and data reductions are described in Section~\ref{sec:observations} of the text.}
\label{fig:sample}
\end{figure*}

All the new spectra presented here cover the H$\alpha$ lines of the respective targets. Examples of spectra are shown in Figure~\ref{fig:sample}. Most of the spectra were taken with the Low-Resolution Spectrograph (LRS) on the Hobby-Eberly Telescope (HET) between 2010 and 2013, and with the red channel of the Double Imaging Spectrograph (DIS) of the Astrophysical Research Consortium (ARC) 3.5m telescope at Apache Point Observatory in 2014. PKS~0921$-$213 was observed with the Cassegrain Spectrograph (CSPEC) on the 1.5m telescope at Cerro Tololo Interamerical Observatory (CTIO) as part of two intensive monitoring campaigns, at a rate of approximately once every two weeks. The first campaign lasted from November 2009 to May 2010 and the second one from February to June 2011. The spectrum of Pictor~A was taken with the Magellan Echellette Spectrograph \citep[MagE;][]{Marshall2008} on the Magellan Baade telescope in 2016 and the spectrum of 1E~0450.3-1817 was taken with the Ohio State Multi-Object Spectrograph \citep[OSMOS;][]{Martini2011} on the 2.4m Hiltner telescope at the MDM Observatory in 2018. The log of new observations is given in Table~\ref{tab:observations}.

The HET+LRS spectra were taken with the G3 grism through a 1\farcs0 or 1\farcs5 slit. They cover the wavelength range 6250--9100~\AA\ with a resolution of 4.2~\AA\ when using the  1\farcs0 slit and or 6.3~\AA\ when using the 1\farcs5 slit. The ARC 3.5m+DIS red-channel spectra were taken with the R300 grating through a 1\farcs5 slit. They cover the spectral range 5150--9850~\AA\ with a spectral resolution of 6.0~\AA. The CTIO 1.5m+CSPEC spectra of PKS~0921$-$213 were taken with grating 35 through a 1\farcs8 slit. They cover the wavelength range 6235--7467~\AA\ with a spectral resolution of 3.5~\AA. The spectrum of 1E~0450.3-1817 was taken with the red VPH grism through a 1\farcs2 slit. It covers the wavelength range 5263--10888~\AA\ with a resolution of 3.6~\AA. The spectrum of Pictor~A was taken through a 1\farcs0 slit and covers the wavelength range 3060-8288~\AA\ in orders 8–20 with a spectral resolution of 0.24--0.57~\AA.   The H$\alpha$ line fell in the 9th order where the spectral resolution was 0.51~\AA.

All the spectra, except for the Magellan spectrum of Pictor A, were reduced and calibrated using common procedures for long-slit spectra, as described in Section~5.1 of \citet{Eracleous2012}. The spectrum of Pictor~A was reduced with the MagE pipeline available in {\tt CarPy} \citep{Kelson2000,Kelson2003}. The first steps included bias subtraction, flat-field division, sky subtraction, scattered light subtraction, and extraction of raw spectra. The wavelength calibration was based on ThAr arc lamp spectra. The derivation of the wavelength scale used 910 arc lines over the entire spectrum, including 60 in the same order as H$\alpha$. The root-mean-squared residuals of the polynomial fit to the wavelength-pixel number relation were less than 0.1 pixel.

To measure the velocity of a peak in the new H$\alpha$ line profiles we fitted the region around it with a Gaussian in order to get the centroid. For this exercise selected fitting windows encompassing each peak that were 50--100~\AA\ wide. We repeated the measurement 500 times, varying our choice of fitting window, and took the average to be the best estimate of the peak wavelength and the standard deviation of all the measurements to be the measurement uncertainty. In spectra where the red or blue peak are difficult to discern by eye (e.g., PKS~1739+184 and CBS~74 in Figure~\ref{fig:sample}), we performed a double Gaussian fit over both regions and determined the two centres simultaneously. 

In the specific case of PKS~0921$-$213, we measure the velocities from all spectra, but only include a subset of the measurements, as we are mainly interested in the long-term behaviour of the radial velocity curve. In doing so, we avoid the effects of short-term jitter (on time-scales of months or less), which could be caused by incorporating numerous data points within a short observing window; long-term velocity variations, on time-scales of years or more, are meant to be described by our orbital model, which we describe in \S\ref{sec:jitter}). We follow the same approach in order to select data for the radial velocity curve of NGC~1097, which was also sampled very frequently. The new velocity measurements are given in Table~\ref{tab:observations}. The radial velocity curves are presented and discussed in Section~\ref{sec:results}.

\section{Velocity Jitter}\label{sec:jitter}

Before we fit the binary model over the observed radial velocity curves, we need to take into account {\it radial velocity jitter}, which we define as small-amplitude fluctuations in the radial velocity curve on time-scales of order a year or less. Jitter can be caused by intrinsic variations in one single BLR that occur on the light-crossing or dynamical time-scale. This view of jitter is bolstered by the values of these time-scales, which we estimate below. Additional corroboration comes from reverberation mapping campaigns that have observed such radial velocity jitter in response to fluctuations in the continuum. Analysis of the data from such campaigns indicates that the velocities of the peaks of the broad Balmer lines fluctuate by a few hundred $\mathrm{km\; s^{-1}}$ over time intervals of 1--6 months (see \citealt{Barth2015} and \citealt{Guo2019}).

To evaluate the time-scales on which we could expect radial velocity jitter in our targets, we estimate the light-crossing and dynamical time-scales of a single broad-line region. We begin with the {\it characteristic} size of the BLR from the empirical radius-luminosity relationship by \cite{Bentz2013}, which we re-cast as:
\begin{equation}
  \log\left({R_{BLR}\over 10\;\textrm{lt-day}}\right) = 0.527 + 0.533 \log\left[\lambda L_{\lambda}(5100\,\textrm{\AA})\over10^{44}\;\textrm{erg\,s}^{-1}\right] \; .
  \label{eq:rblr}
\end{equation}
In the above equation, $\lambda L_{\lambda}(5100\,\textrm{\AA})$ is the monochromatic continuum luminosity at a rest-frame wavelength of 5100~\AA, which we estimate from the $V$-band magnitude of each object. For the range of $\lambda L_{\lambda}(5100\,\textrm{\AA})$ for our objects, $4.3\times 10^{43}$--$6.7\times 10^{44}\;\textrm{erg\;s}^{-1}$, we obtain light-crossing times (i.e., values of $R_{BLR}/c$) of 21--92~days by direct application of equation~(\ref{eq:rblr}). But we must also consider that the observed $V$-band magnitude of an object represents the sum of the luminosities of the two accreting BHs, which we take to be comparable since the strengths of the peaks of the double-peaked Balmer lines are similar. Therefore, the value of $\lambda L_{\lambda}(5100\,\textrm{\AA})$ that enters in equation~(\ref{eq:rblr}) should be a factor of 2 smaller and the resulting value of $R_{BLR}$ of each individual BH should be approximately 1.45 times smaller, leading to a range of light-crossing times of 14--64 days. Moreover, the value of $R_{BLR}$ is likely to be even smaller than that obtained above, as the BLR around each BH would be truncated by tides from the binary companion \citep[see discussion in Section~4 of][and Section~3.1 of \citealt{Nguyen2016}]{Runnoe2017}.

The dynamical time-scale at $R_{BLR}$ can then be estimated from
\begin{equation}
  t_{dyn}(R_{BLR}) = 1.14\,\left(R_{BLR}\over 10\;\textrm{lt-day}\right)^{3/2}
  \left(M_\bullet\over 10^8\,\Msol\right)^{-1/2}\, \textrm{yr,}
  \label{eq:tdyn}
\end{equation}
where $M_\bullet$ is the mass of the BH. To get the range of $t_{dyn}(R_{BLR})$ appropriate for our targets we explore a range of BH masses: we set a lower mass limit by assuming a maximum Eddington ratio of 0.1 and adopt an upper limit of $10^{10} \; \Msol$. Equation~(\ref{eq:tdyn}) then yields $t_{dyn}$ between 8 months and 7~yr for all our targets, before considering tidal truncation of the BLR. Taking tidal truncation into account, we conclude that $t_{dyn}$ is between a few months and a few years and comparable to or slightly longer than the typical time intervals between our observations. This suggests that reverberation or dynamical processes in a single BLR are plausible causes of jitter in the radial velocity curves of our targets.

With the above considerations in mind it is plausible to separate velocity variations into two broad families, the (fast) jitter observed on time-scales of order a year or less and slower variations observed on time-scales of several years or more. The latter family includes the slow variations or undulations that we aim to use to test the SBHB hypothesis. The former family is a source of noise for our purposes so we seek ways of characterising its properties and dealing with its effects while we try to fit orbital models to the slow variations. In the remainder of this section we describe how we characterise jitter empirically for our particular objects and in Section~\ref{sec:fitting} we describe the statistical methods we use to deal with the effects of jitter while fitting models to the radial velocity curves. 

\begin{figure*}
  \hbox to 15cm{
   \hskip -5cm
    \vbox{
      \centerline{\includegraphics[height=5.8cm]{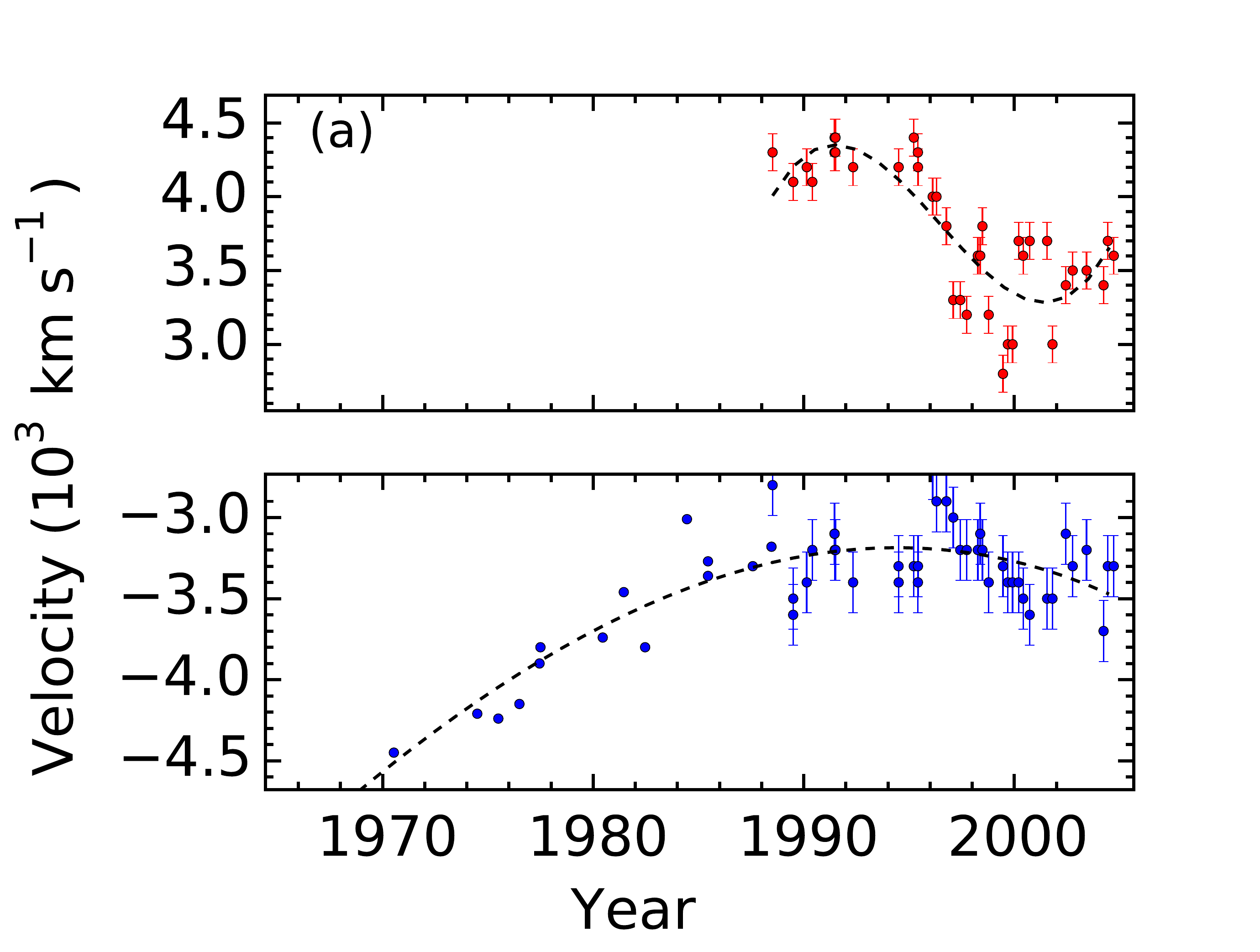}}
      \centerline{\includegraphics[height=5.8cm]{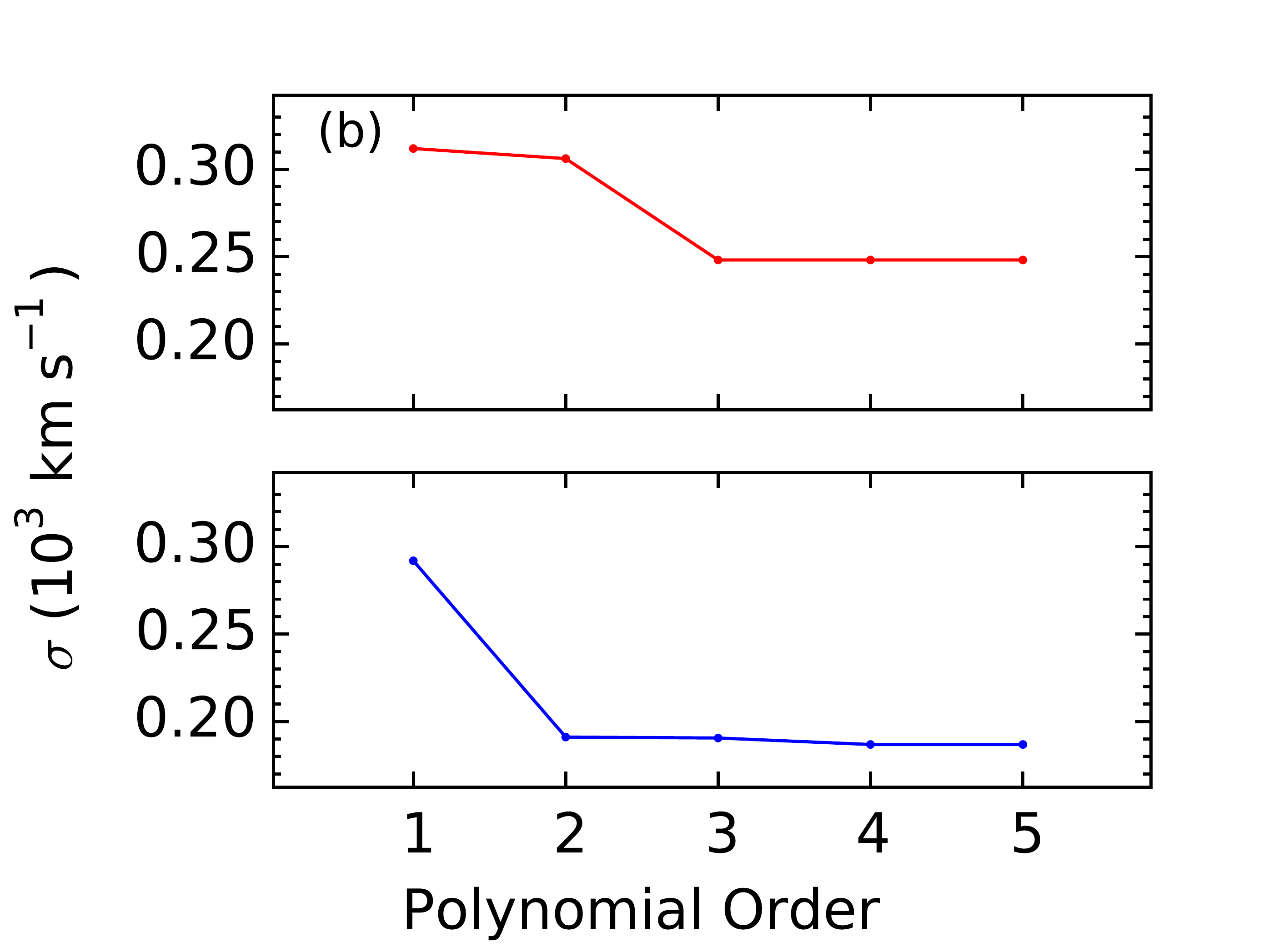}}
      }
    \hskip -4cm
    \includegraphics[height=12.4cm]{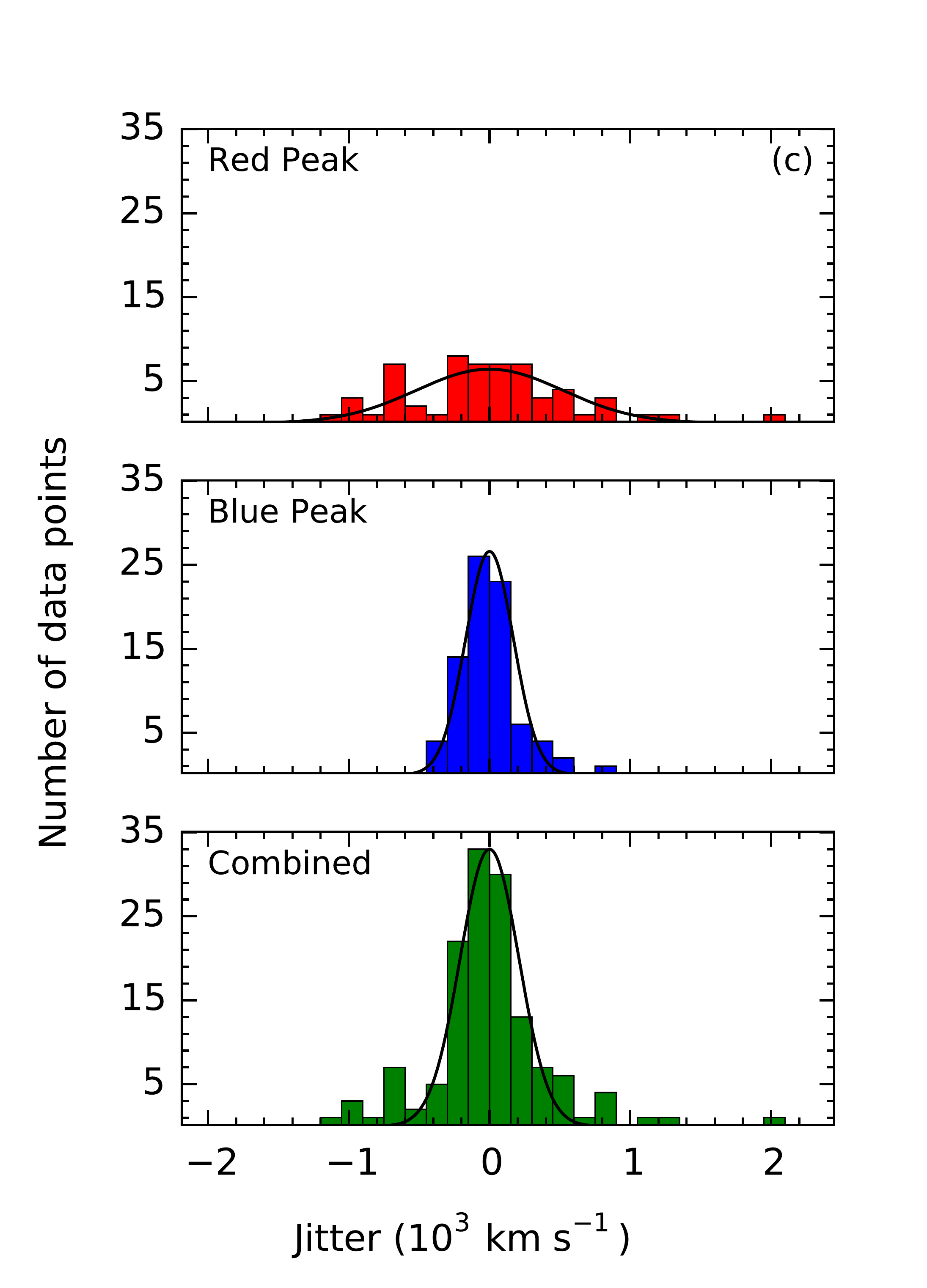}
    
  } 
\caption{Illustration of the method used to determine the jitter properties, described in Section~\ref{sec:jitter} of the text. (a) Fits to the radial velocity curves of the red and blue peaks of the H$\alpha$ line of 3C~390.3 with 3rd-order polynomials. The scatter of the data about the fits makes up jitter that is discussed in Section~\ref{sec:jitter}. (b) The standard deviation of jitter as a function of polynomial order for the red and blue radial velocity curves of 3C~390.3. There is no significant change in the standard deviation when the polynomial order increases above three.  (c) The corresponding jitter (i.e., residual) distribution and Gaussian fit, showing that the Gaussian is a good description of the observed distribution.}
\label{fig:jitter}
\end{figure*}

To characterise jitter empirically, we fit the radial velocity curves with a low-order polynomial, and ascribe the departure of the data points from the polynomial fit to jitter. The polynomial is intended to follow the undulations of the radial velocity curves on times scales longer than the dynamical time-scale of the BLR. After experimenting with fits to the well-sampled radial velocity curves of a few objects we find that polynomials of third order or higher do indeed accomplish this goal. The standard deviation of the data about the best fit decreases as we increase the polynomial order but only up to third order; as we increase the polynomial order further, the standard deviation does not decrease. As an illustration of the method, we plot the standard deviation of the fit residuals (the jitter) for the red and blue peaks of 3C~390.3 as a function of the polynomial order in Figure~\ref{fig:jitter}. The jitter does not change significantly for either peak once the polynomial order reaches three. We thus choose to fit the velocity curves of all our targets with a third order polynomial as it can sufficiently describe the long-term shape of the radial velocity curve. Moreover, the polynomial has the ability to fit the undulations of the radial velocity curve without mimicking the orbital model that we wish to fit eventually with the result that the jitter that we determine is not equivalent to the deviations measured relative to the orbital model. We show an example of a 3rd-order polynomial fit to the radial velocity curves of 3C~390.3 in Figure~\ref{fig:jitter}.

\begin{table}
\begin{minipage}{0.9\linewidth}
\renewcommand{\thefootnote}{\alph{footnote}}
\caption{Jitter Amplitudes}
\label{tab:jitter}
\begin{center}
\begin{tabular}{lcc}
\noalign{\vskip 4pt}
\hline
\hline
\noalign{\vskip 2 pt}
& \multicolumn{2}{c}{Jitter (km~s$^{-1}$)\,\footnotemark[1]} \\
\noalign{\vskip -6 pt}
& \multicolumn{2}{c}{\hrulefill} \\
Object & Red Peak  & Blue Peak \\
\hline
3C 332           & 420 & 640 \\
1E 0450.3$-$1817 & 590 & 610 \\
Arp 102B         & 160 & 470 \\
3C 59            & 250 & 110 \\
CBS 74           & 190 & 310 \\
PKS 1020$-$103   & 220 & 080 \\
PKS 0921$-$213   & 170 & 160 \\
3C 390.3         & 210 & 580 \\
PKS 1739+184     & 480 & 230 \\
Pictor A         & 470 & 510 \\
Mrk 668          & 190 & 490 \\
NGC 1097         & 230 & 280 \\
PKS 0235+023     & 280 & 310 \\
3C 227           & 330 & 520 \\
\hline
\end{tabular}
\end{center}
\footnotetext[1]{Standard deviation of the radial velocity values relative to a 3rd order polynomial fit to the radial velocity curve of each peak. See details in Section~\ref{sec:jitter}.}
\end{minipage}
\end{table}

We find that the jitter distribution for the well-sampled radial velocity curves of 3C~390.3 and Arp~102B is well represented by a Gaussian. This is also the case for the combined jitter distribution for all objects in our sample. Typically, the magnitude of jitter (i.e., the standard deviation of its distribution) is of order a few hundred $\;\mathrm{km\; s^{-1}}$, i.e., a factor of 2--3 greater than the measurement uncertainty. These findings agree with the conclusions of \citet{Guo2019} on the shape and width of the jitter distribution caused by reverberation and they agree with the magnitude of jitter measured in specific objects by \citet{Barth2015}. Since the jitter distribution is Gaussian we treat jitter in the fitting process described in the next section by adding it in quadrature with the uncertainty in individual measurements, effectively reducing the value of $\chi^2$ and resulting in a more conservative lower limit on the SBHB masses (see Section~\ref{sec:fitting}). We also model jitter as a Gaussian process, as we describe in detail in Section~\ref{sec:gaussproc}, in order to test the validity of the simple approach and we get similar results.

In view of the above results, we henceforth make the assumption that the jitter properties are similar among the objects in our sample i.e. that the distribution can be described by a Gaussian. Since some targets (e.g., Pictor~A, 3C~227, PKS~1020$-$103) have sparsely-sampled radial velocity curves, this approach is straightforward to implement in practice: by fitting the radial velocity curves with a cubic polynomial and finding the standard deviation of the residuals, we determine the magnitude of the jitter for that object. The jitter values measured for each object in this manner are listed in Table~\ref{tab:jitter}. We also expect that the long-term radial velocity variations (variations on time-scales much longer than the time interval between successive measurements) are taken out by the polynomial fit and that the jitter that we measure is not influenced by any of the long-term variations that can be interpreted as the result of orbital motion.

\section{Methodology for Fitting Radial Velocity Curves}\label{sec:fitting}

\subsection{Orbital Model}
\label{sec:model} 

We adopt a model of eccentric binary orbits, in contrast to \cite{Liu2016} and earlier works that adopted circular orbits, in order to explore the full parameter space of possible solutions. Thus, our model contains six independent parameters: the (projected) radial velocity at pericentre of the object that produces the red peak (the receding object), $u_{r,pe}$, the orbital period, $P$, the epoch/time of pericentre passage, the phase offset, $t_0$ (defined in equation~\ref{eq:eccanom}), the mass ratio, $q$, the eccentricity, $e$, and the argument of periapsis, $\omega$. For convenience, we define $q \equiv M_r/M_b $, where $M$ is the mass of one of the BHs and the subscripts `$r$' and `$b$' denote the object producing the red and blue peak, respectively (i.e., the receding and approaching object, respectively). With this definition, $q$ can be larger or smaller than unity depending on which peak corresponds to the primary (more massive) BH.

To calculate $u_r$ and $u_b$ in the model, we begin with the radial velocity equation from Keplerian dynamics:
\begin{equation}
  u_r = \left[\frac{GM_{tot}}{a(1 - e^2)}\right]^{1/2} {\sin{\textit{i}}\over 1+q}\, \left[ \cos(\omega + f) + e\cos\,\omega \right]
  \label{eq:ur_dyn}
\end{equation}
where $\textit{i}$ is the inclination angle of the orbital plane (i.e., the angle between the line of sight and the normal to that plane) and $M_{tot}  \equiv M_r + M_b$ is the total mass of the SBHB. We solve for $f$, the true anomaly, numerically from $t_0$, $P$, and $e$ using the set of equations:
\begin{equation}
    \tan^2{\left(\frac{f}{2}\right)} = \frac{1+e}{1-e}  \tan^2{\left(\frac{E}{2}\right)}
    \quad {\rm and} \quad
    E - e\sin{E} = \frac{2\pi}{P} \left( t - t_0 \right)
    \label{eq:eccanom}
\end{equation}
where $E$ is the eccentric anomaly. Since the projected velocity at pericentre, $u_{r,pe}$, is simply the solution of equation~(\ref{eq:ur_dyn}) for the case $f = 0$, we rewrite $u_r$ in terms of $u_{r,pe}$ as follows
\begin{equation}
  u_r = u_{r,pe}\frac{ \cos(\omega + f) + e\cos\,\omega }{\left( 1 + e \right) \cos\,\omega}
  \label{eq:ur_geom}
\end{equation}
We can then determine $u_b$ from $u_r$ through the simple relation $ u_b = -q u_r $.

To compare models and data, we adopt the $\chi^2 $ statistic as our likelihood function:
\begin{equation}
  \chi^2
  = \sum_{k=1}^{N_r} \frac{\left[u_r^{obs}(t_k) - u_r(t_k)\right]^2}{[\sigma_r^{obs}(t_k)]^2}
  + \sum_{k=1}^{N_b} \frac{\left[u_b^{obs}(t_k) - u_b(t_k)\right]^2}{[\sigma_b^{obs}(t_k)]^2} \, ,
\end{equation}
where $u_r^{obs}(t_k)$ and $u_b^{obs}(t_k)$ are the observed radial velocities of the red and blue peaks at time $t_k$, respectively, and $\sigma_r^{obs}(t_k)$ and $\sigma_b^{obs}(t_k)$ are the corresponding uncertainties. As discussed in Section~\ref{sec:jitter}, we can account for jitter in our model by increasing these uncertainties accordingly. In Section~\ref{sec:results} we report results of simulations with and without jitter.

\subsection{Exploration of Parameter Space}
\label{sec:simulation} 

We use an MCMC method to explore parameter space efficiently and find the probability distributions of all orbital parameters. We also use the fundamental model parameters to compute some of the physical parameters of the system, such as the total mass and the decay time of the orbit by gravitational radiation, and examine their distributions as well. We employ the code \textit{emcee}, a Python implementation of Goodman-Weare affine invariant Markov Chain Monte Carlo (MCMC) Ensemble sampler by \cite{Foreman-Mackey2013}. The inputs to the code are the data, the prior distributions of the model parameters, the likelihood function, and initial positions of the random walkers. The code evaluates the likelihood of different solutions, then performs jumps and explores the parameter space for a specified number of steps, before returning as outputs the posterior distributions of model parameters.

For many of our model parameters, we adopted a Jeffreys prior, as it is non-informative and thus presents the least bias in the parameter spaces. The Jeffreys prior prescribes a uniform distribution for location parameters ($t_0$, $\omega$, $e$) and an inverse distribution (uniform in log space) for scale parameters ($u_{r,pe}$, $P$). The mass ratio, $q$, can be calculated as $q = u_b^{obs}(t_k) / u_r^{obs}(t_k)$ for any observation $k$ in which $u_r$ and $u_b$ are measured simultaneously. However, there is a noticeable spread in the observed values of $q$ (most notably in 1E~0450.3$-$1817, where $q$ spans the range from 0.5 to 2.0). In order to be consistent with the binary hypothesis, we attribute the spread to uncertainties in measurements or jitter (see Section~\ref{sec:jitter}), and set the prior for $q$ as a Gaussian distribution, whose mean and standard deviation are determined from the sample of $q$ values for a particular object. The possible values for $u_{r,pe}$ span the range from 0 to $c$, the speed of light. For $P$, we set the lower limit to be $T$, the length of observation; if the binary has completed one revolution within the observation window, both the radial velocity curves for the red and blue peak would have to cross over the zero-point, which is not the case for any of our objects. Since there is no hard constraint on the upper limit on $P$, we set it to be the age of the universe (14~Gyr). We also note that changing the upper limit by one or two orders of magnitude does not appear to affect the posterior distributions of model parameters. The ranges for $t_0$, $e$, and $\omega$ are $\left( 0, P \right) $, $\left( 0, 1 \right)$, and $\left( 0, 2\pi \right)$, respectively. We summarise our choice of priors in Table~\ref{tab:prior}.

\begin{table}
\renewcommand{\thefootnote}{\alph{footnote}}
\caption{Prior distributions of model parameters $^a$}
\label{tab:prior}
\begin{center}
\begin{tabular}{lccc}
\hline
\hline
          &              & Lower & Upper \\
Paramater & Distribution & Bound & Bound \\
\hline
$u_{r,pe}$ \dotfill & Inverse & 0 & $c$\\
$P $ (yr) \dotfill & Inverse & T & $ 1.4 \times 10^{10} $\\
$t_0$ (yr) \dotfill & Uniform & 0 & P\\
$q$ \dotfill & Gaussian & 0 & ...\\
$e$ \dotfill & Uniform & 0 & 1 \\
$\omega$ (rad) \dotfill & Uniform & $ 0 $ & $2 \pi $ \\
\hline
\end{tabular}
\end{center}
\begin{tablenotes}
  \item[1] \footnotetext[1]{See discussion in Section~\ref{sec:simulation} of the text.}
\end{tablenotes}
\end{table}

To run a simulation we initialise a cluster of 1000 random walkers, slightly perturbed from an initial guess for their position ($\delta{Q}/Q \sim 10^{-4}$, where $Q$ is any one of the parameters). We then perform the burn-in phase, where we run the simulations for enough iterations so that the walkers settle into their posterior distributions. Finally, we reset the MCMC chain, run the simulations for 500 further steps, and collect the posterior distributions from the positions of the walkers in the 6-dimensional parameter space. 

\begin{figure}
	\includegraphics[width=\linewidth]{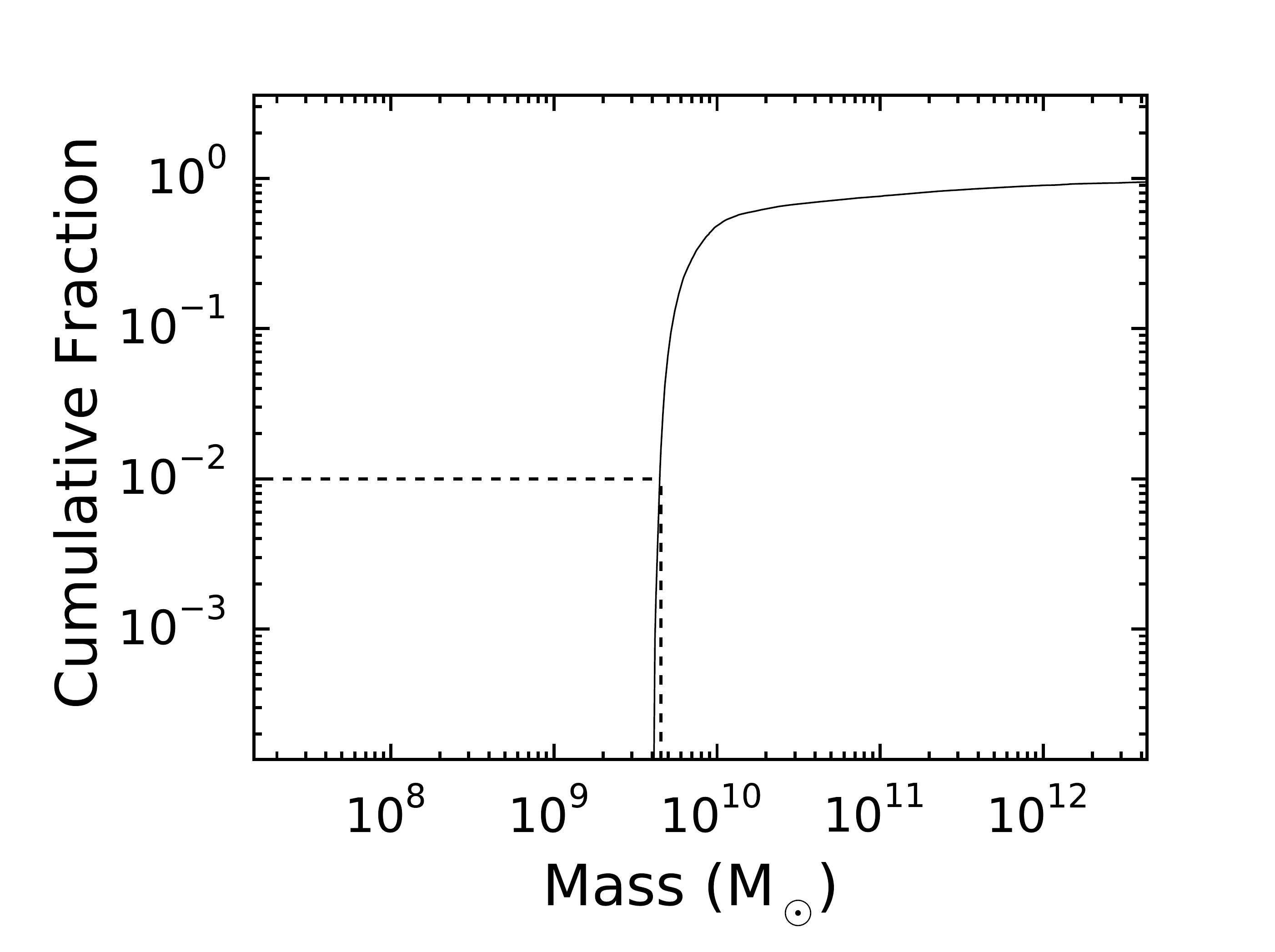}
    \caption{The cumulative mass distribution, $\psi (>\! M)$, of 3C~332. The dashed lines locate $M_{99\%}$, the 99th percentile lower limit on the total mass of the BHs. See discussion in Section~\ref{sec:simulation} of the text.}
    \label{fig:mass_cumm}
\end{figure}

For each set of parameters that we obtain at the end of a single fit in an MCMC simulation, we calculate the corresponding total mass of the SBHB, which is the main parameter of interest, using  
\begin{eqnarray} 
  M_{tot} = 1.42 \times 10^9 \left(1+q\over \cos\omega\,\sin i\right)^3 \left( \frac{1-e}{1+e} \right)^{\frac{3}{2}} \nonumber \\
  \times \left( \frac{P}{300 \, yr} \right) \left( \frac{u_{r,pe}}{5000 \, \mathrm{km\, s^{-1}}} \right)^3\; \Msol
  \label{eq:mtot}
\end{eqnarray}
The above expression is obtained by setting $f=0$ and $u_r=u_{r,pe}$ in equation (\ref{eq:ur_dyn}) and then expressing $a$ in terms of the orbital period via $(P/2\pi)^2=a^3/GM_{tot}$. All of the parameters in equation~(\ref{eq:mtot}), except for the inclination angle, are obtained from the fit to the radial velocity curve. Since the inclination angle is unknown, we set $\sin i=1$ to get a conservative lower limit on $M_{tot}$ and note that, if $i < 28^\circ$ this limit would be raised by an order of magnitude or more (see also Section~\ref{sec:discussion}). From the posterior {\it cumulative} mass distribution $\psi (>\! M)$, we define $M_{x\%}$ such that $\psi(>\! M_{x\%}) = x \% $. Figure~\ref{fig:mass_cumm} shows as an example the cumulative mass distribution $\psi (>\! M)$ of 3C~332. The dashed lines locate $M_{99\%}$, the 99th percentile lower limit on the total mass of the BHs. 

\begin{figure}
	\includegraphics[width=\linewidth]{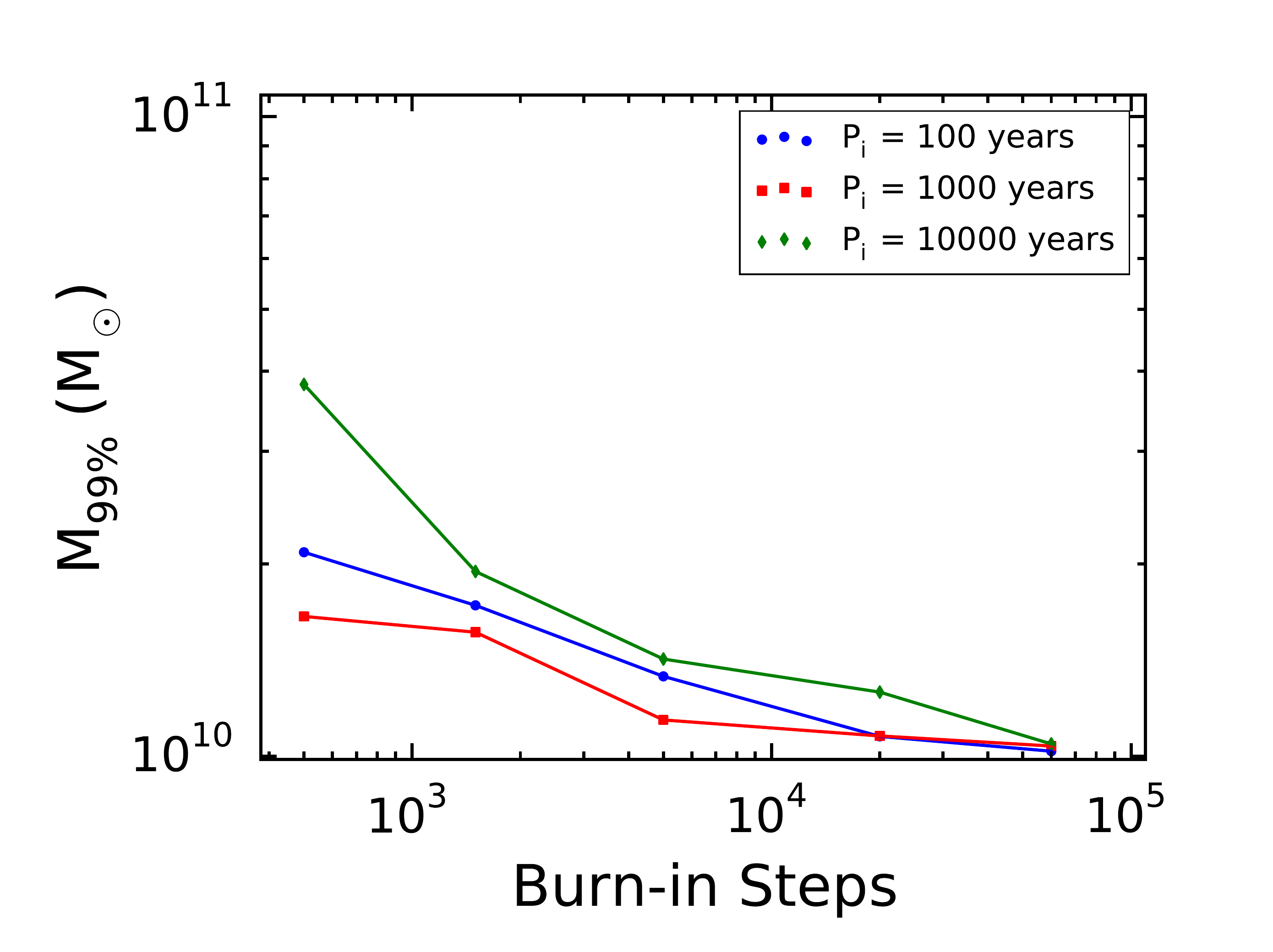}
    \caption{The results of the MCMC simulations for 3C~332. For all three different choices of initial guess period ($10^2$, $10^3$, and $10^4\;$yr), $M_{99\%}$ eventually converge to the same value at $\sim 5 \times 10^{4}$ burn-in steps. The convergence to the same period after a large number of burn-in steps is a universal behaviour among all our objects.}
    \label{fig:burnin_steps}
\end{figure}

It is essential that the burn-in phase is sufficiently long, otherwise the posterior distributions would still be biased by our choice of priors. To that end, for each object we carry out tests using burn-in phases of different lengths, spanning three orders of magnitude, and evaluate how the parameter of interest varies with the number of burn-in steps. We also vary the initial parameter guesses to ensure that the result does not depend on these choices. We then determine and plot the values of $M_{99\%}$ across multiple simulations to evaluate whether the above requirements have been satisfied. As an illustration, we present in Figure~\ref{fig:burnin_steps} the results of the MCMC simulations for 3C~332. For all three different choices of initial period ($10^2$, $10^3$, and $10^4\;$yr), $M_{99\%}$ eventually converges to approximately the same value at $\sim 5 \times 10^{4}$ burn-in steps. The convergence to the same period after a large number of burn-in steps is a universal behavior among all our objects, which leads us to adopt $10^5$ burn-in steps for all of them.

\subsection{Modeling Jitter as a Gaussian Process}\label{sec:gaussproc}

In order to test the simple approach for incorporating jitter described in earlier sections, we also  model it as a Gaussian process, using the code \textit{george}, developed by \cite{Ambikasaran2014}. This approach introduces a covariance matrix into our model. Our primary motivation for the tests described here is not to find the best description of jitter but to assess whether the choice of jitter description influences our results. Since there is no established statistical or physical description of velocity jitter in in quasar broad emission lines, we choose the often used Matern-3/2 covariance function to describe the covariance matrix:
\begin{equation}
C_{3/2}(d) = a \left( 1 + \frac{\sqrt{3}d}{ \tau} \right) \exp \left( - \frac{\sqrt{3}d} {\tau}\right)
\label{eq:matern}
\end{equation}
where $d = \left| t_i - t_j \right|$ is the time interval between two observations, and $a$ and $\tau$ are the amplitude and characteristic time-scale of the Gaussian process, respectively. The Matern-3/2 covariance function is commonly used to define the statistical covariance between measurements made at two different times that are separated from each other by a time interval (or ``distance'') $d$ \citep{Rasmussen2006}. As shown in equation~(\ref{eq:matern}), two additional parameters describing the Gaussian process need to be included into our MCMC model: the amplitude, $a$, and the characteristic time-scale of the process, $\tau$.

To evaluate the degree of agreement between different models, we perform fits for a variety of test cases. Since the simulations involving Gaussian process models are computationally ``expensive'' we only carry out the tests for a subset of the objects in our collection. The test cases we have computed are listed below and the results of the tests are reported in Section~\ref{sec:results}.

\begin{enumerate}

\item 
  In the first set of tests, we assign fixed values for both the amplitude and the characteristic time-scale. We set the amplitude to be the standard deviation of the jitter, and the characteristic time-scale to be 1 year. We choose 1 year because this is typical of the dynamical time-scale, the time-scale on which we expect the jitter to appear (see discussion in Section~\ref{sec:jitter}).

\item \label{it:fixed}
  In our second set of tests, we prescribe the prior distribution of $\tau $ as a Gaussian, centred at 1 year with a standard deviation of 0.5 year.

\item
  In the third set, we modify our prior distribution of  $\tau $ to be a Gaussian centred at 5~yr with a standard deviation of 3~yr to more cover more thoroughly both the regimes of the light-crossing time-scale and the dynamical time-scale.

\item
  In the next set of tests, we proceed to relax also our constraint on the amplitude, allowing it to vary as a Gaussian as well. We set the amplitude to be the amplitude of the jitter listed in Table~\ref{tab:jitter}, and the standard deviation to be half of that value.

\item
  In our last set of tests, we employ a different covariance function, the Ornstein-Uhlenbeck covariance function, another stationary process that describes Brownian motion \citep{Rasmussen2006}:
  \begin{equation}
    C_{OU}(d) = a \exp \left( - \frac{d} {\tau}\right)
    \label{eq:brownian}
  \end{equation}
  Where $d$, $a$, and $\tau$ are defined as in equation~(\ref{eq:matern}). The values of $a$ and $\tau$ are the same as in case~\ref{it:fixed}.
  
\end{enumerate}

\section{Results} \label{sec:results}

\begin{table*}
  \renewcommand{\thefootnote}{\alph{footnote}}
\caption{Orbital Parameters Corresponding to 99\% Lower Limits on BH Masses: Circular Orbits, No Jitter}
\label{tab:circnojit}
\setlength{\tabcolsep}{9pt}
% original is 20.
\begin{center}
\begin{tabu}{lrrrrrrrr}
\noalign{\vskip 4pt}
\hline
\hline
&
\multicolumn{1}{c}{$u_{r,pe}$} &
\multicolumn{1}{c}{$P$} &
\multicolumn{1}{c}{$t_0$} &
&
&
\multicolumn{1}{c}{$\omega$} &
&
\\
Object &
\multicolumn{1}{c}{(km s$^{-1}$)} &
\multicolumn{1}{c}{(yr)}  &
\multicolumn{1}{c}{(yr)} &
\multicolumn{1}{c}{$q$} &
\multicolumn{1}{c}{$e$} &
\multicolumn{1}{c}{(rad)} &
\multicolumn{1}{c}{$\log{\left( M_{min} / \Msol \right)}$\,\footnotemark[1]} &
\multicolumn{1}{c}{$\log{\left( T_{gr}/{\rm yr} \right)}$\,\footnotemark[1]} \\
\noalign{\vskip 2pt}
\hline
3C~59          & 4470 ~~ &   160 &   10 & 0.48 & ... & ... ~~ &  9.3 ~~~~ &  8.6 ~~~ \\
PKS~0235+023   & 3580 ~~ &   140 &   20 & 1.25 & ... & ... ~~ &  9.5 ~~~~ &  8.4 ~~~ \\
NGC~1097       & 3770 ~~ &   380 &   20 & 1.27 & ... & ... ~~ &  9.7 ~~~~ &  7.5 ~~~ \\
1E 0450$-$1817 & 3270 ~~ &   150 &   20 & 1.58 & ... & ... ~~ &  9.0 ~~~~ &  7.4 ~~~ \\
Pictor~A       & 6010 ~~ &  1460 &   30 & 0.95 & ... & ... ~~ &  9.5 ~~~~ &  6.9 ~~~ \\
CBS~74         & 3790 ~~ &   160 &   20 & 0.44 & ... & ... ~~ &  8.4 ~~~~ &  8.4 ~~~ \\  
PKS~0921$-$213 & 2900 ~~ &   170 &   30 & 0.40 & ... & ... ~~ &  9.1 ~~~~ &  8.9 ~~~ \\
PKS~1020$-$103 & 6640 ~~ & 15420 &  120 & 1.40 & ... & ... ~~ & 10.4 ~~~~ &  8.8 ~~~ \\
3C~227         & 1210 ~~ &   180 &   20 & 1.45 & ... & ... ~~ &  8.4 ~~~~ &  9.7 ~~~ \\
Mrk~668        & 2630 ~~ &   780 &   30 & 1.38 & ... & ... ~~ & 10.6 ~~~~ &  7.0 ~~~ \\
3C~332         & 9290 ~~ &   990 &   50 & 0.76 & ... & ... ~~ & 10.2 ~~~~ &  6.4 ~~~ \\
Arp~102B       &10160 ~~ & 2060  &  120 & 1.08 & ... & ... ~~ & 11.9 ~~~~ &  7.2 ~~~ \\
PKS~1739+184   & 3640 ~~ &  4720 &   40 & 0.73 & ... & ... ~~ &  8.7 ~~~~ &  8.3 ~~~ \\
3C~390.3       & 6110 ~~ &  3420 &  480 & 0.77 & ... & ... ~~ & 12.3 ~~~~ &  7.6 ~~~ \\
\hline
\end{tabu}
\end{center}
  \begin{tablenotes}
  \item[1] \footnotemark[1] These values assume $\sin i=1$. See equation~(\ref{eq:mtot}) and discussion in the first paragraph of Section~\ref{sec:results}.
  \end{tablenotes}
\end{table*}

\begin{table*}
  \renewcommand{\thefootnote}{\alph{footnote}}
  \caption{Orbital Parameters Corresponding to 99\% Lower Limits on BH Masses: Circular Orbits, With Jitter}
  \label{tab:circjit}
  \setlength{\tabcolsep}{9pt}  % original is 20pt.
  \begin{center}
    \begin{tabu}{lrrrrrrrr}
      \noalign{\vskip 4pt}
      \hline
      \hline
      &
      \multicolumn{1}{c}{$u_{r,pe}$} &
      \multicolumn{1}{c}{$P$} &
      \multicolumn{1}{c}{$t_0$} &
      &
      &
      \multicolumn{1}{c}{$\omega$} &
      &
      \\
      Object &
      \multicolumn{1}{c}{(km s$^{-1}$)} &
      \multicolumn{1}{c}{(yr)}  &
      \multicolumn{1}{c}{(yr)} &
      \multicolumn{1}{c}{$q$} &
      \multicolumn{1}{c}{$e$} &
      \multicolumn{1}{c}{(rad)} &
      \multicolumn{1}{c}{$\log{\left( M_{min} / \Msol \right)}$\,\footnotemark[1]} &
      \multicolumn{1}{c}{$\log{\left( T_{gr}/{\rm yr} \right)}$\,\footnotemark[1]} \\
      \noalign{\vskip 2pt}
      \hline
      
      3C~59          & 4530 ~~ &   190 &   30 & 0.52 & ... & ... ~~ &  9.1 ~~~~ &  8.8 ~~~ \\			
      PKS~0235+023   & 3460 ~~ &   130 &   20 & 1.18 & ... & ... ~~ &  9.4 ~~~~ &  8.5 ~~~ \\			
      NGC~1097       & 3720 ~~ &   350 &   10 & 1.31 & ... & ... ~~ &  9.5 ~~~~ &  7.3 ~~~ \\			
      1E 0450$-$1817 & 3310 ~~ &   170 &   40 & 1.61 & ... & ... ~~ &  8.8 ~~~~ &  7.5 ~~~ \\			
      Pictor~A       & 5880 ~~ &  1390 &   20 & 0.98 & ... & ... ~~ &  9.5 ~~~~ &  7.1 ~~~ \\			
      CBS~74         & 3750 ~~ &   180 &   30 & 0.42 & ... & ... ~~ &  8.2 ~~~~ &  8.6 ~~~ \\			
      PKS~0921$-$213 & 2970 ~~ &   160 &   20 & 0.38 & ... & ... ~~ &  8.8 ~~~~ &  9.2 ~~~ \\			
      PKS~1020$-$103 & 6490 ~~ & 13380 &  100 & 1.42 & ... & ... ~~ & 10.0 ~~~~ &  8.6 ~~~ \\			
      3C~227         & 1190 ~~ &   190 &   20 & 1.51 & ... & ... ~~ &  8.3 ~~~~ &  9.9 ~~~ \\			
      Mrk~668        & 2730 ~~ &   810 &   50 & 1.35 & ... & ... ~~ & 10.3 ~~~~ &  7.2 ~~~ \\
      3C~332         & 8860 ~~ &   950 &   40 & 0.79 & ... & ... ~~ & 10.1 ~~~~ &  6.1 ~~~ \\
      Arp~102B       &10040 ~~ & 2110  &   90 & 1.11 & ... & ... ~~ & 11.4 ~~~~ &  7.4 ~~~ \\ 		
      PKS~1739+184   & 3570 ~~ &  4690 &   50 & 0.68 & ... & ... ~~ &  8.6 ~~~~ &  8.2 ~~~ \\ 
      3C~390.3       & 5980 ~~ &  3530 &  440 & 0.81 & ... & ... ~~ & 11.9 ~~~~ &  7.8 ~~~ \\
      \hline
    \end{tabu}
  \end{center}
  \begin{tablenotes}
  \item[1] \footnotemark[1] These values assume $\sin i=1$. See equation~(\ref{eq:mtot}) and discussion in the first paragraph of Section~\ref{sec:results}.
  \end{tablenotes}
\end{table*}

\begin{table*}
  \renewcommand{\thefootnote}{\alph{footnote}}
  \caption{Orbital Parameters Corresponding to 99\% Lower Limits on BH Masses: Eccentric Orbits, No Jitter}
  \label{tab:eccnojit}
  \setlength{\tabcolsep}{9pt} % original is 20 pt.
  \begin{center}
    \begin{tabu}{lrrrrrrrr}
      \noalign{\vskip 4pt}
      \hline
      \hline
      &
      \multicolumn{1}{c}{$u_{r,pe}$} &
      \multicolumn{1}{c}{$P$} &
      \multicolumn{1}{c}{$t_0$} &
      &
      &
      \multicolumn{1}{c}{$\omega$} &
      &
      \\
      Object &
      \multicolumn{1}{c}{(km s$^{-1}$)} &
      \multicolumn{1}{c}{(yr)}  &
      \multicolumn{1}{c}{(yr)} &
      \multicolumn{1}{c}{$q$} &
      \multicolumn{1}{c}{$e$} &
      \multicolumn{1}{c}{(rad)} &
      \multicolumn{1}{c}{$\log{\left( M_{min} / \Msol \right)}$\,\footnotemark[1]} &
      \multicolumn{1}{c}{$\log{\left( T_{gr}/{\rm yr} \right)}$\,\footnotemark[1]} \\
      \noalign{\vskip 2pt}
      \hline

      3C~59          & 4500 ~~ &   240 & 40 & 0.49 & 0.25 & $-$0.11 ~~ &  9.0 ~~~~ &  9.2 ~~~ \\			
      PKS~0235+023   & 4370 ~~ & 22200 & 30 & 1.36 & 0.97 & $-$0.16 ~~ &  9.1 ~~~~ & 10.3 ~~~ \\			
      NGC~1097       & 4280 ~~ &   340 & 30 & 1.21 & 0.63 & $-$2.88 ~~ &  9.2 ~~~~ &  9.3 ~~~ \\			
      1E 0450$-$1817 & 1940 ~~ &  8370 & 30 & 1.48 & 0.97 &    2.03 ~~ &  9.3 ~~~~ &  9.4 ~~~ \\			
      Pictor~A       & 6540 ~~ & 32420 & 20 & 0.99 & 0.99 & $-$3.14 ~~ &  9.0 ~~~~ &  9.4 ~~~ \\		
      CBS~74         & 3220 ~~ &   110 & 30 & 0.43 & 0.51 & $-$2.46 ~~ &  8.2 ~~~~ &  9.5 ~~~ \\  
      PKS~0921$-$213 & 1090 ~~ &  2910 & 50 & 1.01 & 0.95 & $-$1.80 ~~ &  8.7 ~~~~ &  9.6 ~~~ \\
      PKS~1020$-$103 & 1750 ~~ & 15200 & 80 & 1.39 & 0.74 & $-$0.51 ~~ &  9.6 ~~~~ & 12.1 ~~~ \\
      3C~227         & 1330 ~~ &  4820 & 40 & 1.34 & 0.93 &    2.83 ~~ &  8.0 ~~~~ &  8.9 ~~~ \\			
      Mrk~668        & 5680 ~~ &  1880 & 40 & 1.19 & 0.95 &    0.12 ~~ &  8.7 ~~~~ &  8.8 ~~~ \\
      3C~332         & 9100 ~~ &   460 & 20 & 0.76 & 0.77 & $-$0.60 ~~ &  9.8 ~~~~ &  7.6 ~~~ \\
      Arp~102B       & 1770 ~~ &  6880 & 80 & 1.11 & 0.96 &    1.41 ~~ & 10.9 ~~~~ &  8.1 ~~~ \\			
      PKS~1739+184   & 3630 ~~ &  1320 & 30 & 0.76 & 0.91 & $-$0.07 ~~ &  8.2 ~~~~ & 10.3 ~~~ \\
      3C~390.3       & 3960 ~~ &  2840 & 50 & 0.84 & 0.36 & $-$0.62 ~~ & 10.4 ~~~~ &  9.2 ~~~ \\
      \hline
    \end{tabu}
  \end{center}
  \begin{tablenotes}
  \item[1] \footnotemark[1] These values assume $\sin i=1$. See equation~(\ref{eq:mtot}) and discussion in the first paragraph of Section~\ref{sec:results}.
  \end{tablenotes}
\end{table*}

\begin{table*}
  \renewcommand{\thefootnote}{\alph{footnote}}
  \caption{Orbital Parameters Corresponding to 99\% Lower Limits on BH Masses: Eccentric Orbits, With Jitter}
  \label{tab:eccjit}
  \setlength{\tabcolsep}{9pt}
  \begin{center}
    \begin{tabu}{lrrrrrrrr}
      \noalign{\vskip 4pt}
      \hline
      \hline
      &
      \multicolumn{1}{c}{$u_{r,pe}$} &
      \multicolumn{1}{c}{$P$} &
      \multicolumn{1}{c}{$t_0$} &
      &
      &
      \multicolumn{1}{c}{$\omega$} &
      &
      \\
      Object &
      \multicolumn{1}{c}{(km s$^{-1}$)} &
      \multicolumn{1}{c}{(yr)}  &
      \multicolumn{1}{c}{(yr)} &
      \multicolumn{1}{c}{$q$} &
      \multicolumn{1}{c}{$e$} &
      \multicolumn{1}{c}{(rad)} &
      \multicolumn{1}{c}{$\log{\left( M_{min} / \Msol \right)}$\,\footnotemark[1]} &
      \multicolumn{1}{c}{$\log{\left( T_{gr}/{\rm yr} \right)}$\,\footnotemark[1]} \\
      \noalign{\vskip 2pt}
      \hline
      3C~59          & 3920 ~~ &    320 &   60 & 0.52 & 0.45 &    0.72 ~~ &  9.1 ~~~~ &  9.2 ~~~ \\			
      PKS~0235+023   & 3590 ~~ &   1950 &   30 & 1.29 & 0.85 &    2.92 ~~ &  9.0 ~~~~ & 10.1 ~~~ \\			
      NGC~1097       & 4340 ~~ &    360 &   40 & 1.14 & 0.61 & $-$2.94 ~~ &  9.1 ~~~~ &  9.0 ~~~ \\			
      1E 0450$-$1817 & 1900 ~~ &   8430 &   20 & 1.38 & 0.96 &    1.99 ~~ &  9.0 ~~~~ &  9.6 ~~~ \\			
      Pictor~A       & 5020 ~~ &   1660 &   20 & 0.96 & 0.95 &    2.14 ~~ &  9.2 ~~~~ &  7.9 ~~~ \\		
      CBS~74         & 1880 ~~ &     80 &   30 & 0.46 & 0.64 &    1.97 ~~ &  8.0 ~~~~ &  9.0 ~~~ \\  
      PKS~0921$-$213 & 2960 ~~ &   3790 &   50 & 1.00 & 0.90 & $-$2.80 ~~ &  8.6 ~~~~ & 10.8 ~~~ \\
      PKS~1020$-$103 & 2440 ~~ &   1150 &   50 & 1.30 & 0.96 &    1.22 ~~ &  8.7 ~~~~ &  7.9 ~~~ \\
      3C~227         & 1480 ~~ &    850 &   10 & 1.17 & 0.80 & $-$0.59 ~~ &  7.9 ~~~~ & 11.3 ~~~ \\			
      Mrk~668        & 1440 ~~ &   1010 &   40 & 2.02 & 0.97 &    1.33 ~~ &  8.7 ~~~~ &  8.0 ~~~ \\
      3C~332         & 8730 ~~ &   1980 &   20 & 0.77 & 0.92 &    2.54 ~~ &  9.6 ~~~~ &  8.1 ~~~ \\
      Arp~102B       & 4950 ~~ & 173120 &  900 & 1.02 & 0.93 &    0.74 ~~ & 11.1 ~~~~ &  10.7 ~~~ \\			
      PKS~1739+184   & 3510 ~~ &  28080 &   30 & 0.76 & 0.98 &    0.09 ~~ &  8.4 ~~~~ & 11.2 ~~~ \\
      3C~390.3       & 2750 ~~ &   1300 &   60 & 0.84 & 0.21 &    1.11 ~~ & 10.6 ~~~~ &  8.7 ~~~ \\
      \hline
    \end{tabu}
  \end{center}
  \begin{tablenotes}
  \item[1] \footnotemark[1] These values assume $\sin i=1$. See equation~(\ref{eq:mtot}) and discussion in the first paragraph of Section~\ref{sec:results}.
  \end{tablenotes}
\end{table*}

\begin{table*}
  \renewcommand{\thefootnote}{\alph{footnote}}
  \caption{Goodness of Fit Statistics for Different Orbital Models}
  \label{tab:chisq}
  \setlength{\tabcolsep}{9pt}
  \begin{center}
    \begin{tabu} to \textwidth{lccccc}
      \noalign{\vskip 4pt}
      \hline
      \hline
      \noalign{\vskip 2pt}
      & & \multicolumn{4}{c}{Reduced $\chi^2$ Values for Best Fits\,\footnotemark[1]} \\
      \noalign{\vskip -4pt}
      & & \multicolumn{4}{c}{\hrulefill}\\
      & Degrees    & Circular  & Circular    & Elliptical & Elliptical  \\
      Object & of Freedom \footnotemark[2] & No Jitter & With Jitter & No Jitter  & With Jitter \\
      \hline      
      3C~59          & 28  & ~7.91 & 6.44 & ~5.44 & 4.47 \\      
      PKS~0235+023   & 24  & ~8.17 & 1.86 & ~4.55 & 0.84 \\      
      NGC~1097       & 32  & 31.63 & 3.17 & 22.34 & 2.26 \\      
      1E 0450$-$1817 & 38  & 53.81 & 3.93 & 37.48 & 2.42 \\      
      Pictor~A       & 22  & 65.95 & 5.51 & 46.76 & 3.82 \\      
      CBS~74         & 30  & 18.77 & 3.15 & 11.50 & 1.93 \\      
      PKS~0921$-$213 & 32  & 14.36 & 3.52 & ~7.85 & 2.06 \\      
      PKS~1020$-$103 & 18  & 11.19 & 3.22 & ~6.72 & 1.91 \\      
      3C~227         & 32  & 71.01 & 3.65 & 32.65 & 1.55 \\      
      Mrk~668        & 23  & 18.82 & 8.19 & ~4.78 & 2.17 \\
      3C~332         & 69  & 28.11 & 2.19 & 21.47 & 1.67 \\
      Arp~102B       & 184 & ~9.62 & 2.10 & ~6.17 & 1.35 \\
      PKS~1739+184   & 32  & 13.04 & 3.01 & ~5.78 & 1.46 \\
      3C~390.3       & 97  & 64.02 & 4.04 & 39.84 & 1.67 \\
      \hline
    \end{tabu}
  \end{center}
  \begin{tablenotes}
  \item[1] \footnotemark[1] The orbital models are those described in the first paragraph of Section~\ref{sec:results} This is the degree of freedom corresponding to the elliptical cases. To get the degree of freedom for circular cases, add two to these numbers.
  \item[2] \footnotemark[2] Test
  \end{tablenotes}
\end{table*}

Using the methods of Section~\ref{sec:fitting} we first perform four sets of simulations to explore the effects of orbital eccentricity and the simple jitter prescription. Specifically we explore the following scenarios: (a) circular orbits without jitter, (b) circular orbits with jitter, (c) eccentric orbits without jitter, and (d) eccentric orbits with jitter. From the output of the simulations we calculate the SBHB total mass for each solution using equation~(\ref{eq:mtot}) and assuming that $\sin i = 1$. We also calculate $T_{gr}$, the orbital decay time-scale due to gravitational radiation, by numerically integrating equation~(5.14) of \citet{Peters1964}. Since $T_{gr}\propto M_{tot}^{-3}$ and $M_{tot}\propto (\sin i)^{-3}$ the gravitational radiation decay time is a very sensitive function of the (unknown) inclination of the binary, $T_{gr}\propto (\sin i)^9$. For the four sets of simulations described above, we report the 99\% lower limit on the mass, together with corresponding values of orbital parameters and the orbital decay time-scale, in Tables~\ref{tab:circnojit}--\ref{tab:eccjit}, respectively. We note that the resulting minimum masses and decay time-scales do not change significantly as we take jitter into consideration i.e. from case (a) to case (b), and from case (c) to case (d). However, they do change by an order of magnitude or more as we relax the assumption of circular orbits i.e. from case (a) to (c) and from case (b) to (d). This is reasonable as the former change is a change in the effective uncertainty, whereas the latter change significantly enlarges the space of possible orbits and parameters. The goodness of fit statistics (the reduced $\chi^2$ value, $\chi^2_\nu=\chi^2/\nu$, where $\nu$ is number of degrees of freedom, and $\nu$) for the best-fitting models in all of the above cases are reported in Table~\ref{tab:chisq}. As this table shows, the $\chi^2_\nu$ values are quite large in cases (a) and (c) where jitter is not included. This is not a surprise since the amplitude of the jitter is typically a few times larger than the measurement errors. The fits of elliptical orbit models including jitter yield the lowest $\chi^2_\nu$ values, as expected. However, a substantial fraction of objects have $\chi^2_\nu > 2$ even with this family of models.

\begin{table*}
  \renewcommand{\thefootnote}{\alph{footnote}}
  \caption{BH Mass Limits for Gaussian Process Description of Jitter}
  \label{tab:gaussproc}
  \setlength{\tabcolsep}{9pt}
  \begin{center}
    \begin{tabu} to \textwidth{lccccccc}
      \noalign{\vskip 4pt}
      \hline
      \hline
      \noalign{\vskip 2pt}
      & \multicolumn{6}{c}{$\log{\left( M_{min} / \Msol \right)}$}\\
      \noalign{\vskip -4pt}
      & \multicolumn{6}{c}{\hrulefill}\\
      & Fixed $\tau$\,\footnotemark[1] & $\tau = \mathcal{N}(1,0.5)$\,\footnotemark[2]  & $\tau = \mathcal{N}(5,3)$\,\footnotemark[2] & $\tau = \mathcal{N}(5,3)$\,\footnotemark[2] & \hskip -1.5em Ornstein-Uhlenbeck & \hskip -.5em Simple Jitter \\
      Object & Fixed $a$\,\footnotemark[1] & Fixed $a$\,\footnotemark[1] & Fixed $a$\,\footnotemark[1] & $a = \mathcal{N}(j,j/2)$\,\footnotemark[3] & \hskip -1.5em Description\,\footnotemark[4] & \hskip -.5em Description\,\footnotemark[5]  \\
      \hline
      3C~59          &  ... & ~9.4 &  ... &  ... & ...  & ~9.1 \\      
      PKS~0235+023   &  ... & ~8.7 &  ... &  ... & ...  & ~9.0 \\      
      NGC~1097       &  ... & ~9.4 &  ... &  ... & ...  & ~9.1 \\     
      1E 0450$-$1817 & ~9.6 & ~9.4 & ~9.3 & ~9.1 & ~9.3 & ~9.0 \\     
      Pictor~A       & ~9.3 & ~9.0 &  ... &  ... & ...  & ~9.2 \\    
      CBS~74         &  ... & ~7.8 &  ... &  ... & ...  & ~8.0 \\	  
      PKS~0921$-$213 &  ... & ~8.8 &  ... &  ... & ...  & ~8.6 \\
      PKS~1020$-$103 &  ... & ~8.5 &  ... &  ... & ...  & ~8.7 \\    
      3C~227         &  ... & ~7.7 &  ... &  ... & ...  & ~7.9 \\   
      Mrk~668        &  ... & ~8.5 &  ... &  ... & ...  & ~8.7 \\
      3C~332         & ~9.8 & ~9.7 &  ... &  ... & ...  & ~9.6 \\
      Arp~102B       & 10.4 & 10.6 & 11.2 & 11.3 & 10.8 & 11.1 \\
      PKS~1739+184   & ~8.6 & ~8.7 & ~8.6 & ~8.3 & ~8.4 & ~8.4 \\
      3C~390.3       & 10.4 & 10.7 &  ... &  ... & ...  & 10.6 \\
      \hline
    \end{tabu}
  \end{center}
  \begin{tablenotes}
  \item[1] \footnotemark[1] The fixed value of $\tau=1\;$yr. The fixed value of $a$ is the measured jitter for each object listed in Table~\ref{tab:jitter}.  See details in Section~\ref{sec:jitter}.
  \item[2] \footnotemark[2] $\tau={\cal N}(m,s)$ denotes $\tau$ drawn from a normal distribution of mean $m$ and standard deviation $s$, both expressed in years.
  \item[3] \footnotemark[3] In the $a = \mathcal{N}(j,j/2)$ case we draw the value of $a$ from a normal distribution with a mean equal to the jitter in Table~\ref{tab:jitter} and a standard deviation equal to half that value. 
  \item[4] \footnotemark[4] We use a fixed value $\tau=1\;$yr and the value of $a$ for each object, as listed in Table~\ref{tab:jitter}.
  \item[5] \footnotemark[5]  In this description we expand the error bars on the radial velocity measurements to match the jitter measured by the method of Section~\ref{sec:jitter}. These are the same results reported in Table~\ref{tab:eccjit} and are reproduced here for easy comparison.
  \end{tablenotes}
\end{table*}

Our results of modeling jitter via Gaussian processes are presented in Table~\ref{tab:gaussproc}, where we quote the value of the 99\% lower limit on the total mass of the SBHB. Because of the high computational cost of running Gaussian process simulations, we perform a limited number of representative tests. Nevertheless, comparison of the results of these tests with the results presented above show that there are minimal differences between the results of the Gaussian process methods and those from our fiducial treatment of jitter in the previous paragraph. The mass limits resulting from the various methods differ by a factor of a few, which leads us to conclude that our simple description of jitter is fairly robust. Thus we adopt the results of the simple jitter description in our later discussion.

We focus our attention on the orbital parameters of case (d) that involves eccentric orbits and includes jitter, as it is the most inclusive scenario and yields the most conservative constraint on BH masses. Henceforth, we illustrate the results of case (d), unless noted otherwise. As shown in Table~\ref{tab:eccjit}, the lower limits on the SBHB masses are in the range $10^8$--$10^{11}\; \Msol$. However, the eccentricity corresponding to the minimum mass tends to be very high, as more than half the cases have eccentricity values greater than 0.9. We also note that the decay time-scale falls in the range $10^7$--$10^{12}\;$yr. We re-iterate that the realistic mass limits should be higher than those of Table~\ref{tab:eccjit} by a factor of $(\sin i)^3$ and the orbital decay times should be shorter by a factor of $(\sin i)^9$; we discuss this issue further in Section~\ref{sec:discussion}.

In Figure~\ref{fig:radvel} we present the radial velocity curves along with three solutions for each object, corresponding to the 99\%, 90\% and 68\% lower limits on the period; these are shown respectively as solid blue, dashed green, and dotted red lines superimposed on the data. The minimum $\chi^2_\nu$ values corresponding to these fits are in the last column of Table~\ref{tab:chisq}. Some of the fits are particularly poor, which is reflected in the $\chi^2_\nu$ values. Some noteworthy examples of poor fits are: (a) 3C~59, 1E~0450.3$-$1817 and Pictor~A where there are very large gradients in the peak velocities on time-scales of a decade that are inconsistent with orbital motion, (b) Mrk~668 where the well-sampled radial velocity curve of the red peak has the opposite curvature from what orbital motion would dictate, and (c) NGC~1097, CBS~74, 3C~332 and PKS~1739+184 where the radial velocities in last 5--10~yr of the monitoring period depart systematically from the fit that describes well the radial velocities of the first 15--20~yr. There are also examples of fits that appear reasonable, such as PKS~1020$-$103, Arp~102B and 3C~390.3.

Figure~\ref{fig:periodecc} shows a projection of the space of model parameters in the period--eccentricity plane. The colour indicates the relative density of solutions within a specific area in the diagram. As shown in Figure~\ref{fig:periodecc}, some of the objects (e.g. 3C~59, 1E~0450.3$-$1817, NGC~1097) have solutions most densely concentrated in an arc in the diagram; however, not all objects show this behavior. We also note that in many cases, the distribution of eccentricity is almost uniform from 0 to 1 (e.g. PKS~1020$-$103, PKS~0921$-$213), implying that there is no specific preference for or constraint on the value of eccentricity. Similarly, the periods tend to vary across many orders of magnitude. However, in specific cases such as Arp~102B or Pictor~A, the periods tend to concentrate at $\sim$ $10^{10}\;$yr, the upper boundary in our simulation. We attribute this behavior to the inability of the MCMC simulation to find satisfactory solutions at periods lower than the age of the universe, and we take it to suggest that the SBHB hypothesis is incompatible with the data.

Figure~\ref{fig:massdecay} shows the main scientific result of the simulations, the distribution of solutions in the 2-dimensional parameter space of total mass versus orbital decay time. The colours have the same meaning as in Figure~\ref{fig:periodecc}. The vertical dashed line indicates a BH total mass of $10^{10} M_{\odot}$, above which we deem the binary hypothesis unfavourable (see Section~\ref{sec:discussion}). As shown in the Figure~\ref{fig:massdecay}, the general behavior is that the solutions are concentrated most densely along a ridge or an arc in the parameter space, and the density of solutions decreases outward in all directions. The minimum masses resulting from this analysis are close to $10^{11}\;\Msol$ for two objects (Arp~102B, 3C~390.3), in the range (0.4--4)$\times 10^{9}\;\Msol$ for nine objects, and in the range (0.8--3)$\times 10^{8}\;\Msol$ for the remaining three objects.

\section{Discussion}\label{sec:discussion}

We begin our discussion of the results by considering whether the orbital model provides a reasonable description of the radial velocity curves on statistical grounds. We then consider constraints on the orbital inclination based on the radio properties of the targets that increase the minimum SBHB masses we reported in Table~\ref{tab:eccjit}. We also provide a set of additional arguments disfavouring the SBHB hypothesis for many members of our sample. We conclude with a summary of the status of each object and a brief discussion of one of the alternative explanations for double-peaked line profiles. 

The goodness-of-fit tests for many of the objects in our sample indicate that the binary orbit model provides a poor description of the observed radial velocity curves. This conclusion holds even if we use the jitter as a measure of the uncertainty in our calculations of $\chi^2$. The probability that the orbital model provides an adequate description of the radial velocity curves for the six cases with $\chi^2_\nu > 2$ in Table~\ref{tab:eccjit} (3C~59, NGC~1097, 1E~0450.3$-$1817, Pictor~A, CBS~74, PKS~0921$-$213, Mrk~668) is $p<10^{-4}$. An inspection of Figure~\ref{fig:radvel} shows that the fits to the radial velocity curves of those six objects do not capture the full behavior of the data. For 3C~59, the upward trend of the blue peak from 1997--2015 strongly contradicts the model, which predicts a downward trend. In NGC~1097, the model fits the radial velocities of the blue peak but not those of the red peak. In 1E~0450.3$-$1817, the sudden drop in the velocity curve of the red peak  during the period 1995--2000 cannot be explained by a binary model. For CBS~74, the velocity of the blue peak rises and drops sharply within a few years in an erratic manner. The radial velocity curve of the red peak of Mrk~668 stays relatively flat for a long period of time, before experiencing a sudden rise, forcing the eccentricity of the fit to be very high as a result. Moreover, the radial velocity curve of the red peak of Mrk~668 is convex, which is inconsistent with orbital motion (this was also noted by \citealt{Marziani1993} and \citealt{Gezari2007}). In the case of Pictor~A, both the red and the blue peak radial velocity curves display a rise during the same time period, which is inconsistent with the binary hypothesis. Finally, regarding PKS~0921$-$213, the binary model cannot explain the velocity fluctuations of the red and the blue peak during the period 1995--2000. These considerations argue against the orbital model for these six objects. 

The lower limits on SBHB masses that we obtain by adopting the circular orbit model without velocity jitter (Table~\ref{tab:circnojit}) are very similar to those obtained by \cite{Liu2016} for nine of the 13 objects in common. For the remaining four objects, our limits are a few orders of magnitude lower than theirs. We attribute this difference to the different approach used to obtain the minimum mass: we calculate the SBHB mass for every solution that our algorithm finds and then take the minimum value while \cite{Liu2016} find the minimum period based on their fit to the radial velocity curve and use it to compute the minimum mass. For the eccentric orbit models with velocity jitter included (see Table~\ref{tab:eccjit}), we obtain minimum masses for most objects that are a few orders of magnitude lower than in \cite{Liu2016} and previous works. This difference is reasonable, since there is more flexibility in the models that we employ here due to the addition of eccentricity. Nevertheless, we are still able to confidently disfavour the SBHB hypothesis for two targets (Arp~102B, 3C~390.3) on the basis of a very high mass; the 99\% lower limits on the total masses are greater than the most massive BH measured so far \citep[$2 \times 10^{10}\;\Msol$; e.g.,][]{McConnell2012, Thomas2016}. 

We can refine the mass limits of Table~\ref{tab:eccjit} further by considering constraints on  the inclination of the orbital plane of the hypothesized SBHB -- $M_{tot}\propto \left(\sin i\right)^3$ as discussed in Section~\ref{sec:fitting}. Since the vast majority of our objects are radio loud, we can obtain constraints on the inclination angle of the radio jet and connect the direction of the radio jet to the axis of the binary orbit by assuming that the jet is emitted along the spin axis of one of the BHs and that the spins of the two BHs in an SBHB are aligned with the axis of the orbit. The former assumption is justified by the the theory and recent simulations of jet formation \citep[e.g.][and many references therein]{Narayan2014,Tchekhovskoy2010}. The latter assumption is justified in the context of evolutionary scenarios for SBHBs at separations of a few pc or less: a massive circumbinary disc extracts angular momentum from the orbit causing it to decay, supplies accretion fuel to the BHs, and exerts torques to align the spins of the two BHs with the orbital angular momentum vector \citep[see][]{Bogdanovic2007,Dotti2010,Miller2013}. Thus, we can use the spectral index of the radio core and the brightness contrast between the radio core and the radio lobes, which are regarded as indicators of relativistic beaming effects in the jet \citep[e.g.,][]{Orr1982}, to constrain the jet inclination. 

We collected values of the spectral index of the radio core, $\alpha_{\rm core}$ (where $f_\nu\propto \nu^{+\alpha}$), and the core-to-lobe ratio, $\log R$ (where $R\equiv S_{core}/S_{lobes}$, with $S_{core}$ and $S_{lobes}$ the core and lobe fluxes at a rest-frame frequency of 5~GHz), from the literature in Table~\ref{tab:radio} along with references to the sources of the data and relevant observational details. In cases where the value of $S_{core}$ and/or $S_{lobes}$ was measured at a frequency other than 5~GHz, we inferred the 5~GHz values assuming core and lobe spectral indices of $\alpha_{core}=0$ and $\alpha_{lobes}=-0.7$, respectively. In most cases only $\alpha_{core}$ or $\log R$ was readily available (not both) so we used the empirical correlation between $\alpha_{core}$ and $\log R$ by \citet{Runnoe2012}, $\alpha_{core} = - 0.433 + 0.206\,\log R$, to deduce the missing quantity. As Table~\ref{tab:radio} shows, we were unable to find useful information on four of the 14 objects (NGC~1097, 1E~0450-3$-$1817, CBS~74, and Mrk~668). In contrast, in the case of 3C~390.3 a trustworthy upper limit on the inclination is available from superluminal motions in the radio jet.
We compare the data in Table~\ref{tab:radio} with detailed analyses of the emission properties of jets in a variety of radio-loud AGNs by various authors. \citet{Ghisellini1993} and \citet{Guerra1997} have studied the radio morphologies, spectral energy distributions, and superluminal motions of samples of radio-loud quasars and radio galaxies of different types (with and without broad lines). They concluded that in core-dominated quasars ($\log R > 0$) the jet inclination angles are $i\,\la\, 10^\circ$, while lobe-dominated quasars ($-1.36 < \log R < 0$ in their samples) have $10^\circ\,\la\, i \,\la\, 40^\circ$ with a median of $26^\circ$. Similarly, \citet{Hovatta2009} have considered the variability properties and jet beaming factors of a large sample of radio-loud quasars and BL~Lac objects with $\log R > -0.5$ and concluded that in almost all cases the jet inclinations are $i < 20^\circ$ and \citet{DiPompeo2012} carried out a statistical study of radio core spectral indices and concluded that values of $\alpha_{core} > -1$ are associated with inclination angles $i < 26^\circ$, albeit with some scatter.

Considering the radio properties of the objects in Table~\ref{tab:radio} in this light, we find that two of the 10 objects with useful radio information (PKS~0921--213, PKS~1020--103) are likely to have $i\,\la\, 10^\circ$, four (Pictor~A, Arp~102B, PKS~1739+184, 3C~390.3) are likely to have $i\,\la\, 26^\circ$, and the remaining four (3C~59, PKS~0235+023, 3C~227, 3C~332) are likely to have $i\,\la\, 40^\circ$. Taken at face value, these inclination angle limits change the minimum masses of Table~\ref{tab:eccjit} as follows: for $i\,\la\, 40^\circ$, $\log(M_{min}/\Msol)$ increases by 0.6, for $i\,\la\, 26^\circ$, $\log(M_{min}/\Msol)$ increases by 1.1, and for $i\,\la\, 10^\circ$, $\log(M_{min}/\Msol)$ increases by 2.3. As a result, the minimum masses of four more objects from our sample rise above the threshold of $10^{10}\;\Msol$ (Pictor~A, PKS~0921--213, PKS~1020--103, 3C~332) and the minimum masses of two other objects get close to this threshold (3C~59, PKS~0235+023, PKS~0921--213). Similarly, the orbital decay times due to gravitational radiation decrease substantially once the orbital inclination angle is taken into account: for $i\,\la\, 40^\circ$, $\log(T_{gr}/{\rm yr})$ decreases by 1.7, for $i\,\la\, 26^\circ$, $\log(T_{gr}/{\rm yr})$ decreases by 3.2, and for $i\,\la\, 10^\circ$, $\log(T_{gr}/{\rm yr})$ decreases by 6.8. This change makes the orbital decay time uncomfortably short for some objects; $T_{\rm gr}$ becomes shorter than 100~yr for PKS~1020--103 and shorter than $10^6\;$yr for Pictor~A, PKS~0921--213, and 3C~390.3.

\begin{table}
\begin{minipage}{\linewidth}
\renewcommand{\thefootnote}{\alph{footnote}}
\caption{Radio Properties of Sample Objects}
\label{tab:radio}
\begin{center}
\begin{tabular}{lccl}
\hline
\hline
Object & $\alpha_{\rm core}$\,\footnotemark[1] & $\log R$\,\footnotemark[2] & References\,\footnotemark[3] \\
\hline
3C~59                      & [$-0.74$] &  $-1.47$  &  1 \\
PKS~0235+023               &  $-0.72$  & [$-1.39$] &  2 \\
NGC~1097\,\footnotemark[4]  &  $-0.58$  &   ...     &  3, 4 \\
1E 0450$-$1817             &   ...     &   ...     &  \\
Pictor~A                   & [$-0.71$] &  $-1.34$  &  1 \\
CBS~74                     &   ...     &   ...     &  \\
PKS~0921$-$213             &  $-0.38$  & [$+0.26$] &  5 \\
PKS~1020$-$103             &  $-0.43$  &  $-0.07$  &  5, 6 \\
3C~227                     & [$-0.78$] &  $-1.68$  &  1 \\
Mrk~668\,\footnotemark[5]   &   ...     &   ...     &  \\
3C~332                     & [$-0.85$] &  $-2.04$  &  7 \\
Arp~102B\,\footnotemark[6]  &  $+0.08$  &   ...     &  8, 9 \\
PKS~1739+184               &  $+0.07$  &  $-0.85$  &  10, 11 \\
3C~390.3\,\footnotemark[7]  &  $+0.39$  &  $-1.31$  &  12, 13 \\
\hline
\end{tabular}
\end{center}
\footnotetext[1]{The spectral index of the radio spectrum of the core under the convention $f_\nu\propto\nu^{+\alpha}$. Values in square brackets were inferred from $\log R$, as described Section~\ref{sec:discussion}.}
\footnotetext[2]{Measured at 5~GHz for 3C~59, PKS1020--103, and PKS~1739+184, at 5~GHz (core) and 408~MHz (lobes) for 3C~332, at 2.3~GHz for Pictor A and 3C~227, and at 1.48~GHz for 3C~390.3. All values of $\log R$ were converted to 5~GHz assuming $\alpha_{\rm core}=0$ and $\alpha_{\rm lobes}=-0.7$. Values in square brackets are inferred from the value of $\alpha_{\rm core}$ as described in Section~\ref{sec:discussion}.}
\footnotetext[3]{{\it References to sources of data:}
  (1) \citet{Bondi1993},
  (2) \citet{Downes1986},
  (3) \citet{Orienti2010},
  (4) \citet{Hummel1987},
  (5) \citet{Drinkwater1997},
  (6) \citet{Lister1994},
  (7) \citet{Giovannini1988}
  (8) \citet{Helmboldt2007},
  (9) \citet{Healey2007},
 (10) \citet{Swarup1984},
 (11) \citet{Gower1984},
 (12) \citet{Leahy1991},
 (13) \citet{Leahy1995}.
}
\footnotetext[4]{NGC~1097 hosts a compact nuclear radio source. Although the spectral index of the core is available, we do not consider this a useful constraint on the presence of a jet or its possible orientation.}
\footnotetext[5]{Mrk~668 hosts a compact, GHz-peaked radio source \citep[see][]{ODea1998}. We have not found any information that can constrain the jet orientation.}
\footnotetext[6]{The radio maps of Arp~102B show a core with an elongation on one side, reminiscent of a jet, but no lobes.}
\footnotetext[7]{The orientation of the radio jet of 3C~390.3 has been examined by \citet{Eracleous1996} who constrained the inclination angle to be in the range $i=19$--26$^\circ$ based on the observed superluminal motion and the morphology of the double-lobed radio source.}
\end{minipage}
\end{table}

\begin{table}
\begin{minipage}{\linewidth}
\renewcommand{\thefootnote}{\alph{footnote}}
\caption{Arguments Against SBHB Hypothesis}
\label{tab:arguments}
\begin{center}
\begin{tabular}{lccc}
\hline
\hline
       &                                    & Large                       & Other\\
Object & $\chi^2_\nu > 2$?\,\footnotemark[1] & $M_{min}$?\,\footnotemark[2] & Arguments?\,\footnotemark[3] \\
\hline
3C~59                      & yes &     &     \\ 
PKS~0235+023               &     &     & yes \\ 
NGC~1097                   & yes &     & yes \\ 
1E 0450$-$1817             & yes &     &     \\ 
Pictor~A                   & yes & yes & yes \\ 
CBS~74                     & yes &     &     \\ 
PKS~0921$-$213             & yes & yes & yes \\ 
PKS~1020$-$103             &     & yes & yes \\ 
3C~227                     &     &     &     \\ 
Mrk~668                    & yes &     &     \\ 
3C~332                     &     & yes & yes \\ 
Arp~102B                   &     & yes & yes \\ 
PKS~1739+184               &     &     & yes \\ 
3C~390.3                   &     & yes & yes \\ 
\hline
\end{tabular}
\end{center}
\footnotetext[1]{The radial observed velocity curves are not described well by an orbital model. See discussion at the beginning of Section~\ref{sec:discussion}.}
\footnotetext[2]{The minimum BH mass exceeds $ 10^{10}\;\Msol$ after considering constraints on the inclination angle. See discussion in the middle of Section~\ref{sec:discussion}.}
\footnotetext[3]{See discussion in Section~\ref{sec:discussion}.}
\end{minipage}
\end{table}

We also re-iterate additional general arguments against the binary interpretation of double-peaked emitters.

\begin{enumerate}

\item
As \cite{Liu2016} pointed out, the Ly$\alpha$ lines of double-peaked emitters are in fact single-peaked \citep[Arp~102B, 3C~390.3, NGC~1097, PKS~0921--213 -- see][respectively]{Halpern1996, OBrien1998, Storchi-Bergmann2005, Eracleous2009}. This is a strong argument against the binary hypothesis, as one would expect the Ly$\alpha$ line profile to mirror that of the Balmer lines (i.e. double-peaked) if the peaks originate from separate and physically distinct regions.

\item
\cite{Liu2016} noted that the minimum masses inferred from fitting the radial velocity curves are substantially larger than the masses determined from the stellar velocity dispersions of the host galaxies by \citet{Lewis2006} for six objects in our sample (NGC~1097, 1E~0450.3--1817, Pictor~A, PKS~0921--213, Arp~102B, 3C~390.3).

\item
  \cite{Storchi-Bergmann2003} analysed spectra of NGC~1097 spanning 11~yr and noted that the relative separation of the two peaks in the line profile change with the luminosity of the AGN. They reported a decrease of the broad-line flux as the red and blue peak moved further apart from one another (see also \citealt{Schimoia2012,Schimoia2015}). Additionally, \cite{Shapovalova2001} monitored 3C~390.3 during the period 1995-2000 and reported similar behavior and \citet{Zhang2013} reached the same conclusion after studying a different set of observations of the same object. This behavior is a manifestation of the ``breathing'' of a single BLR, observed in variability studies of Seyfert galaxies and quasars \citep[e.g.,][]{Guo2014, Barth2015} and suggests strongly that the two sides of double-peaked emission lines originate in the same BLR.

\item
In a binary scenario, the velocity of the gas orbiting individual BHs is at least several times greater than the velocity of the BHs orbiting one another. Thus, we would expect the two peaks that make up the line profiles to be separated by less than their width, instead of well-separated peaks as is the case for eight objects (PKS~0235+023, NGC~1097, 1E~0450.3--1817, PKS~0921$-$213, PKS~1020--103, Arp~102B, 3C~332, 3C~390.3; see footnote 3 in \citealt{Chen1989a}, the heuristic models of \citealt{Shen2010}, and the detailed models of \citealt{Nguyen2019}). 

\item
Reverberation mapping projects offer additional evidence against the SBHB hypothesis. In the SBHB picture, the two BHs are typically at parsec or sub-parsec separation, so it would take years, or at least months, for changes in the continuum of one BH to reach the BLR of the other. Thus, one should expect a corresponding time lag between the two peaks as they respond to changes in the continuum. However, \cite{Dietrich1998} performed reverberation mapping studies on 3C~390.3 and report that the blue and the red peaks in the double-peaked line profiles vary simultaneously within 3 days. Similarly, \cite{Sergeev2000} and more recently \cite{Shapovalova2013} who performed reverberation mapping studies on Arp~102B found no significant time lag between the two peaks.

\end{enumerate}

We conclude our evaluation of the SBHB hypothesis by summarizing the Table~\ref{tab:arguments} the arguments against this hypothesis for each of the objects in our sample. Included in this table are columns on the goodness of the fit to the radial velocity curves with orbital models, the minimum SBHB mass (after constraining $\sin i$), and additional arguments from other observations presented just above. We populate each of these four columns depending on whether an argument applies to a particular object in our sample. As Table~\ref{tab:arguments} shows there is only one object that passes all the tests (3C~227) and only four objects fail only one test (3C~59, PKS~0235+023, CBS~74, and Mrk~668). Therefore, we disfavour the SBHB hypothesis for 2/3 of the objects in our sample; for Arp~102B and 3C~390.3, the hypothesis can be rejected with great confidence based on the minimum masses alone, even for $\sin i =1$.  Several other objects identified earlier in this section have minimum masses very close to the threshold of $10^{10}\;\Msol$, especially when the orbital inclination angle is taken into account. If we insist on $\sin i =1$, four more objects fail the $\chi_\nu^2$ test {\it and} are disfavoured by additional arguments. Finally, it is important to point out that, if double-peaked emitters are a homogeneous sub-class of AGNs, as their general properties suggest, and the SBHB hypothesis is disfavoured for a substantial fraction of them, then we should not expect {\it any} of them to host SBHBs. Future observational work can test this conclusion further. Specifically, continued spectroscopic monitoring will extend the radial velocity curves of all objects and lead to better constraints on the SBHB mass. Additional constraints on the orbital inclination angle through additional radio observations will help improve the mass limits from the radial velocity curves. Other tests involving UV spectroscopy \citep[see][]{Eracleous2009} and reverberation mapping \citep[see][]{Dietrich1998} will also be helpful.

As the binary hypothesis is unlikely for most of our targets, it is necessary to focus on other hypotheses for objects with double-peaked Balmer lines. To explain the double-peaked profiles of 3C~390.3 and Arp~102B, \citet{Chen1989a} and \cite{Chen1989b} proposed a model consisting of only one BH at the centre, and a disc illuminated by a central source. \cite{Schimoia2015} argue that the radial velocity changes of the Balmer line peaks could be explained by variations in the illumination of such a disc. The model and its extensions can also fit many of the other double-peaked profiles in our sample and the physical picture associated with this model has passed all the observational tests carried out so far \citep{Eracleous1994, Eracleous1995, Eracleous2003}.

\section*{Acknowledgements}

We thank Jason Wright for useful discussions on modeling jitter as a Gaussian process and Jupes Halpern for a careful reading of teh manuscript. 

JCR acknowledges support from the National Geographic Society under grant no. W455-16.  She thanks the staff at Las Campanas observatory for their expert help in carrying out the Magellan/MageE observations and the University of Michigan Astronomy department for access to this instrument. This paper includes data gathered with the 6.5 meter Magellan Telescopes located at Las Campanas Observatory, Chile.

We thank the staff at MDM Observatory, Cerro-Tololo INteramerican Observatory, Apache Point Observatory, and the Hobby-Eberly Telescope for their expert help in carrying out the observations. 

The Hobby-Eberly Telescope (HET) is a joint project of the University of Texas at Austin, the Pennsylvania State University, Stanford University, Ludwig-`Maximillians-Universit\"at M\"unchen, and Georg-August-Universit\"at G\"ottingen. The HET is named in honor of its principal benefactors, William P. Hobby and Robert E. Eberly.

The Marcario Low-Resolution Spectrograph is named for Mike Marcario of High Lonesome Optics, who fabricated several optics for the instrument but died before its completion; it is a joint project of the Hobby-Eberly Telescope partnership and the Instituto de Astronom\'{\i}a de la Universidad Nacional Aut\'onoma de M\'exico.

The OSMOS spectrograph has been generously funded by the National Science Foundation and the Center for Cosmology and AstroParticle Physics at The Ohio State University. Additional support has also been provided by the Department of Astronomy at The Ohio State University and the Department of Physics and Astronomy at Ohio University.

This work is based on observations obtained with the Apache Point Observatory 3.5m telescope, which is owned and operated by the Astrophysical Research Consortium. 

This research has made use of the NASA/IPAC Extragalactic Database (NED) which is operated by the Jet Propulsion Laboratory, California Institute of Technology, under contract with the National Aeronautics and Space Administration.

%%%%%%%%%%%%%%%%%%%%%%%%%%%%%%%%%%%%%%%%%%%%%%%%%%



\bibliographystyle{mnras}
\bibliography{reference}

%%%%%%%%%%%%%%%%%%%% LARGE FIGURE MOSAICS %%%%%%%%%%%%%%%%%%

\begin{figure*}
  \hbox{
    \includegraphics[height=7.cm]{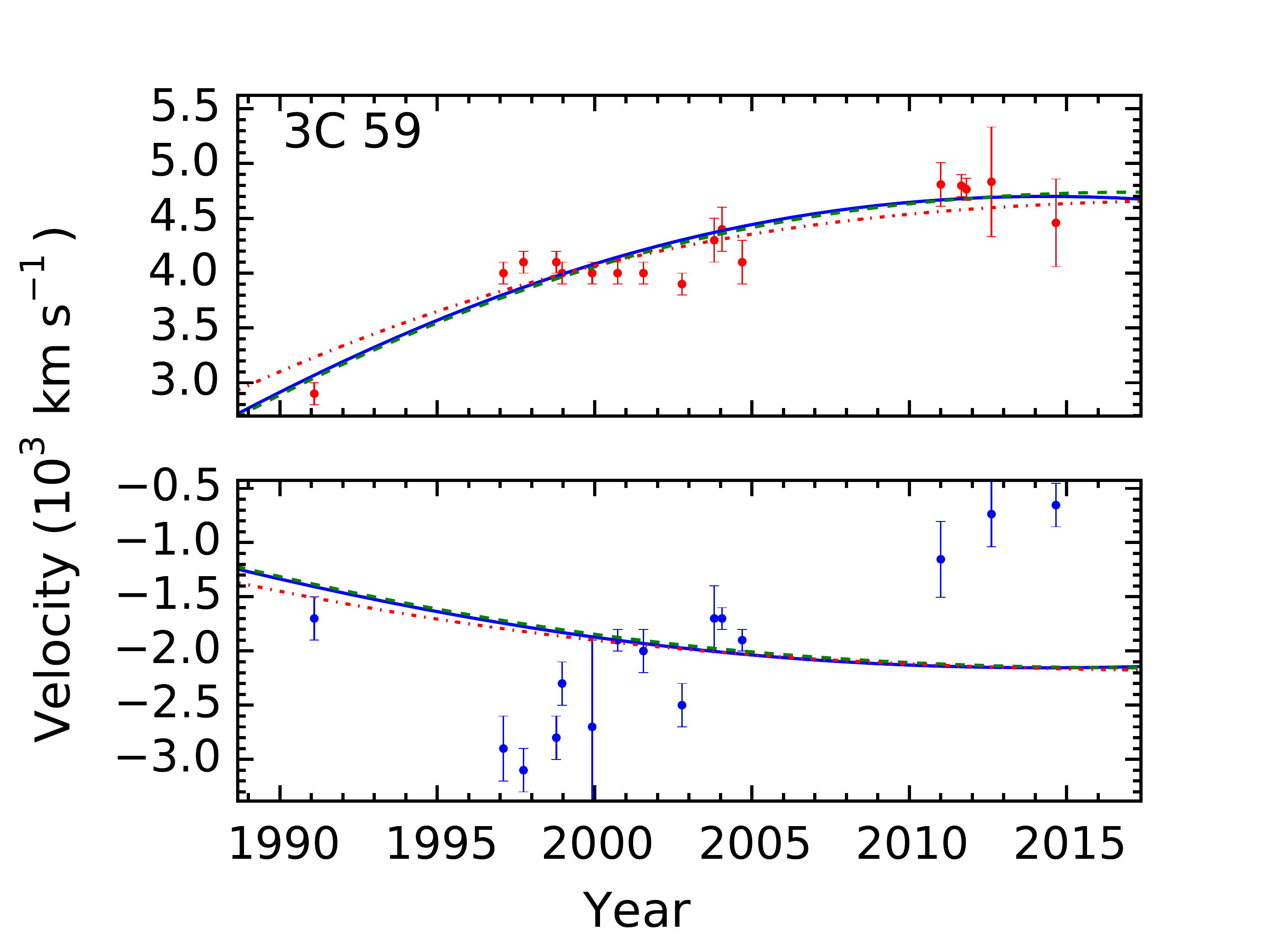}
    \hfill
    \includegraphics[height=7.cm]{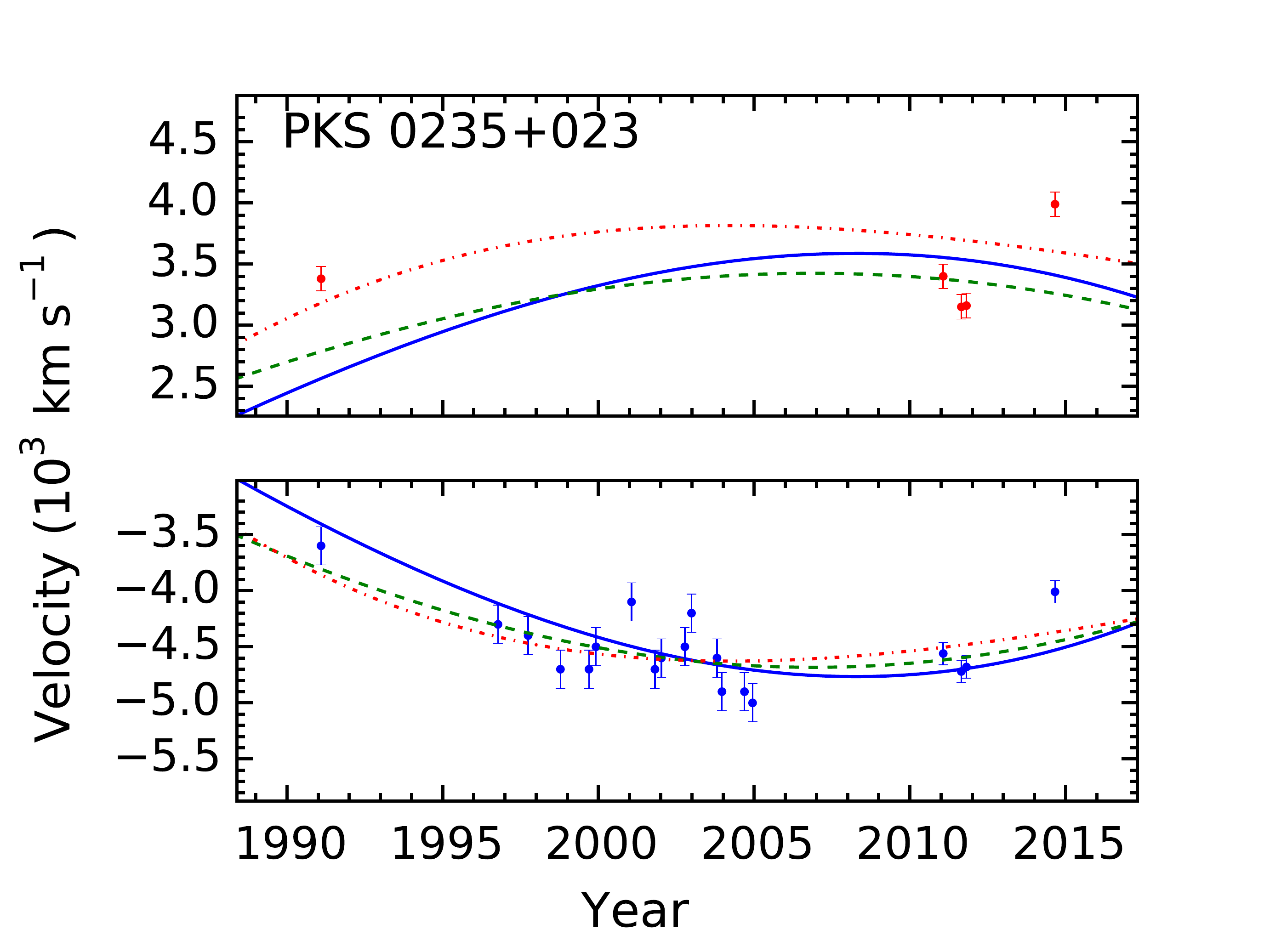}
  }
  \hbox{
    \includegraphics[height=7.cm]{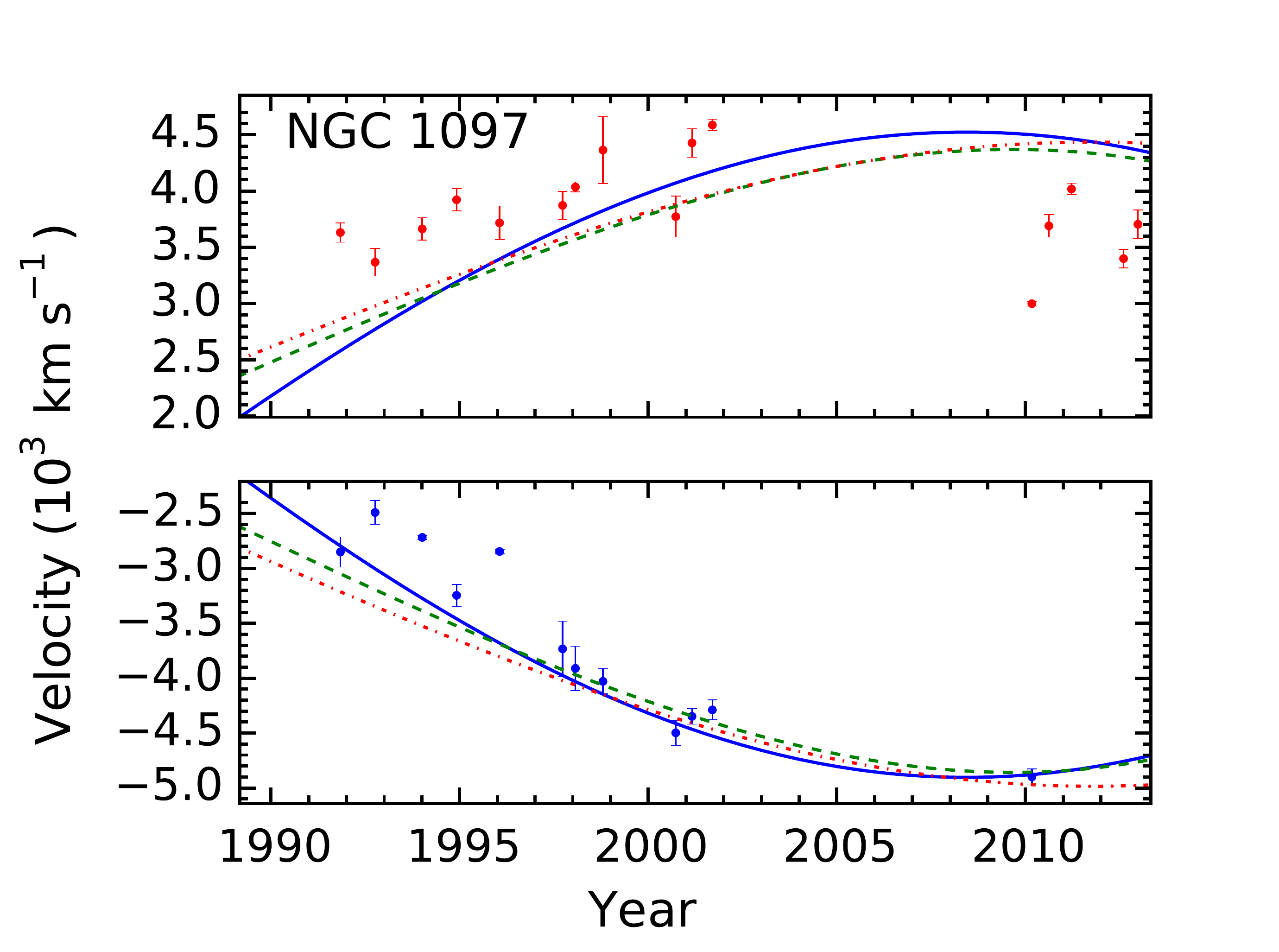}
    \hfill
    \includegraphics[height=7.cm]{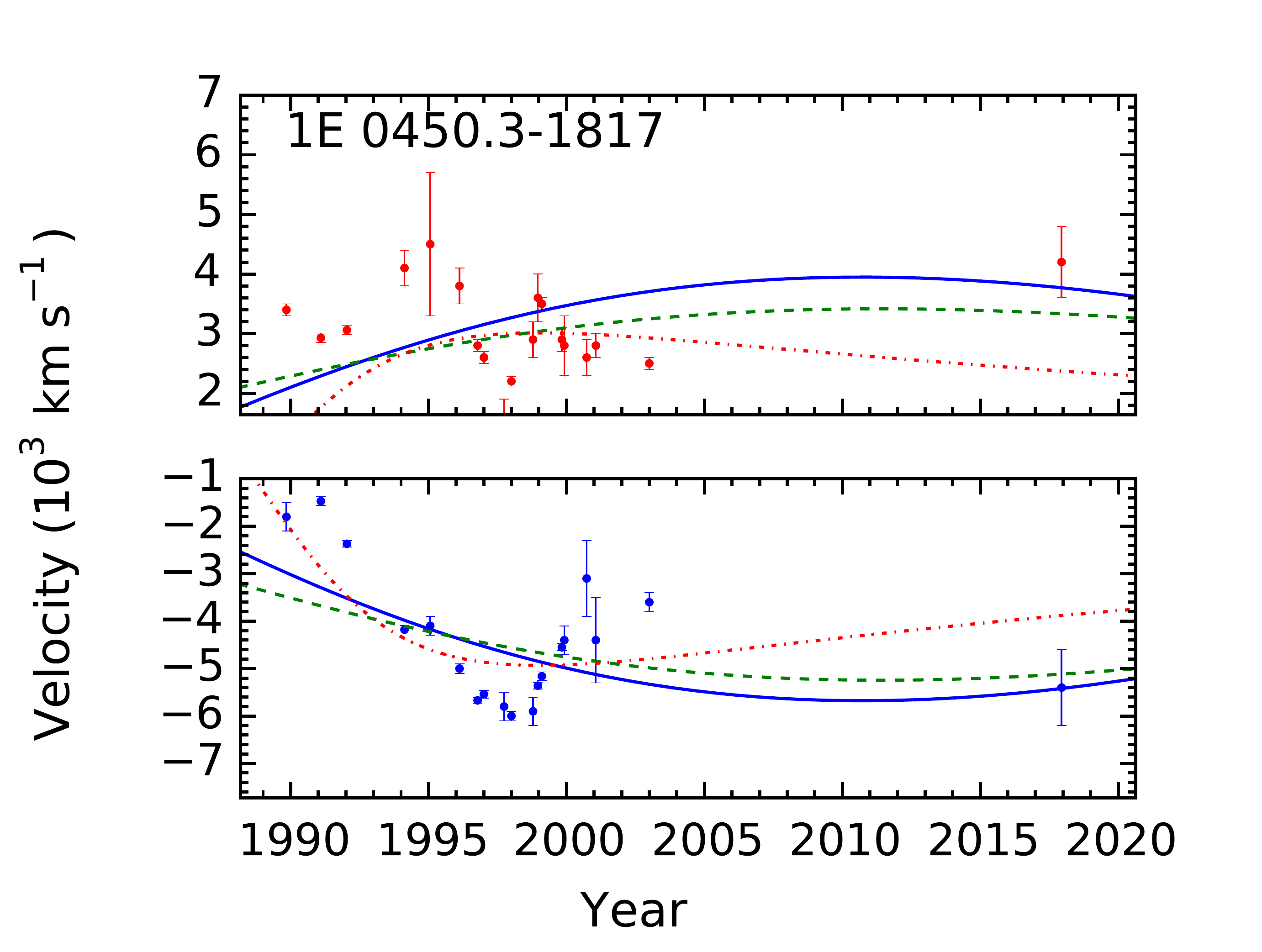}
  }
  \hbox{
    \includegraphics[height=7.cm]{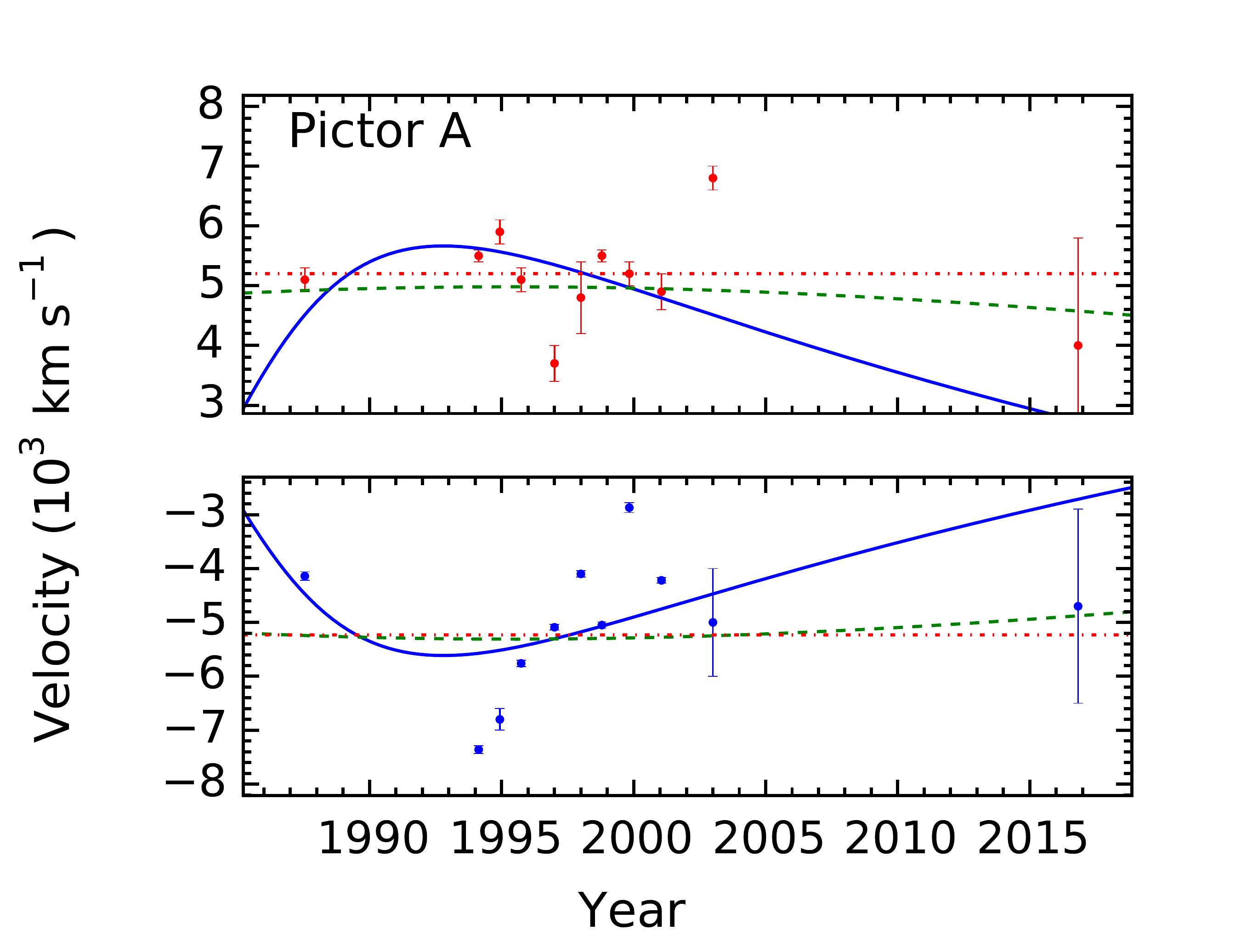}
    \hfill
    \includegraphics[height=7.cm]{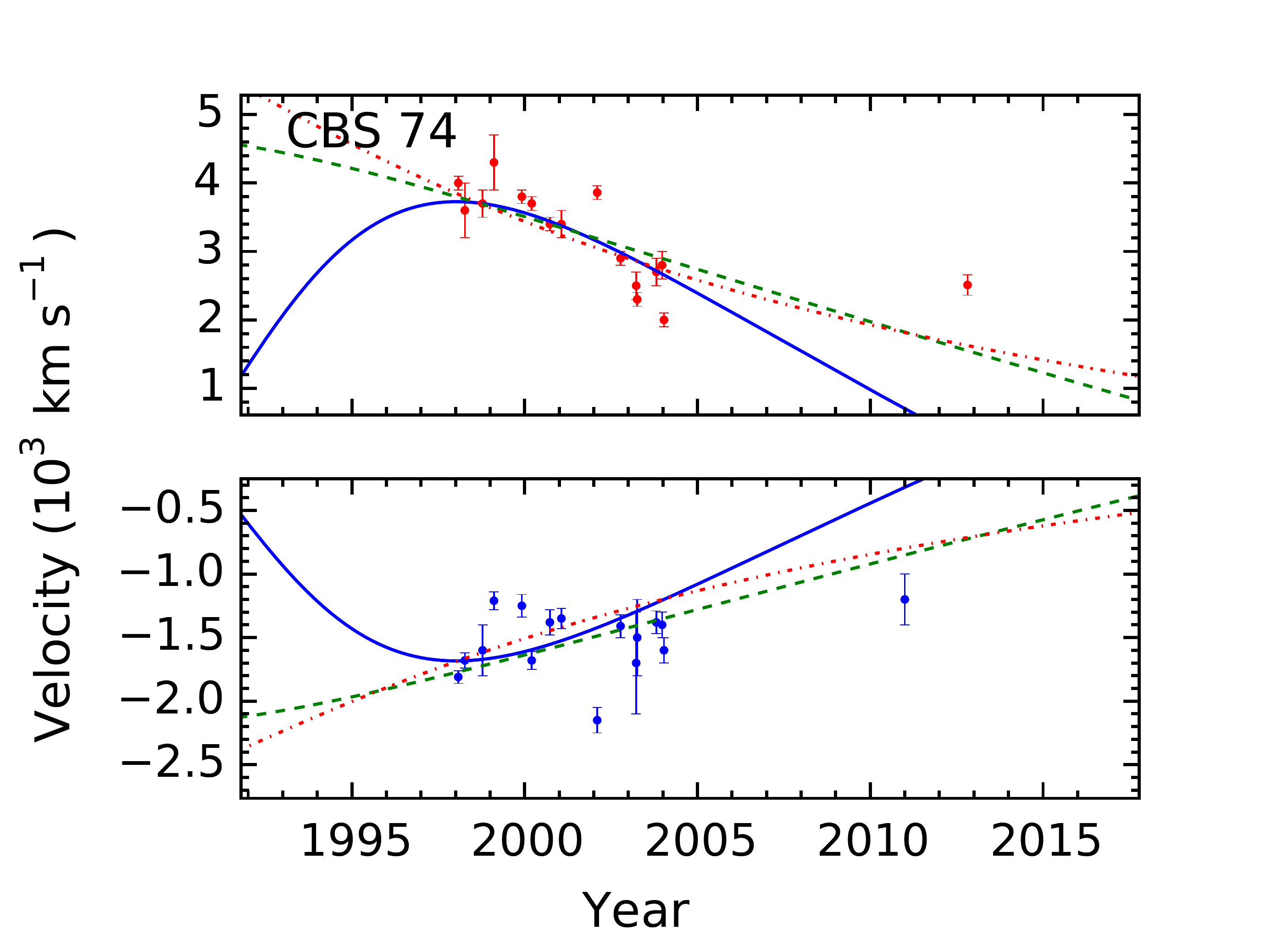}
  }
  \caption{The radial velocity curves of the red and blue peaks of all the targets. Three sample solutions, representing the 99\% , 90\% and 68\% lower limit on period, are also plotted as blue, green, and red lines respectively, superimposed on the data.}
  \label{fig:radvel}
\end{figure*}

\begin{figure*}
  \ContinuedFloat
  \hbox{
    \includegraphics[height=7.cm]{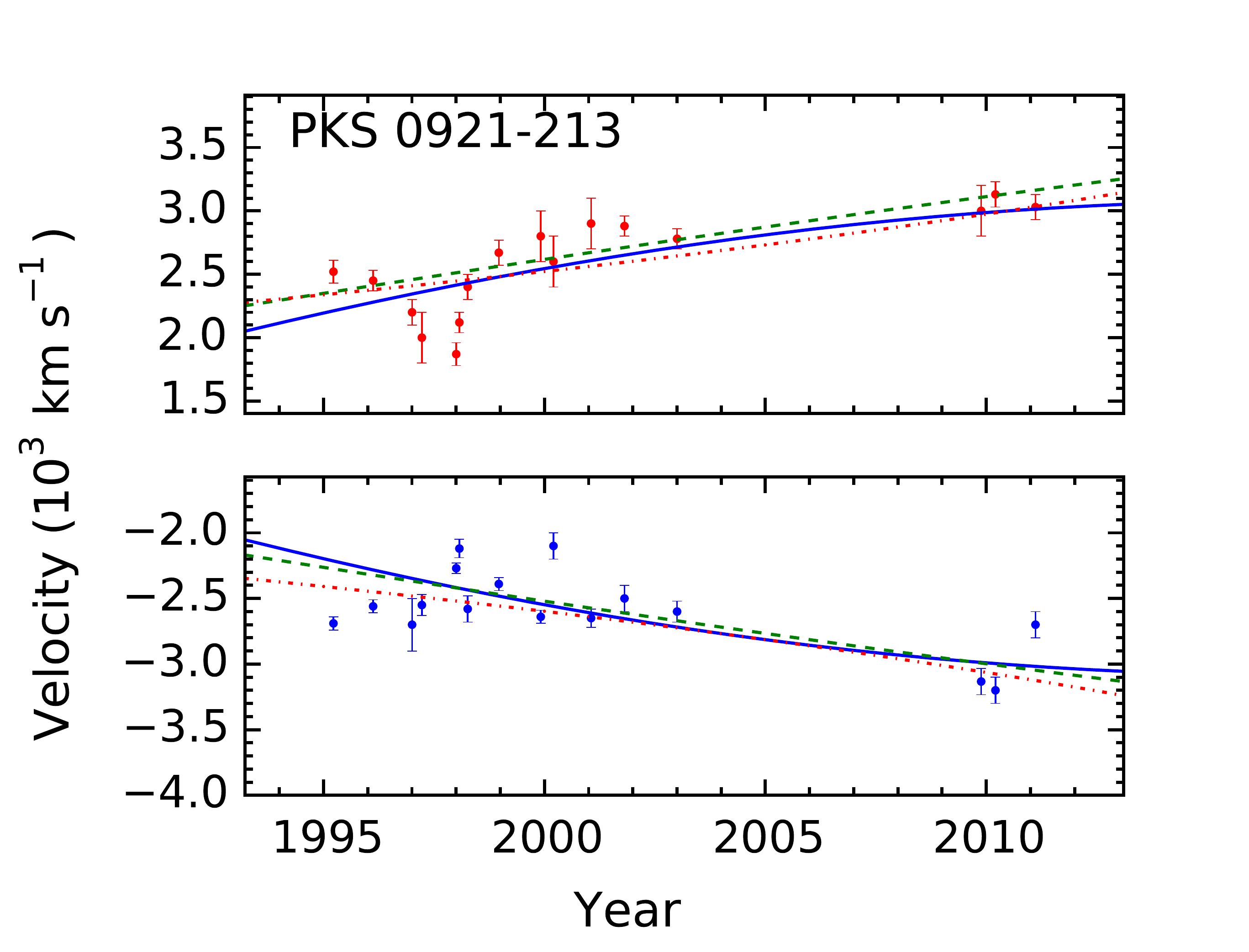}
    \hfill
    \includegraphics[height=7.cm]{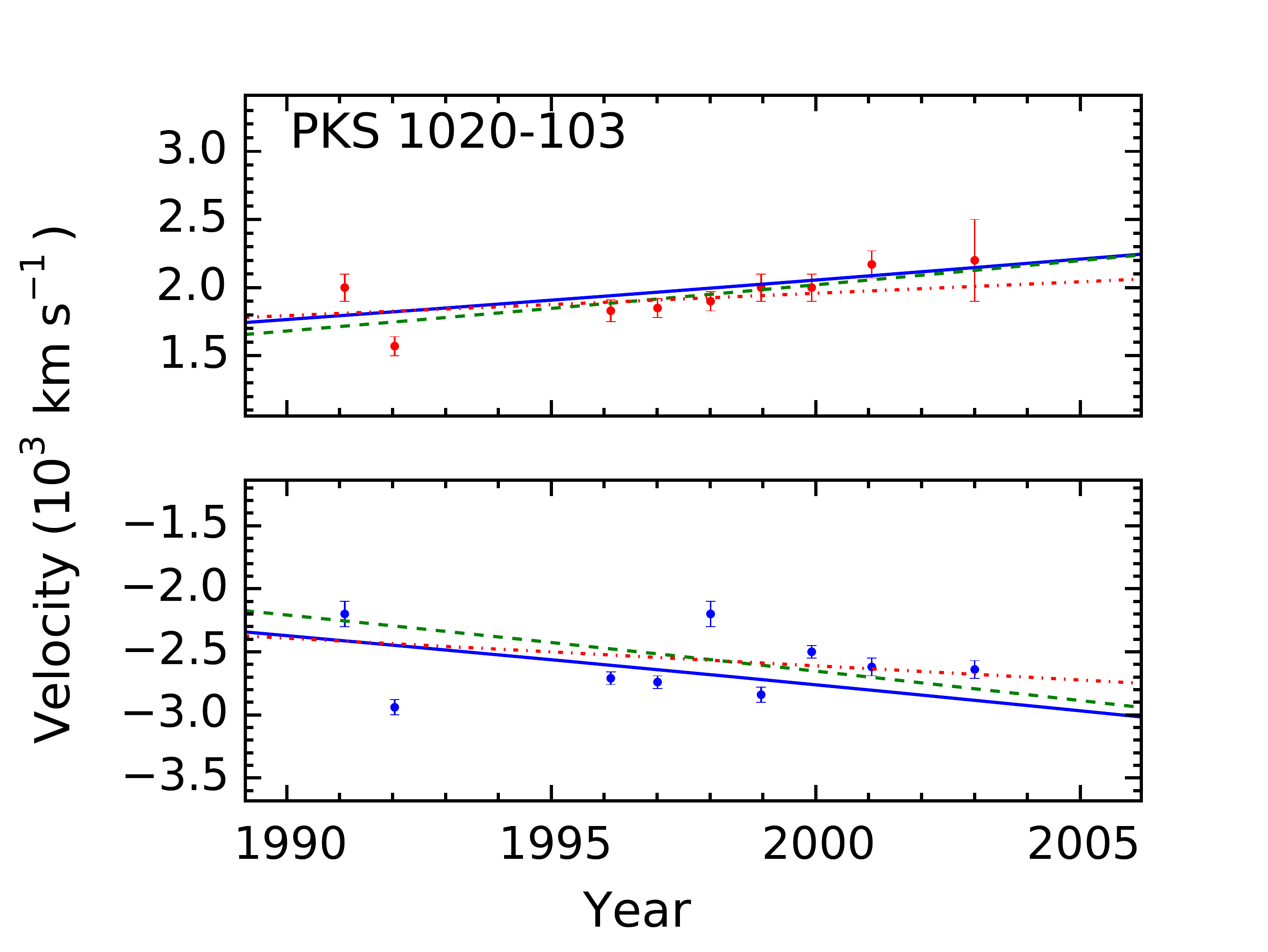}
  }
  \hbox{
    \includegraphics[height=7.cm]{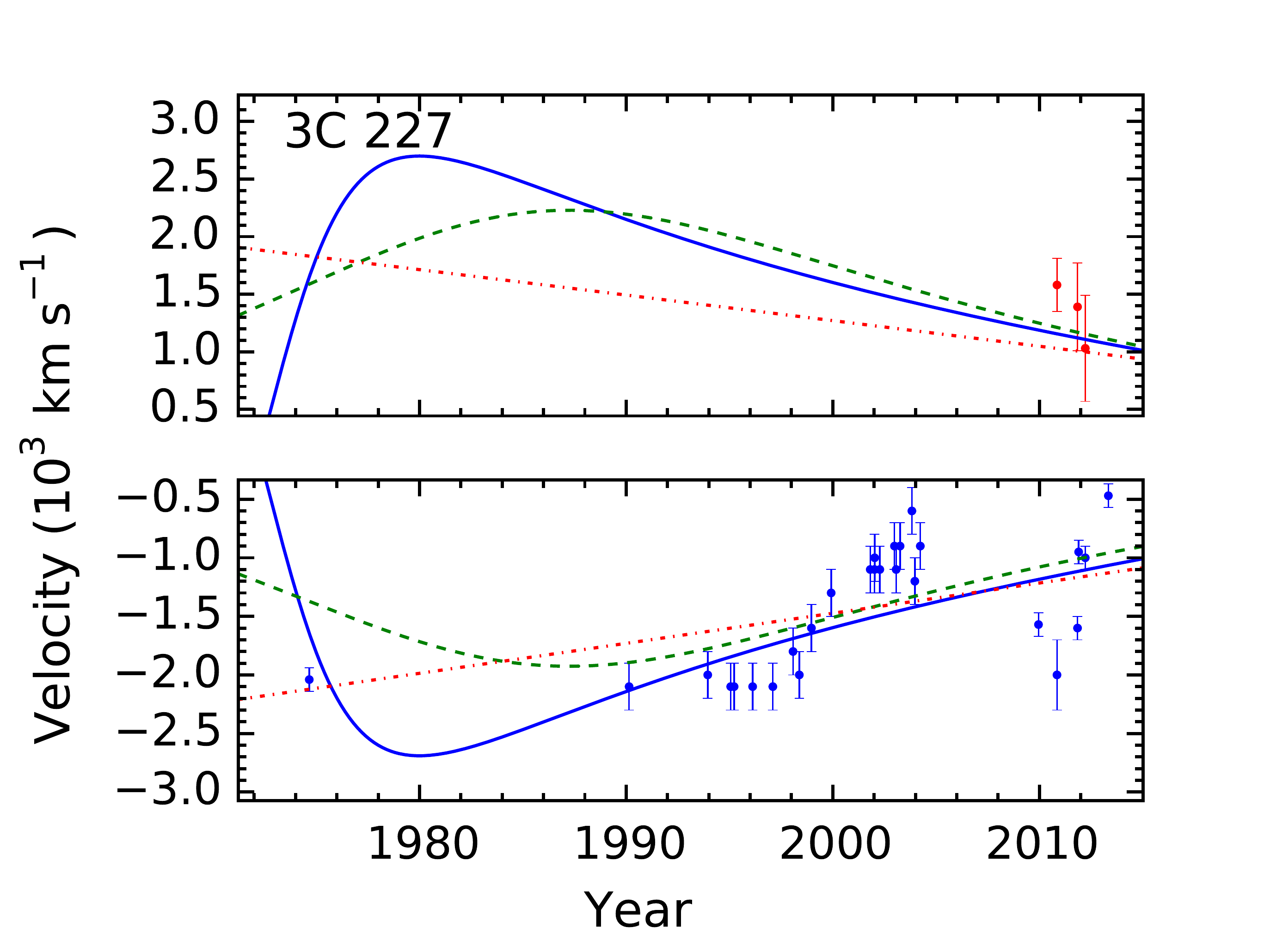}
    \hfill
    \includegraphics[height=7.cm]{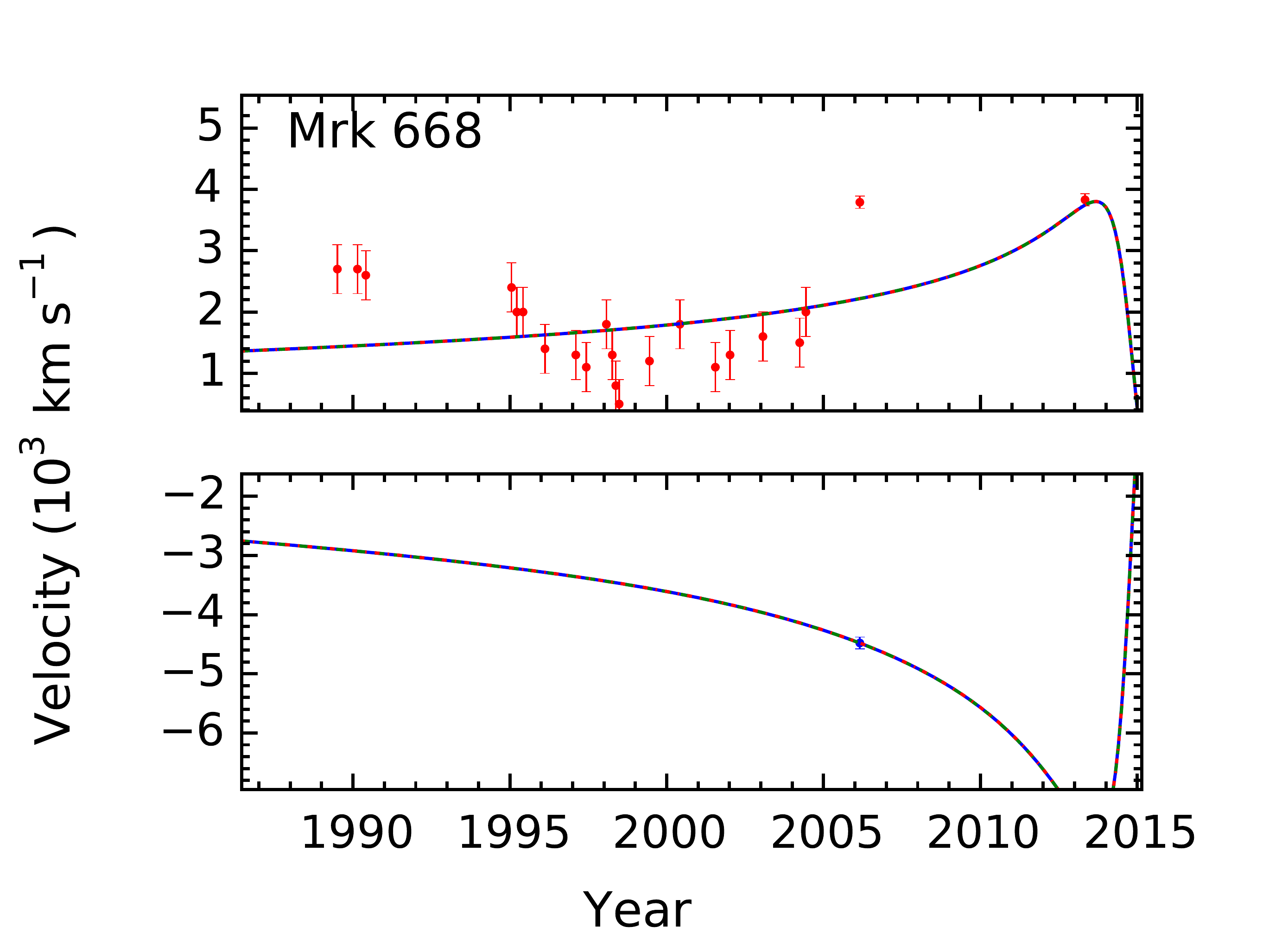}
  }
  \hbox{
    \includegraphics[height=7.cm]{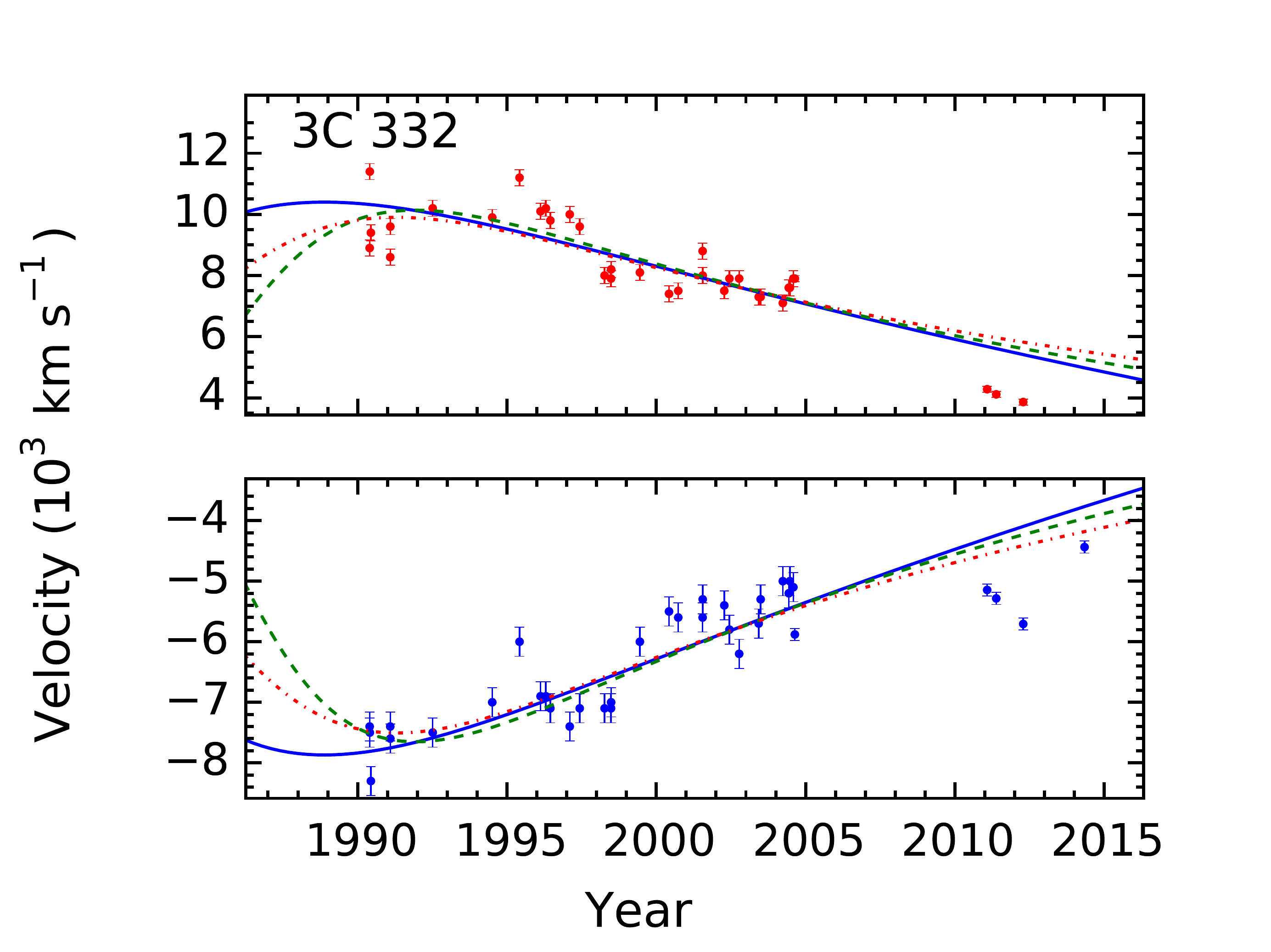} 
    \hfill
    \includegraphics[height=7.cm]{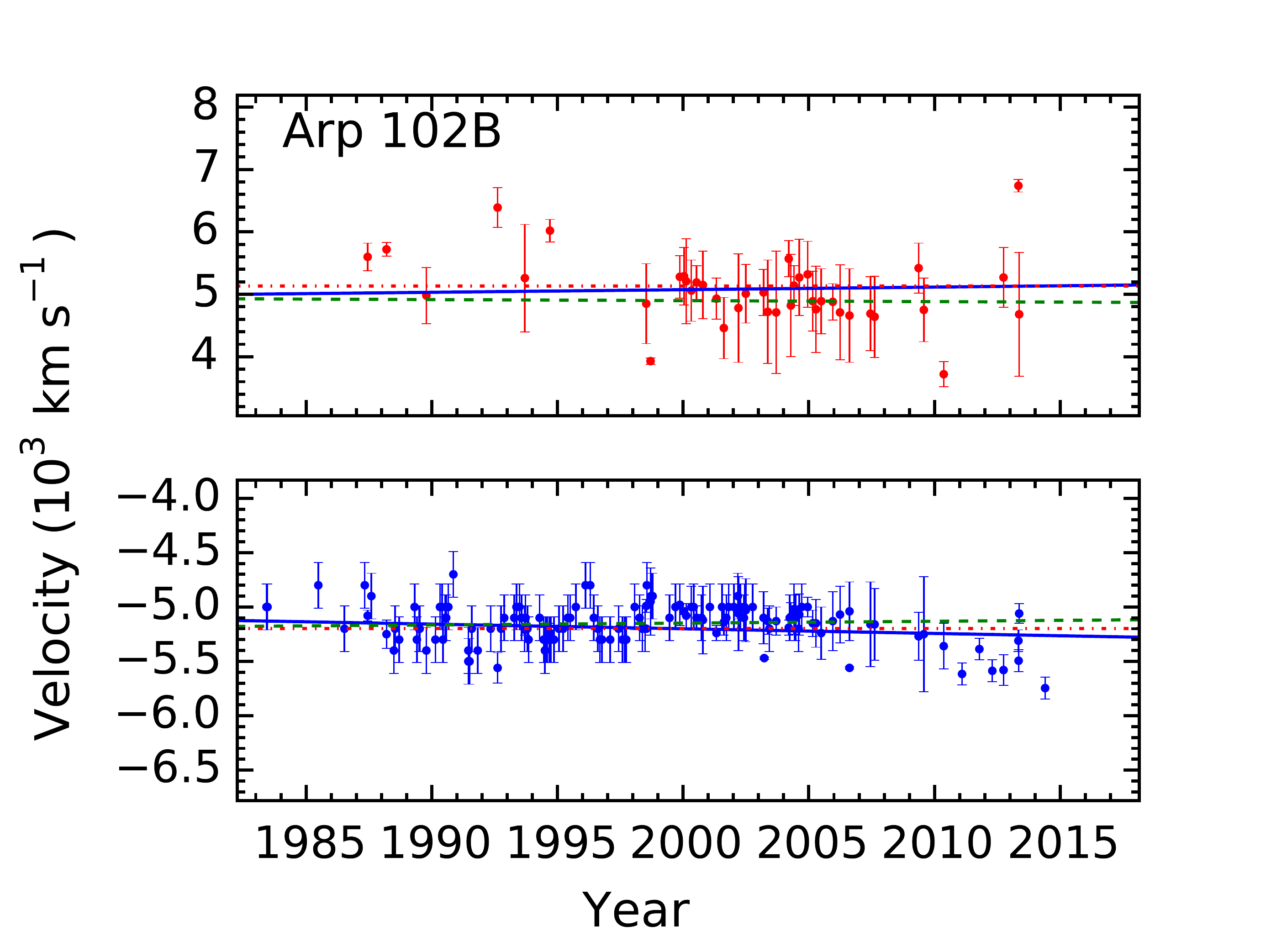}
  }
  \caption{ (continued)}
\end{figure*}

\begin{figure*}
  \ContinuedFloat
  \hbox{
    \includegraphics[height=7.cm]{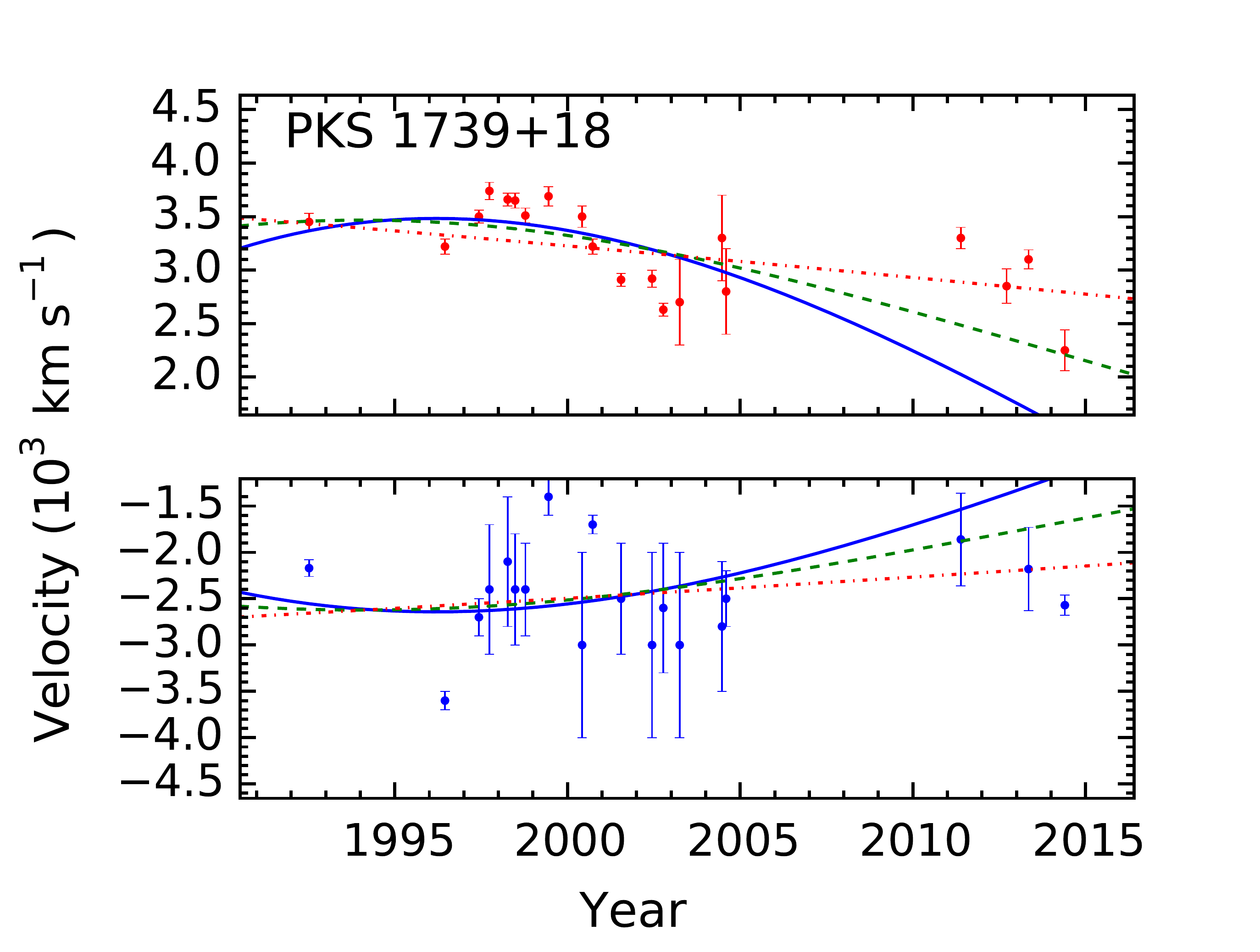}
    \hfill
    \includegraphics[height=7.cm]{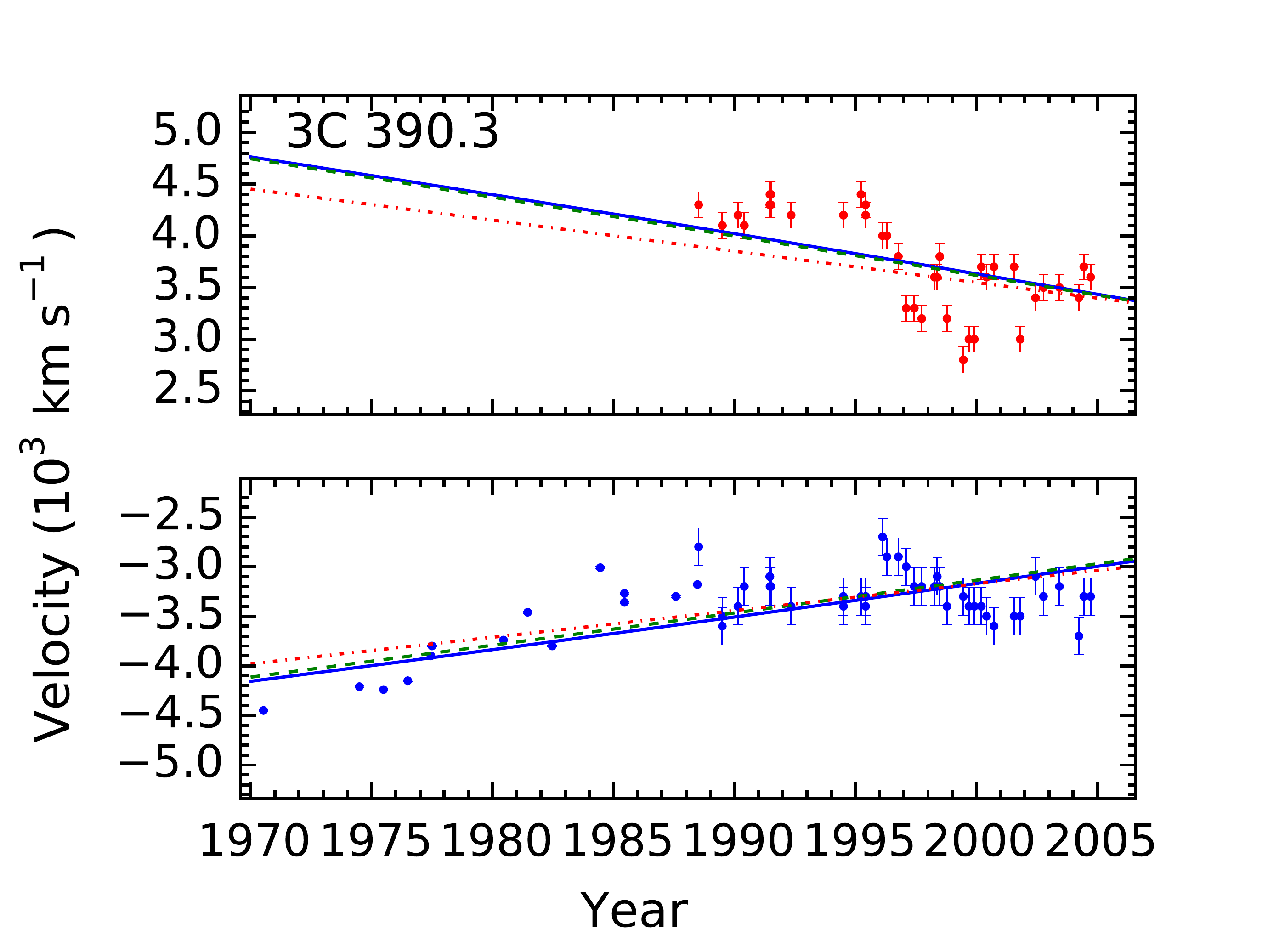}
  }
  \caption{(continued)}
\end{figure*}

%----------------------------------------------------------------

\begin{figure*}
  \hbox{
    \includegraphics[height=7.cm]{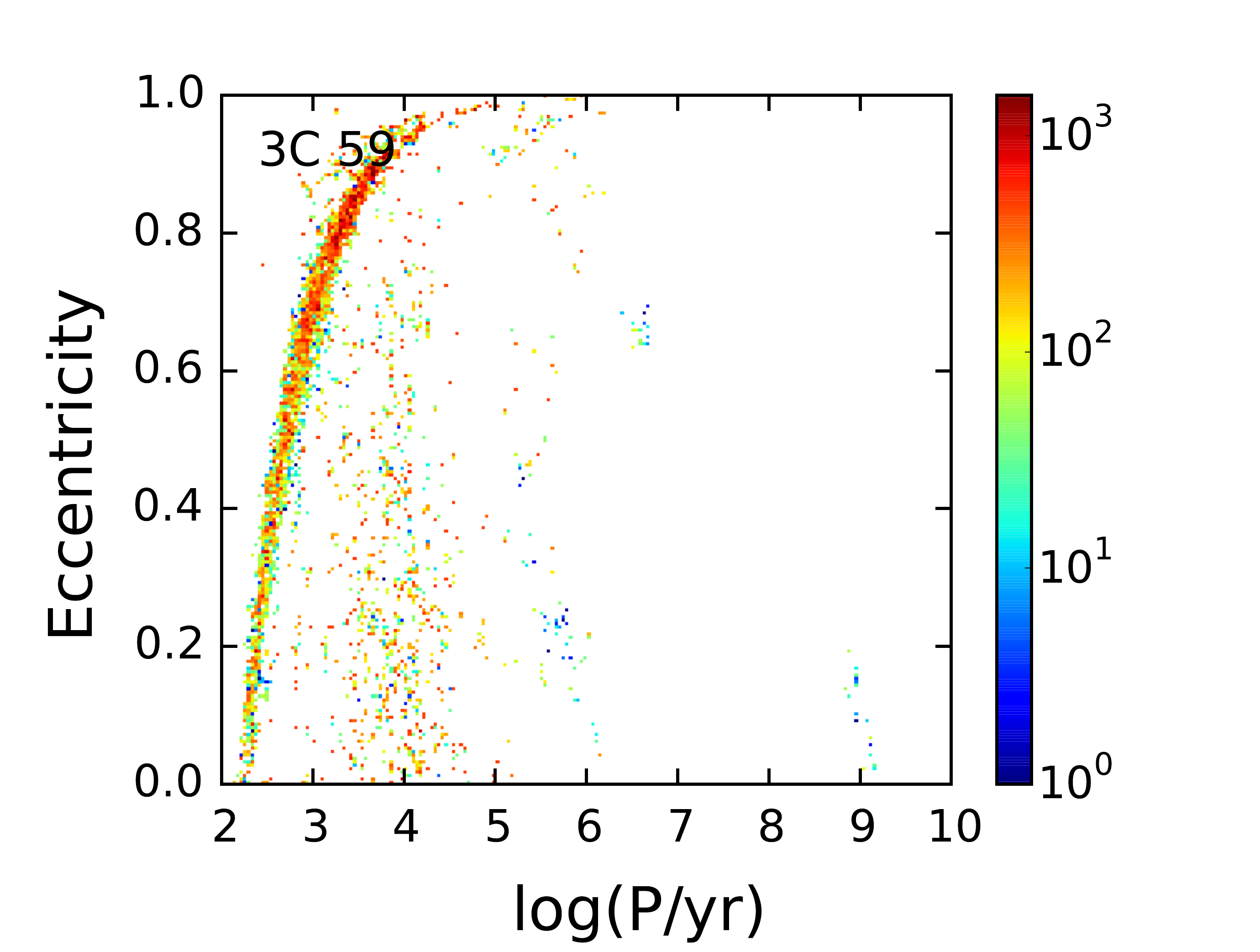}
    \hfill
    \includegraphics[height=7.cm]{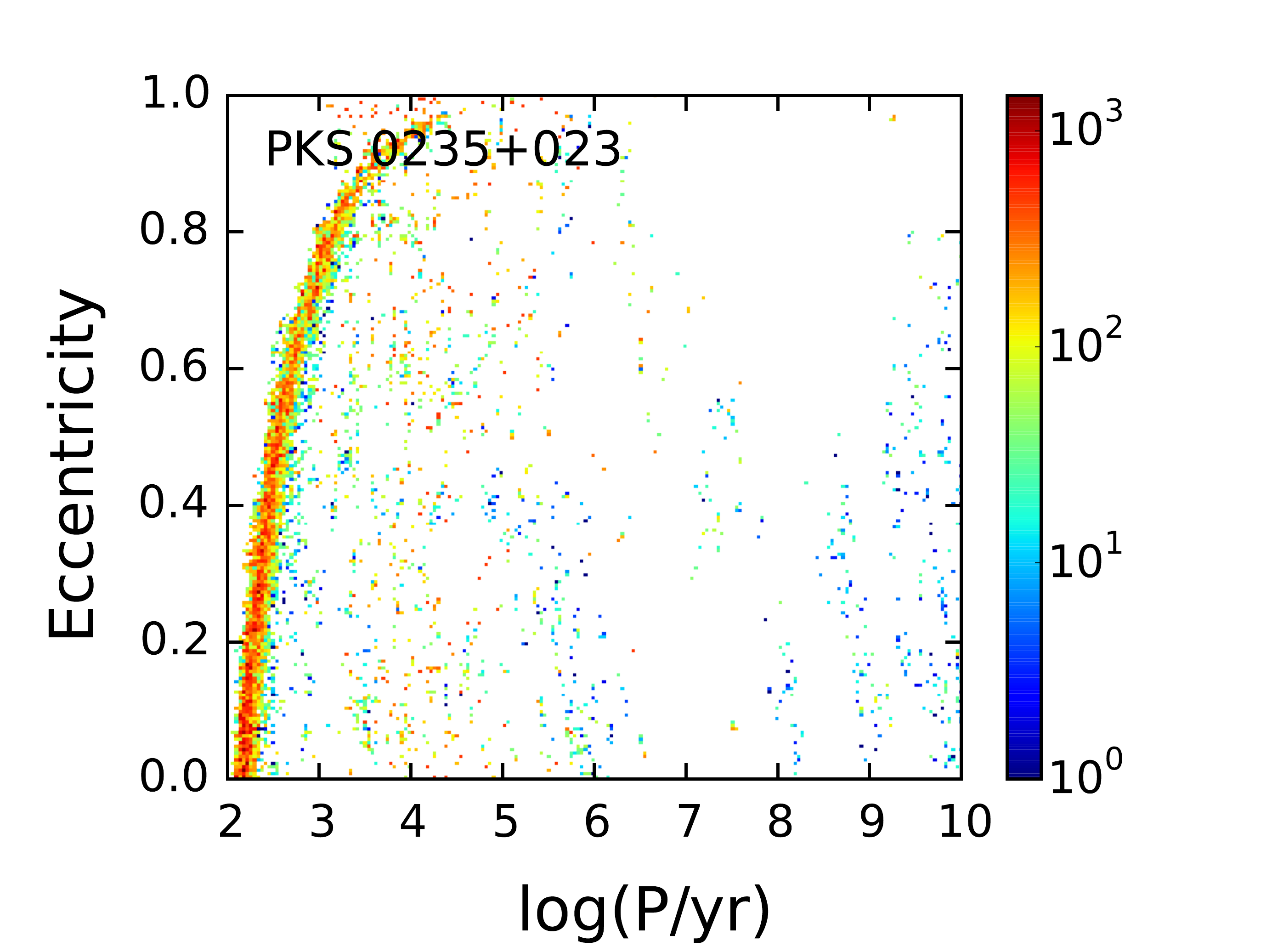}
  }
  \hbox{
    \includegraphics[height=7.cm]{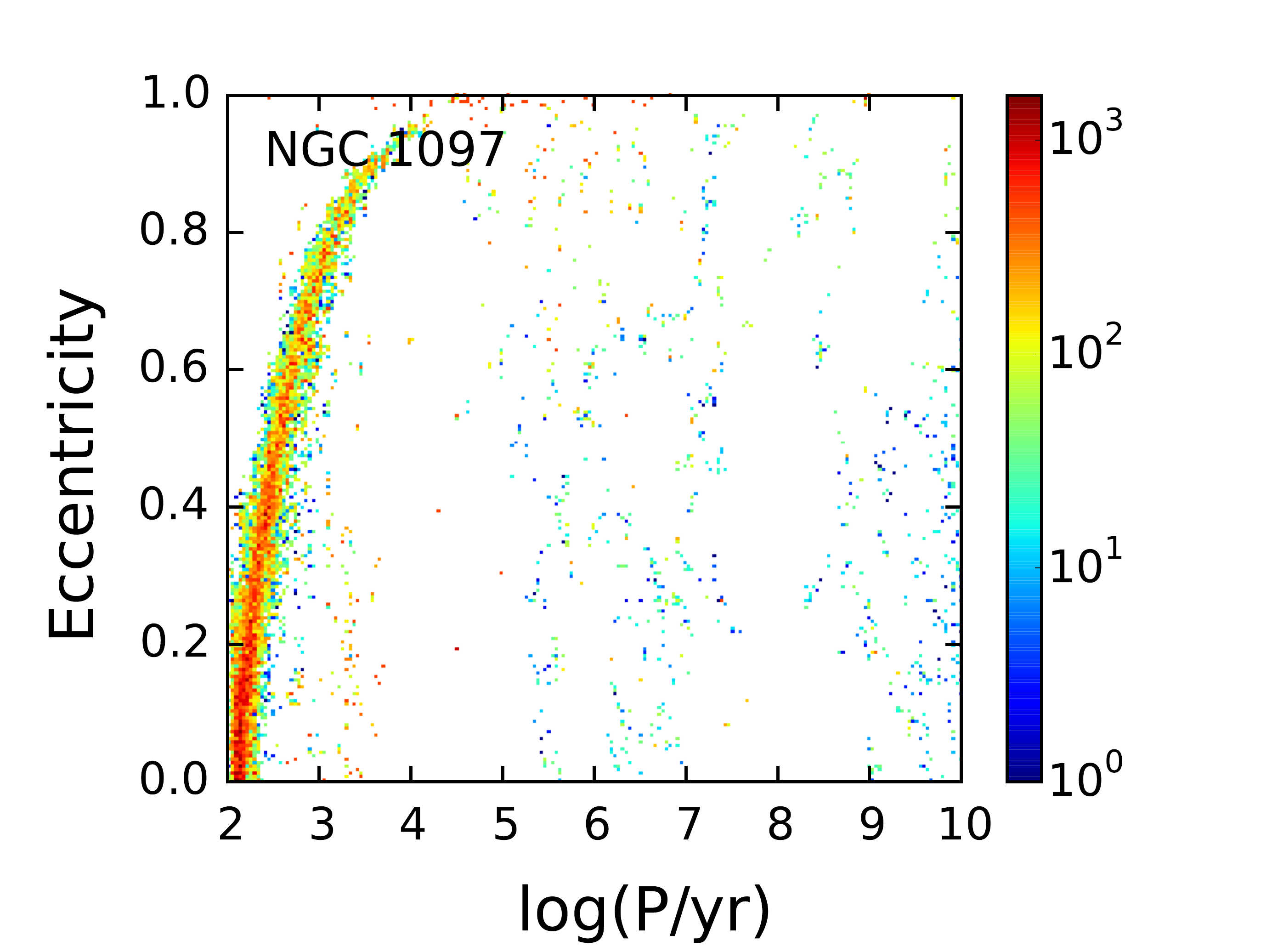}
    \hfill
    \includegraphics[height=7.cm]{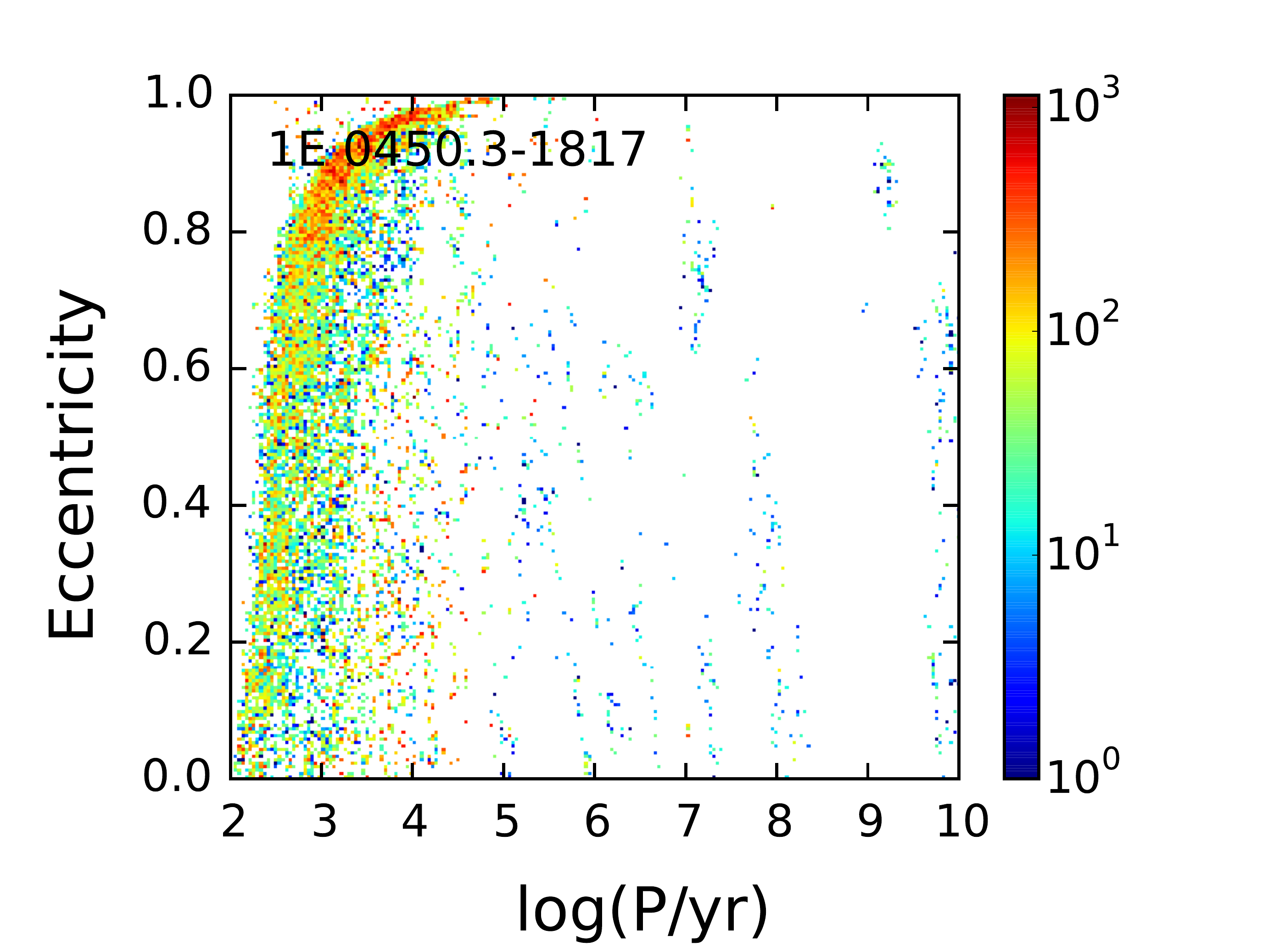}
  }
  \hbox{
    \includegraphics[height=7.1cm]{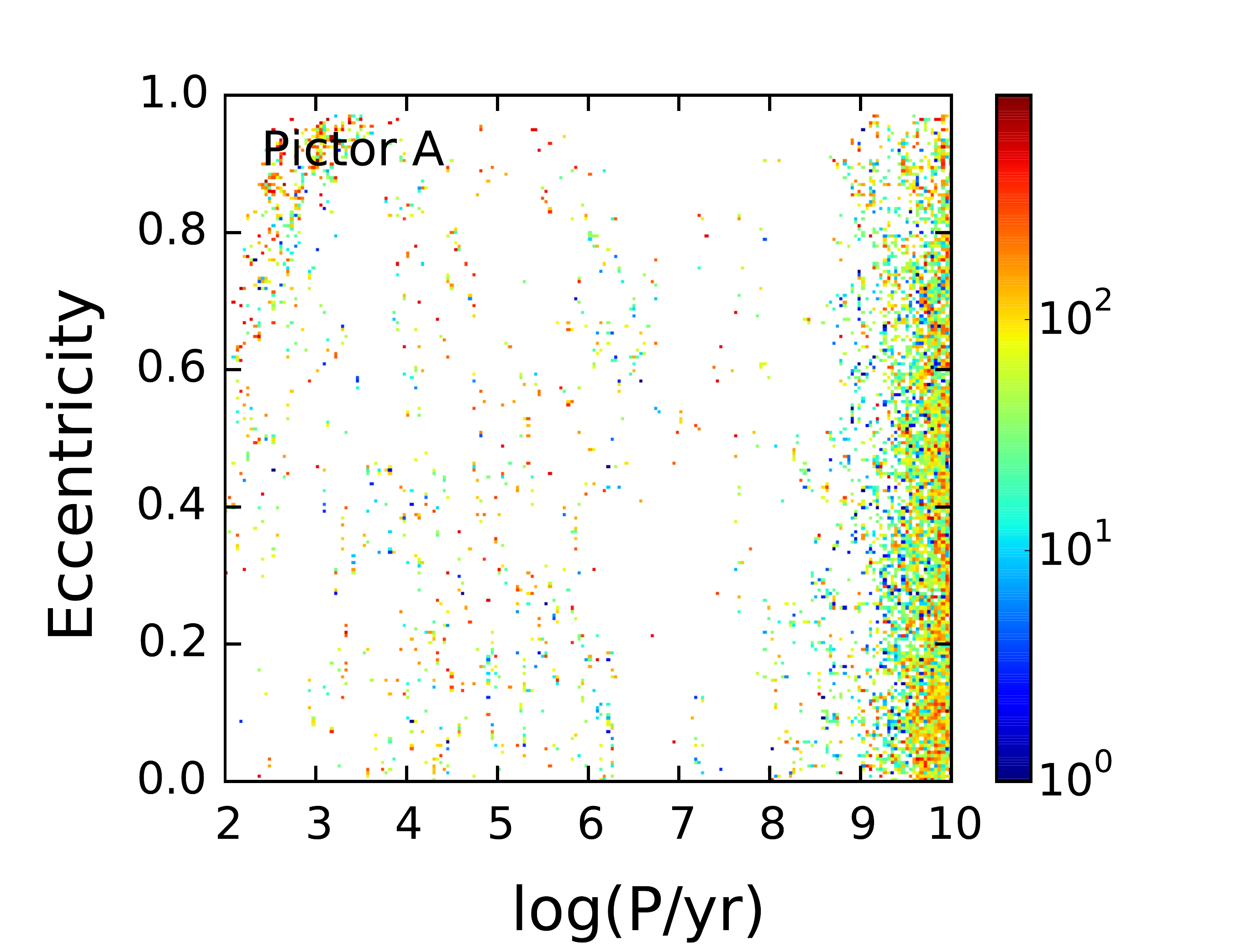}
    \hfill
    \includegraphics[height=7.1cm]{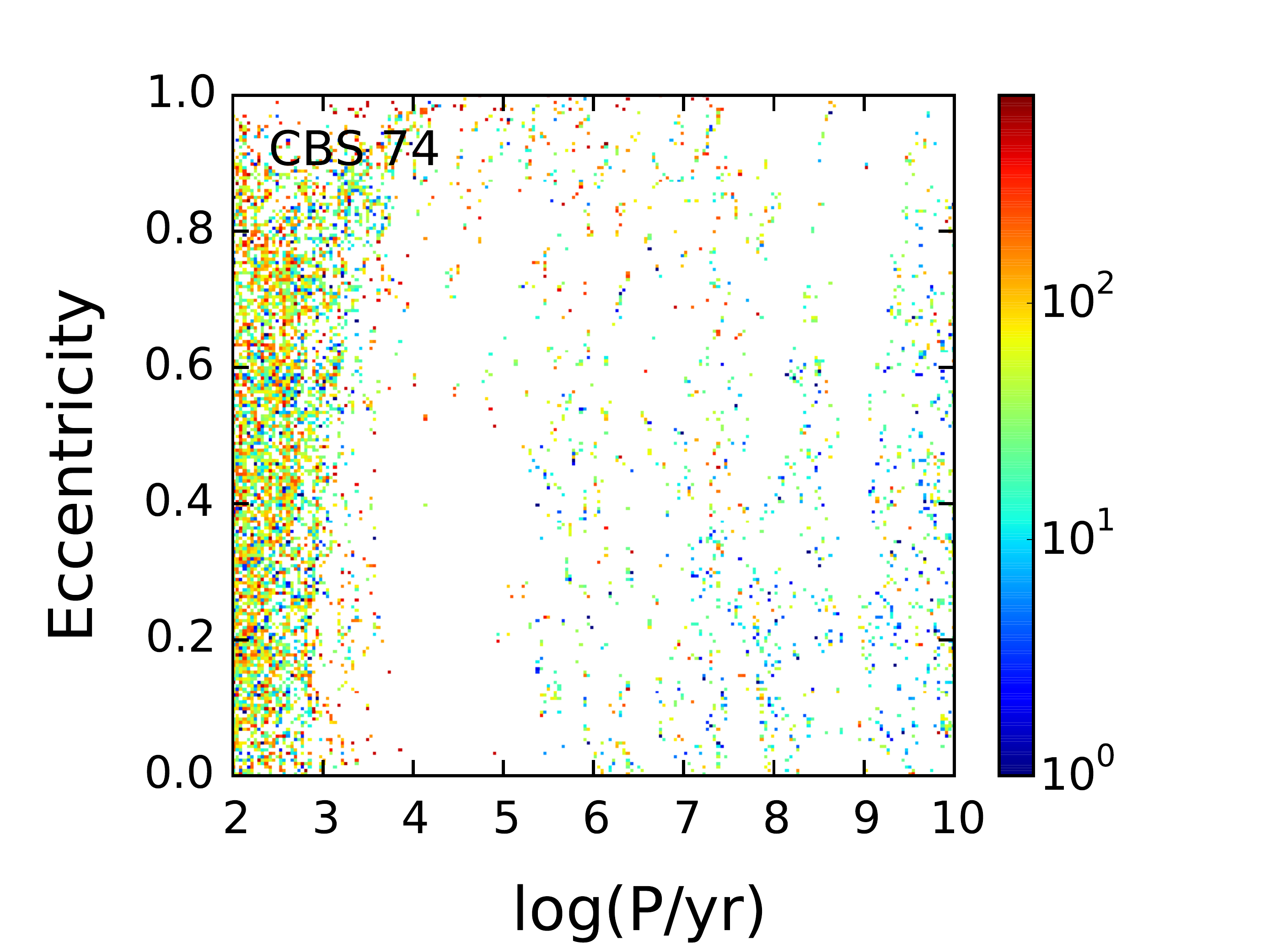}
  }  
  \caption{Projection of the parameter space of solutions into the 2-dimensional period-eccentricity plane. The colour indicates the density of solutions within a specific area in the diagram. Red indicates areas densely populated with solutions, and blue indicates areas sparsely populated. See Section~\ref{sec:results} of the text for details.}
	\label{fig:periodecc}
\end{figure*}

\begin{figure*}
  \ContinuedFloat
  \hbox{
    \includegraphics[height=7.cm]{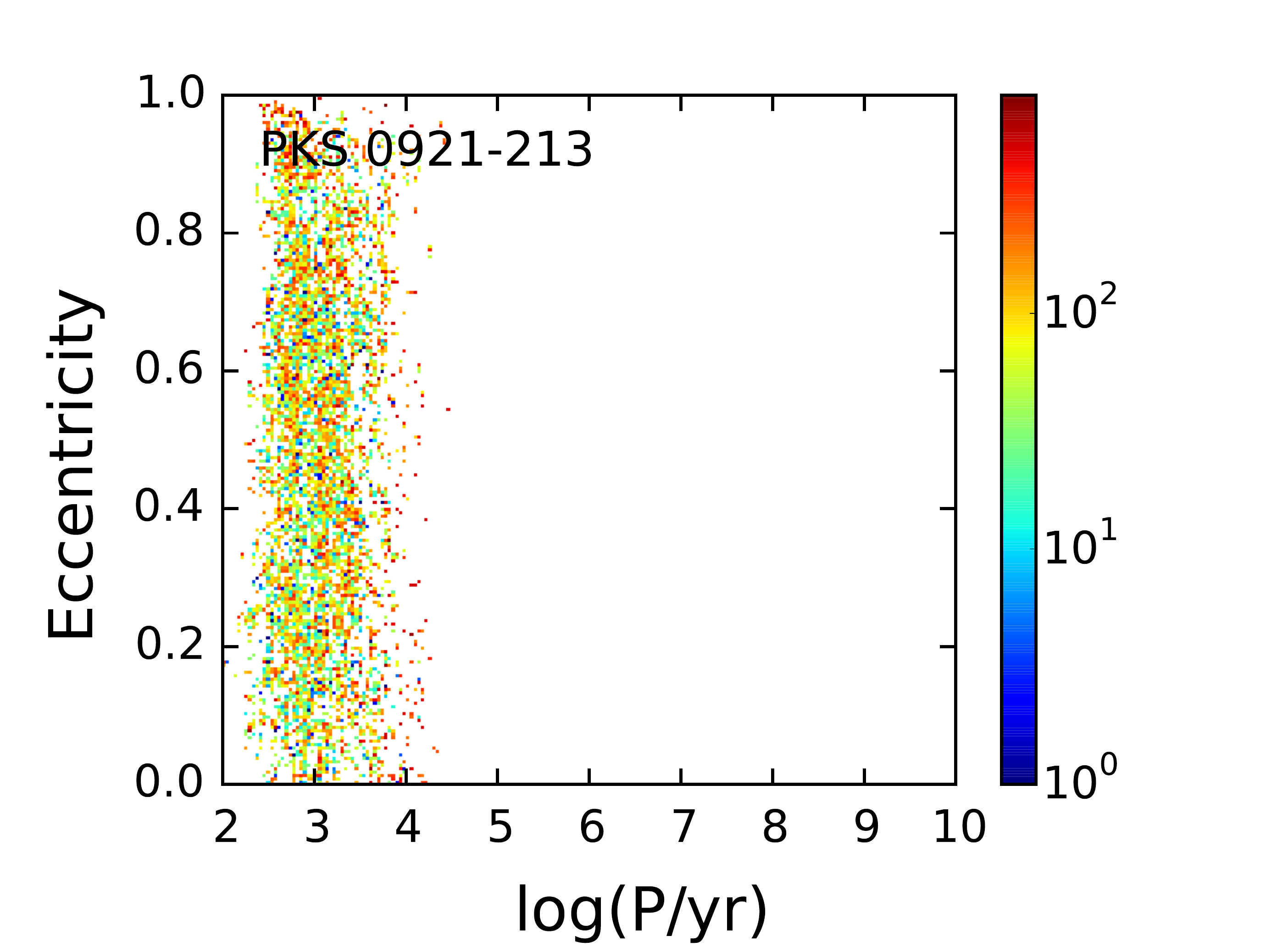}
    \hfill
    \includegraphics[height=7.cm]{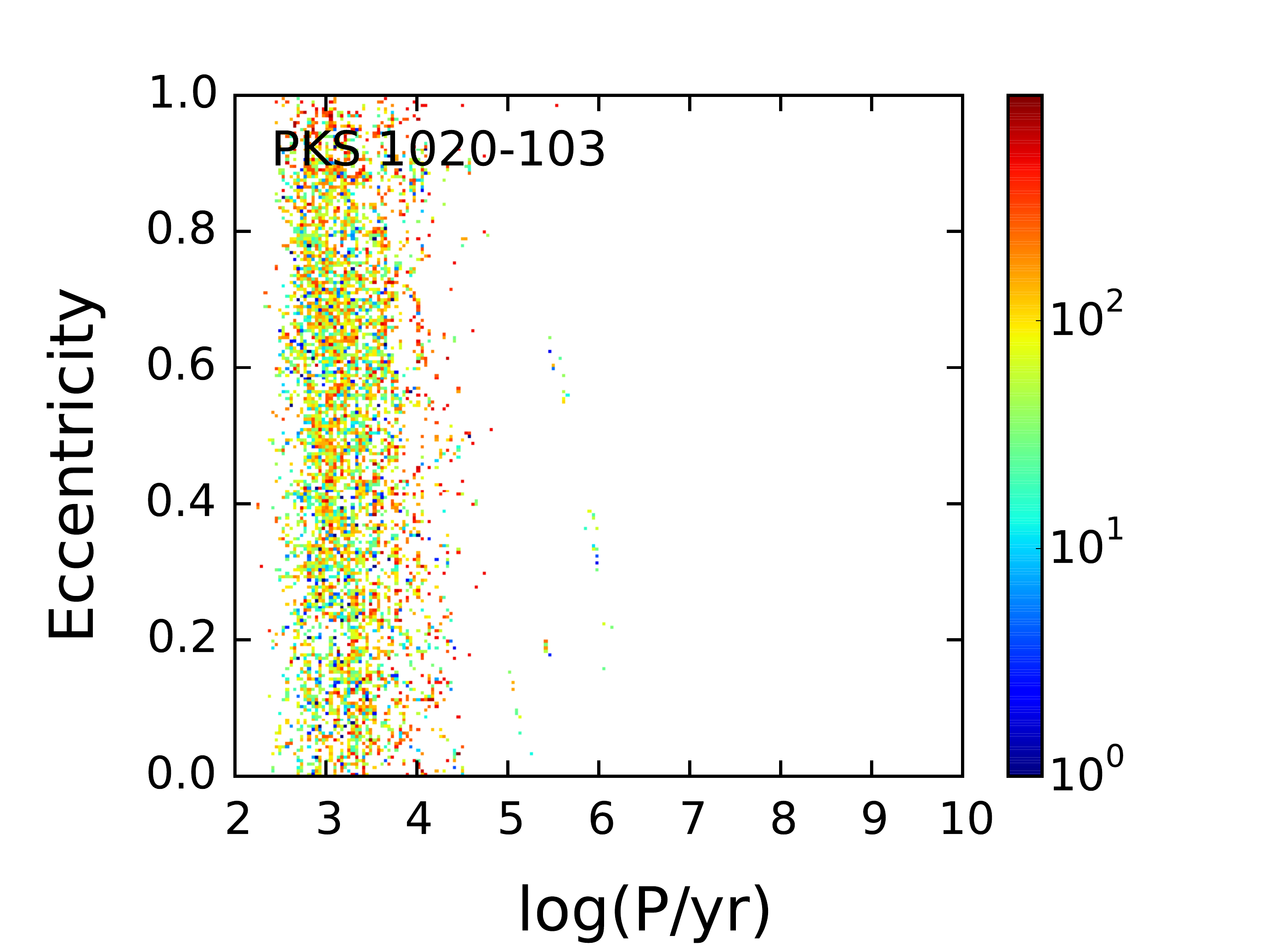}
  }
  \hbox{
  \includegraphics[height=7.cm]{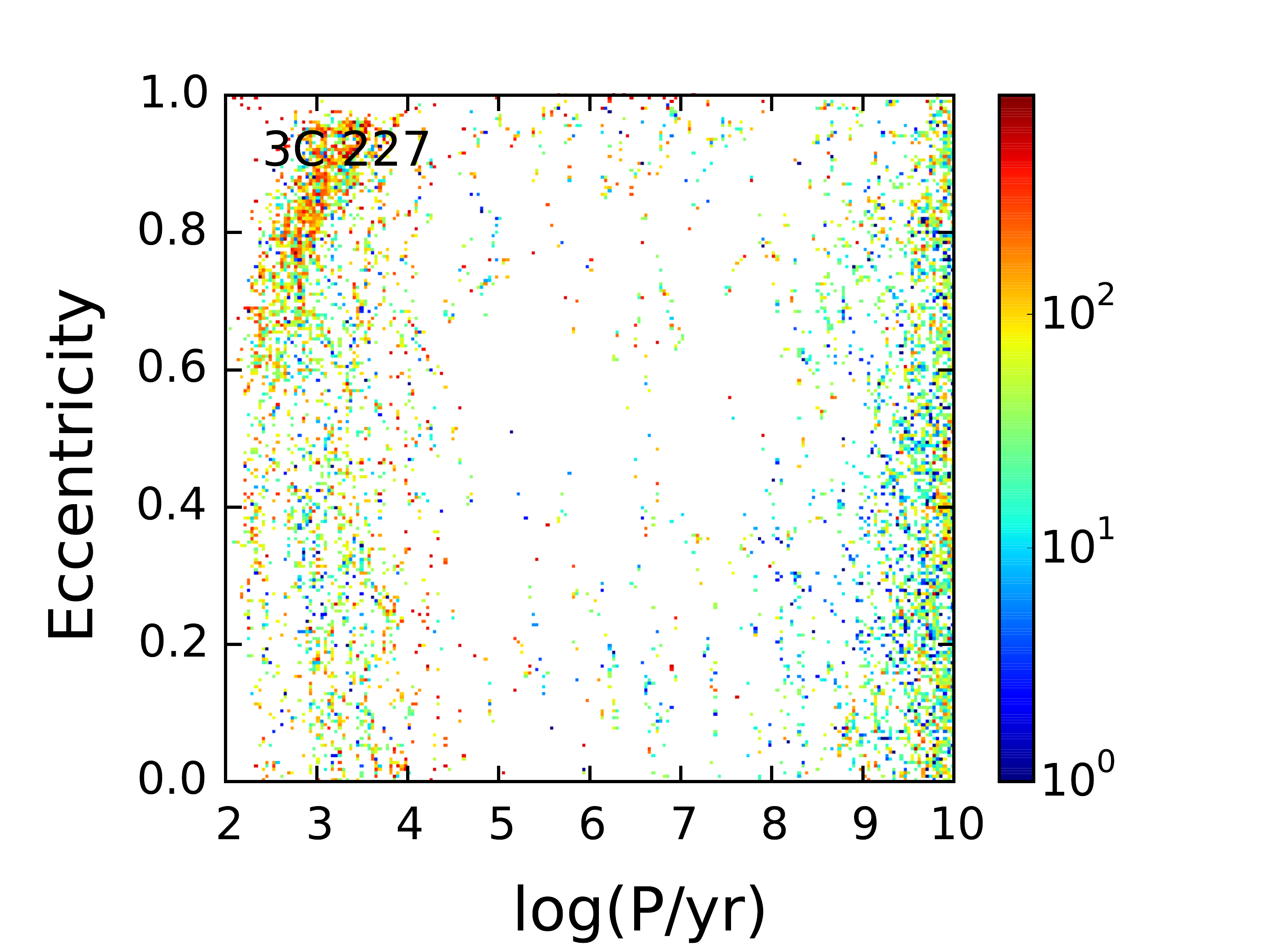}
    \hfill
  \includegraphics[height=7.cm]{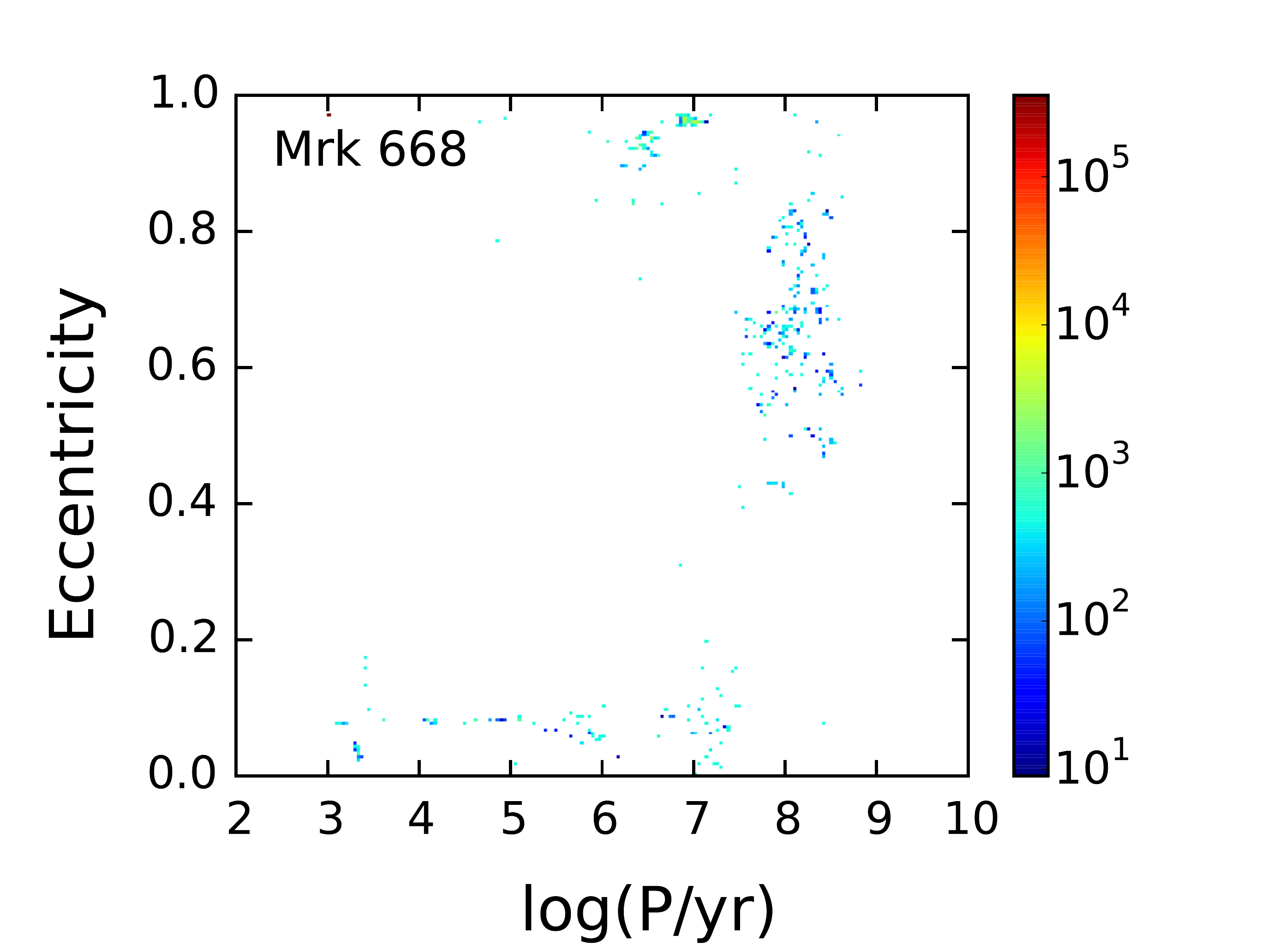}
  }
  \hbox{
  \includegraphics[height=7.cm]{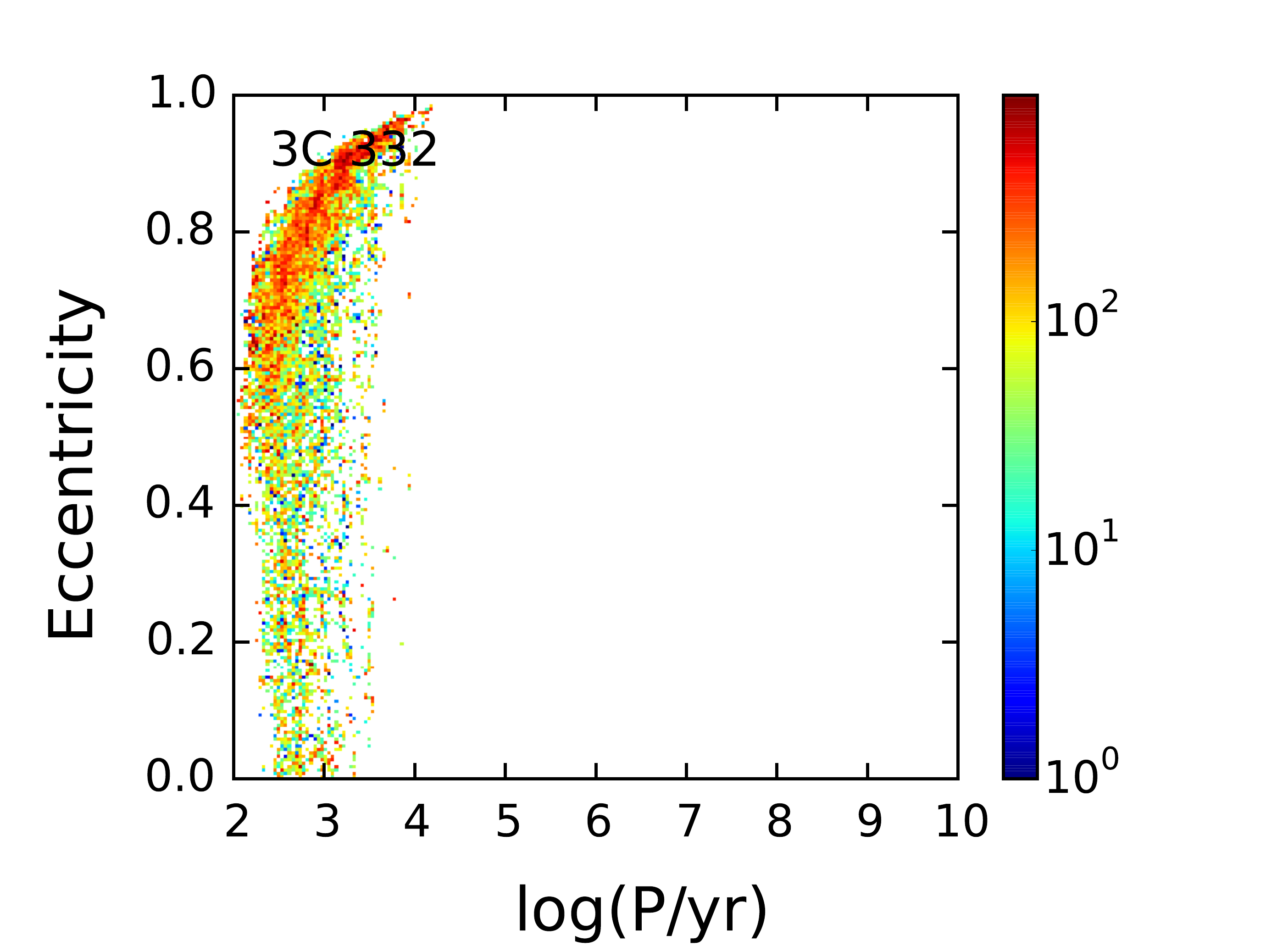}
    \hfill
  \includegraphics[height=7.cm]{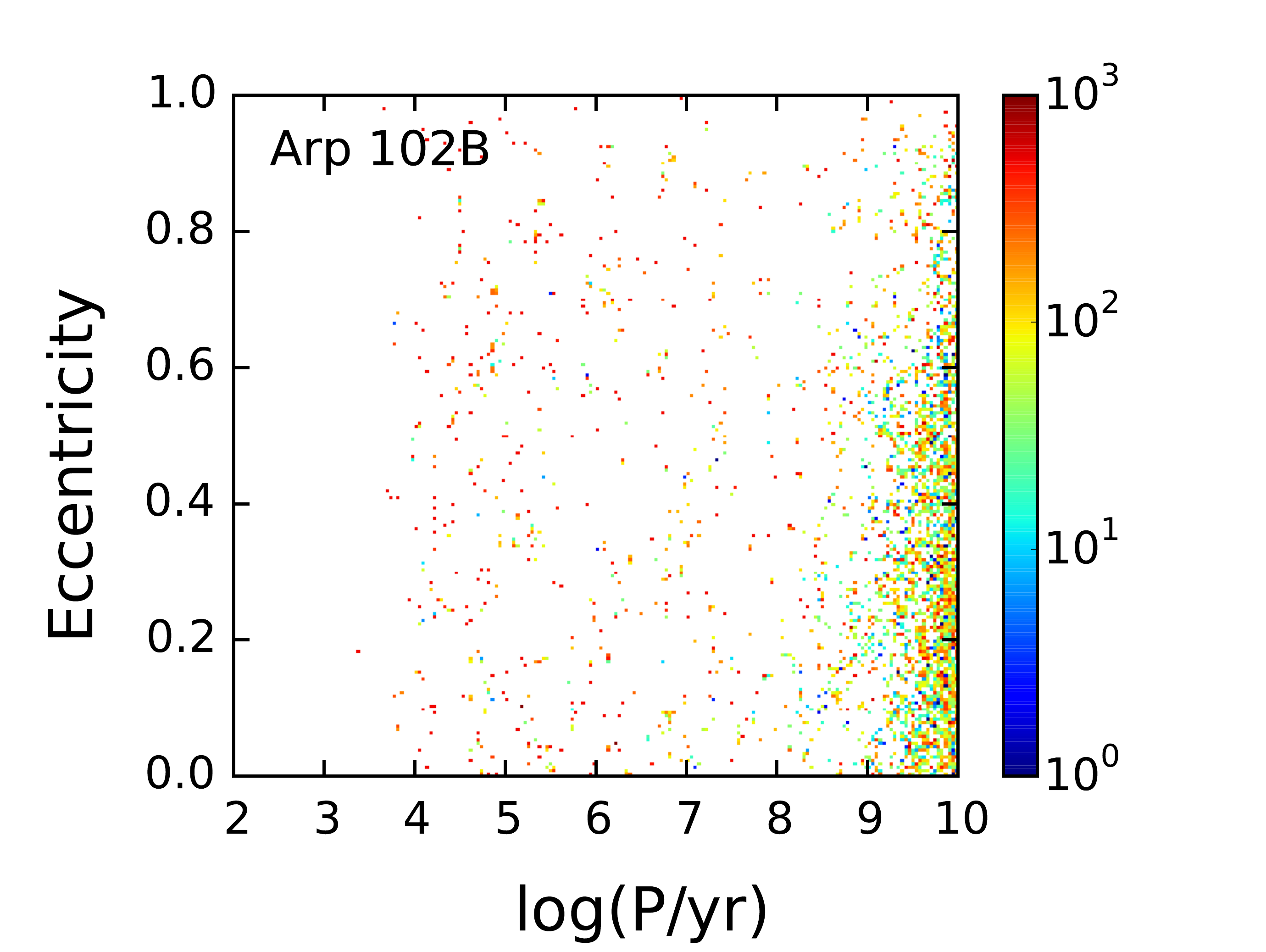}
  }
  \caption{ (continued)}
  \label{}
\end{figure*}

\begin{figure*}
  \ContinuedFloat
  \hbox{
    \includegraphics[height=7.cm]{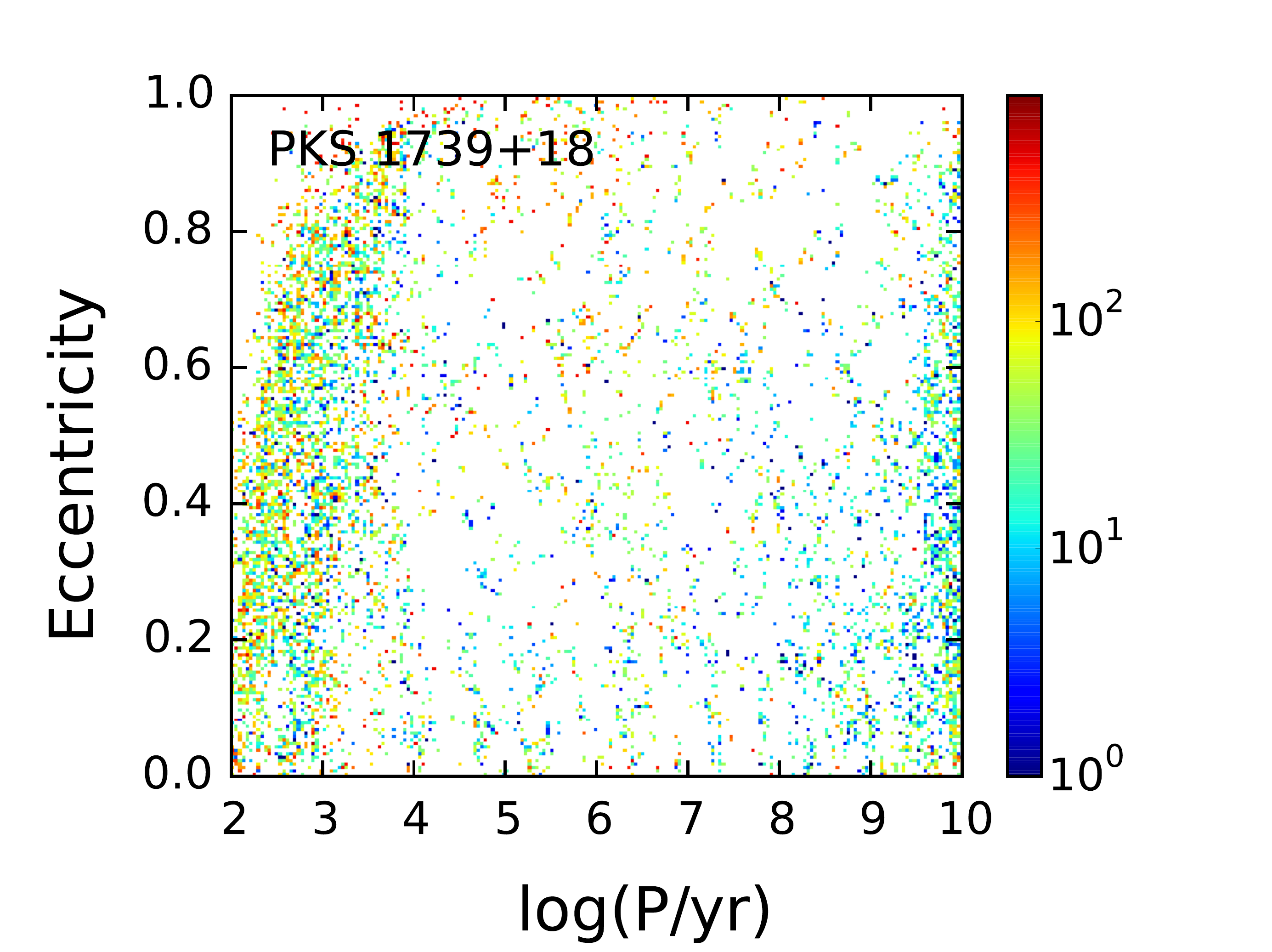}
    \hfill
    \includegraphics[height=7.cm]{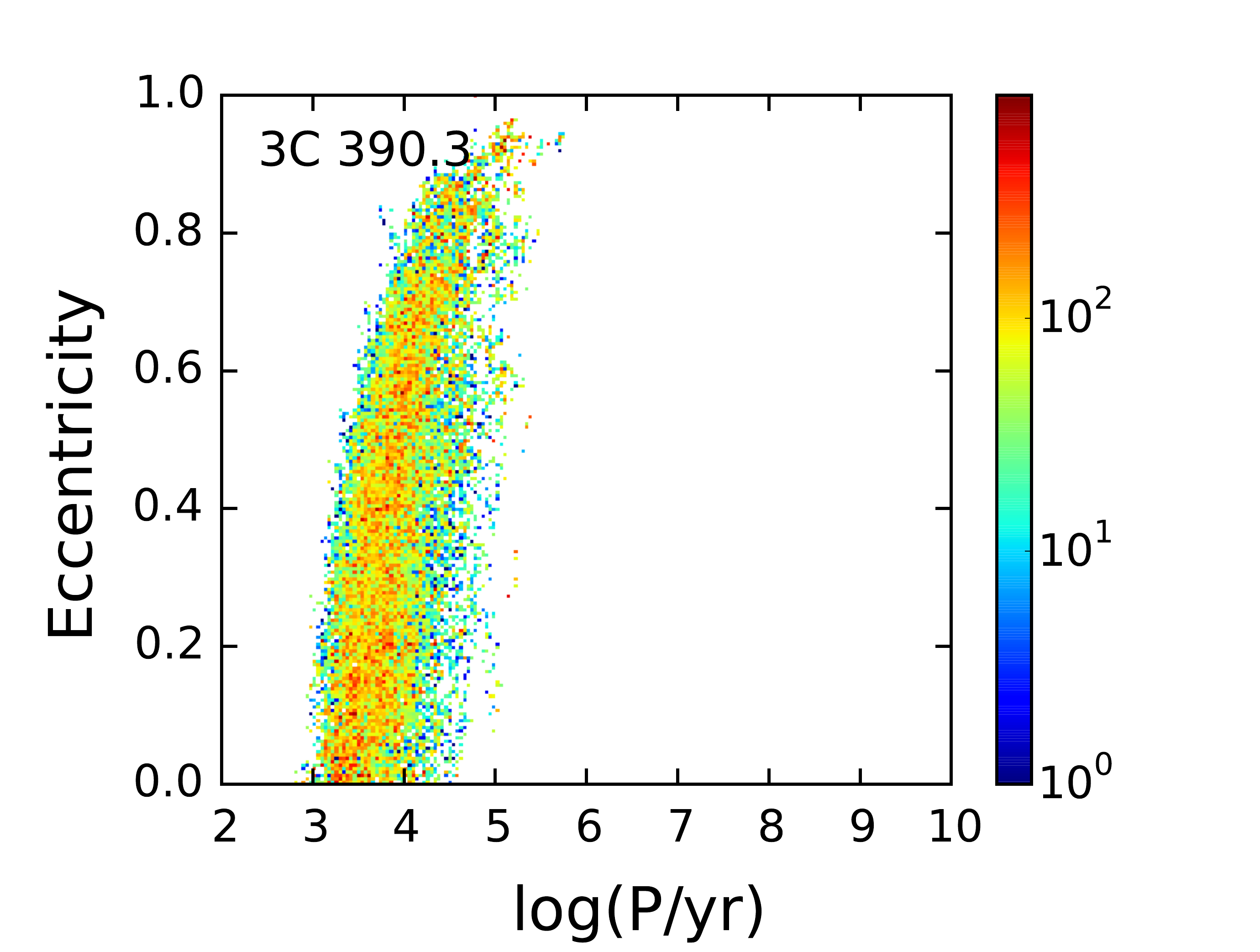}
  }
  \caption{ (continued)}
\end{figure*}

%----------------------------------------------------------------

\begin{figure*}
  \hbox{
    \includegraphics[height=7.cm]{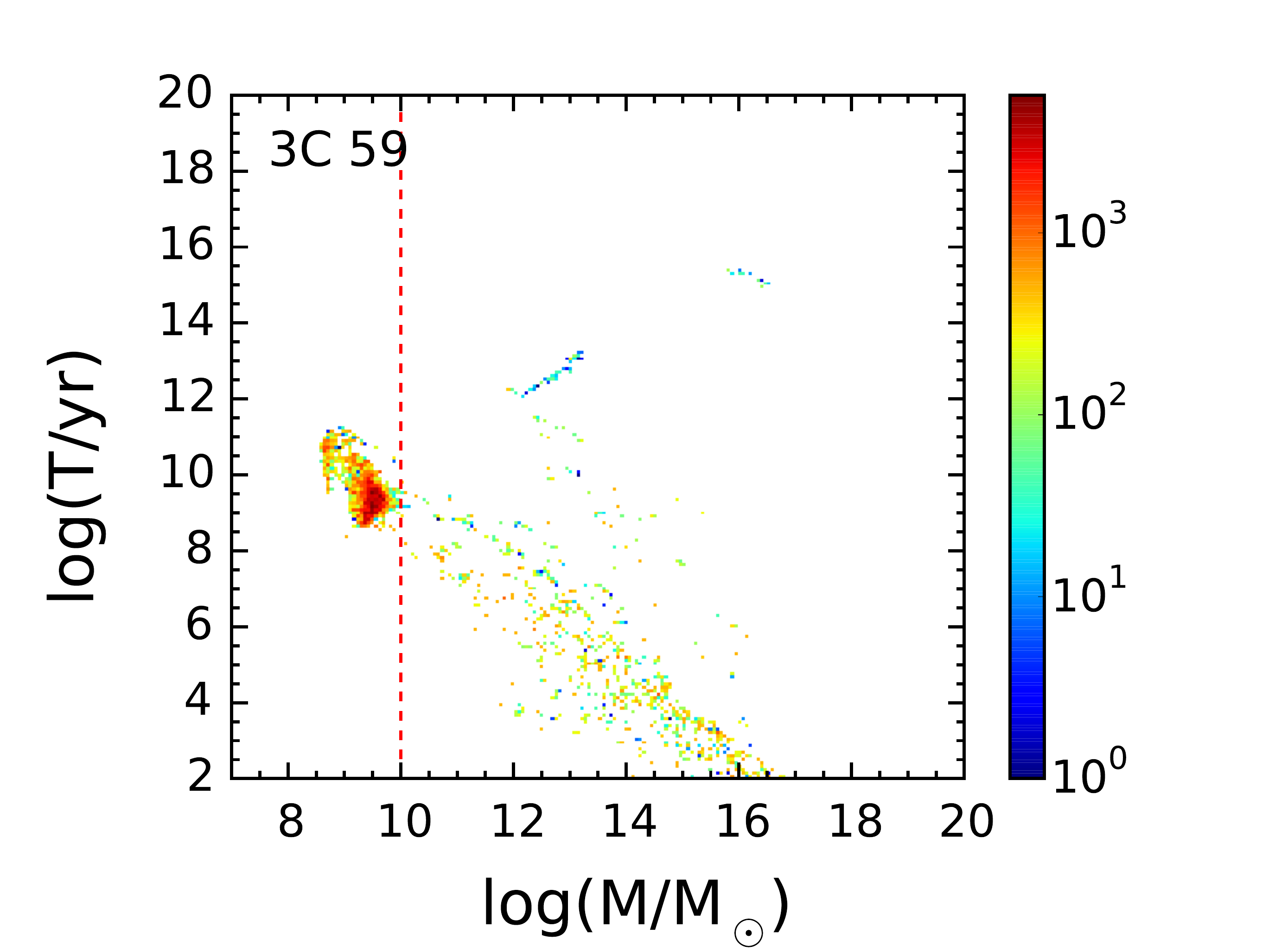}
    \hfill
    \includegraphics[height=7.cm]{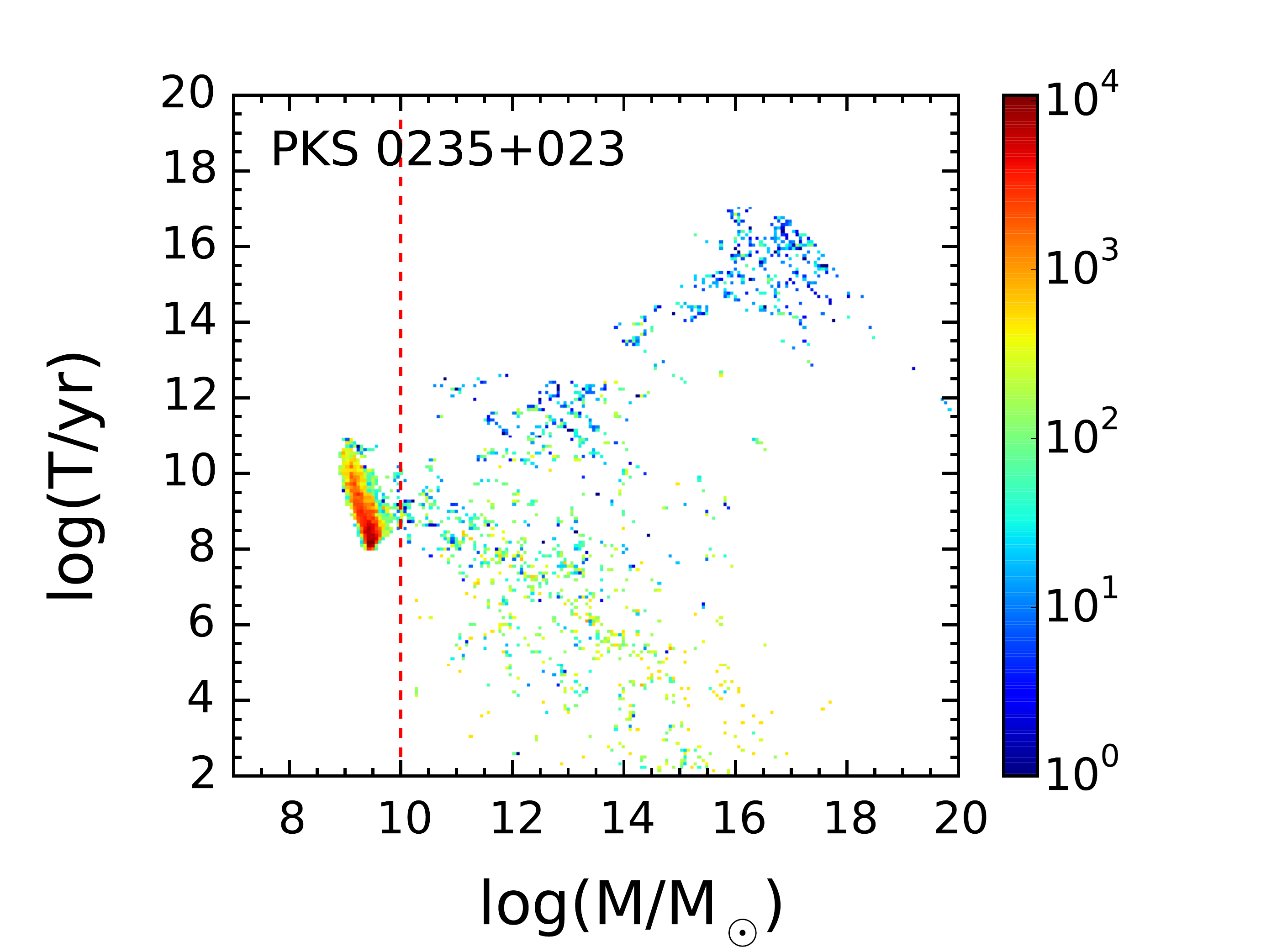}
  }
  \hbox{
    \includegraphics[height=7.cm]{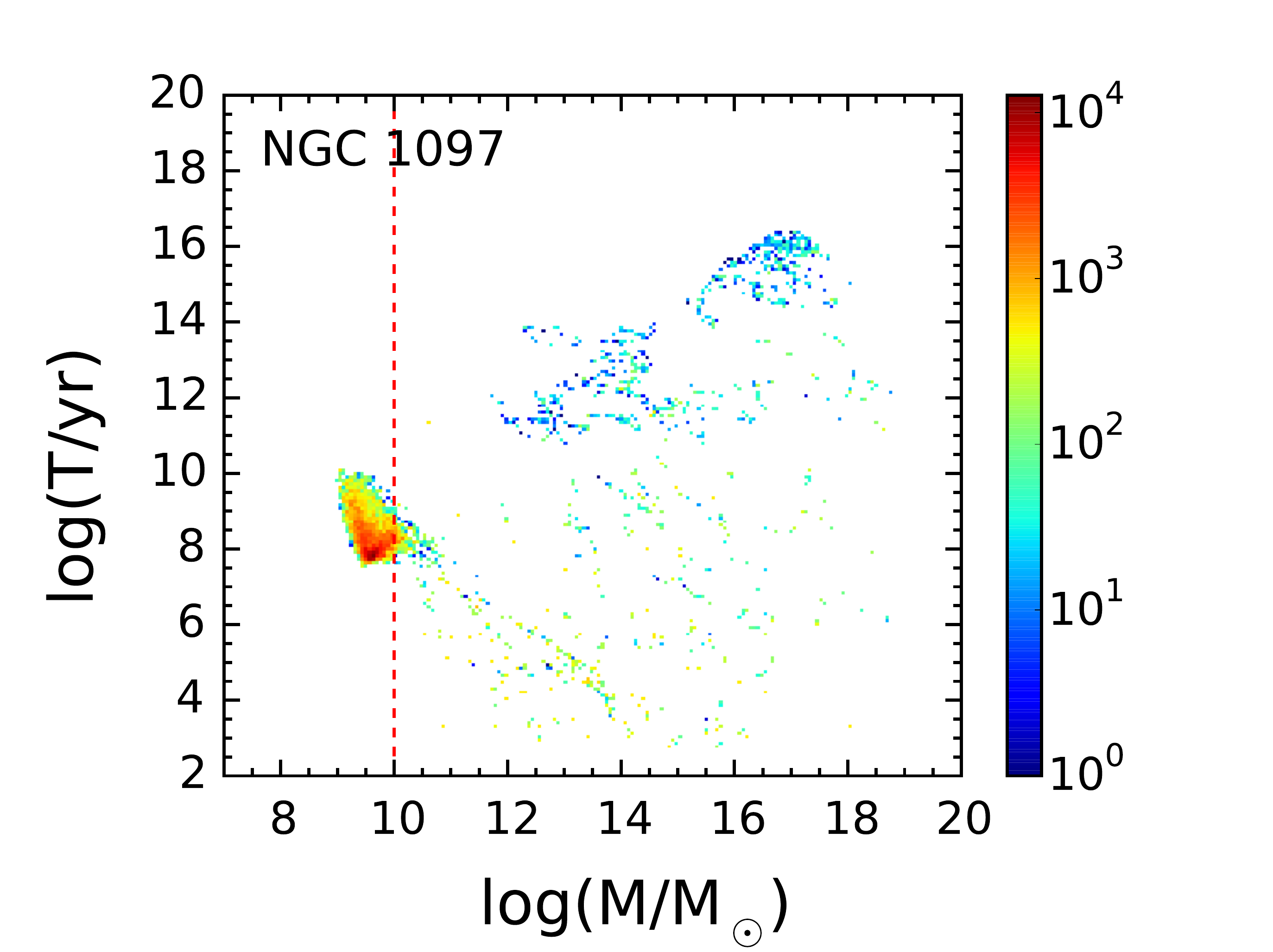}
    \hfill
    \includegraphics[height=7.cm]{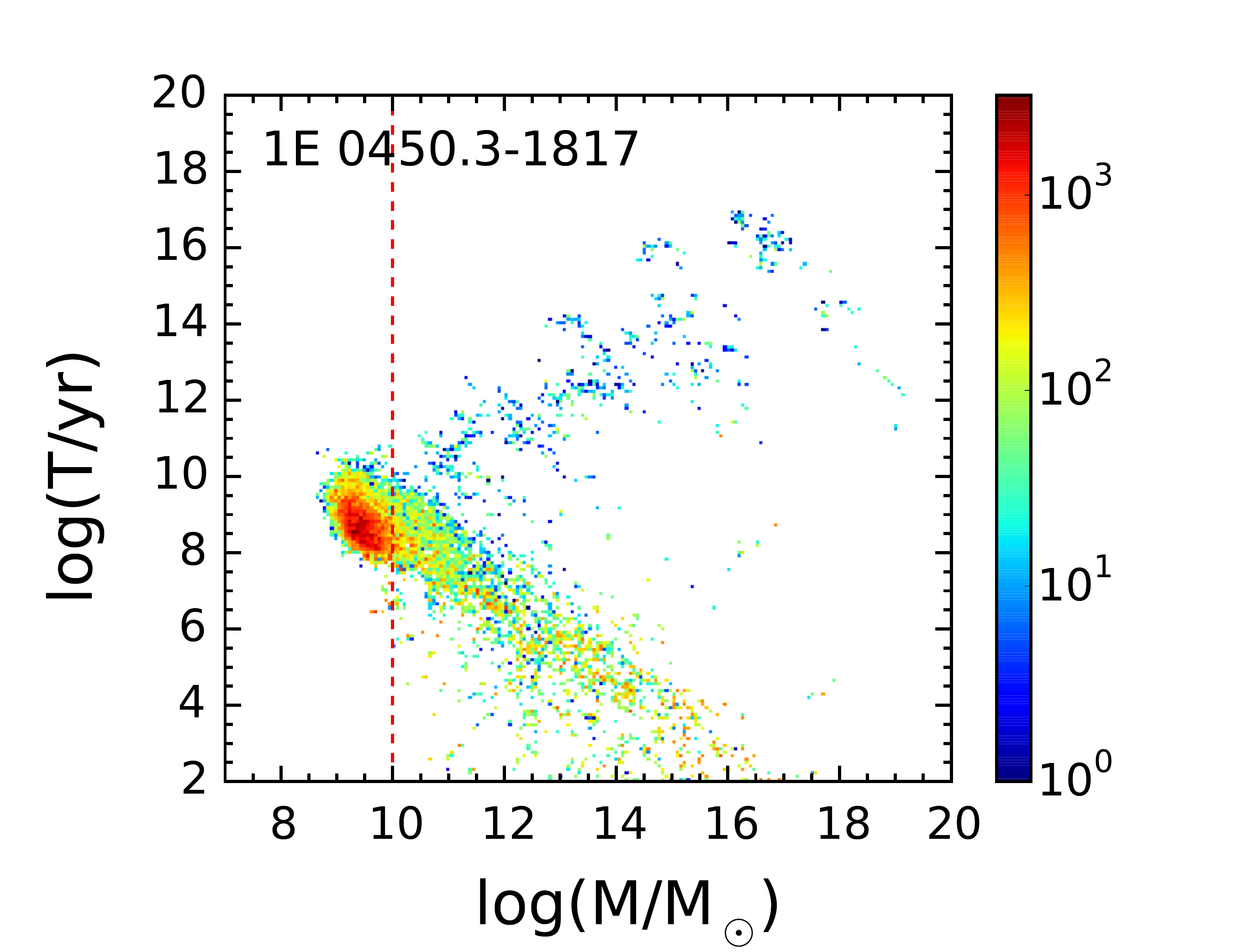}
  }
  \hbox{
    \includegraphics[height=7.cm]{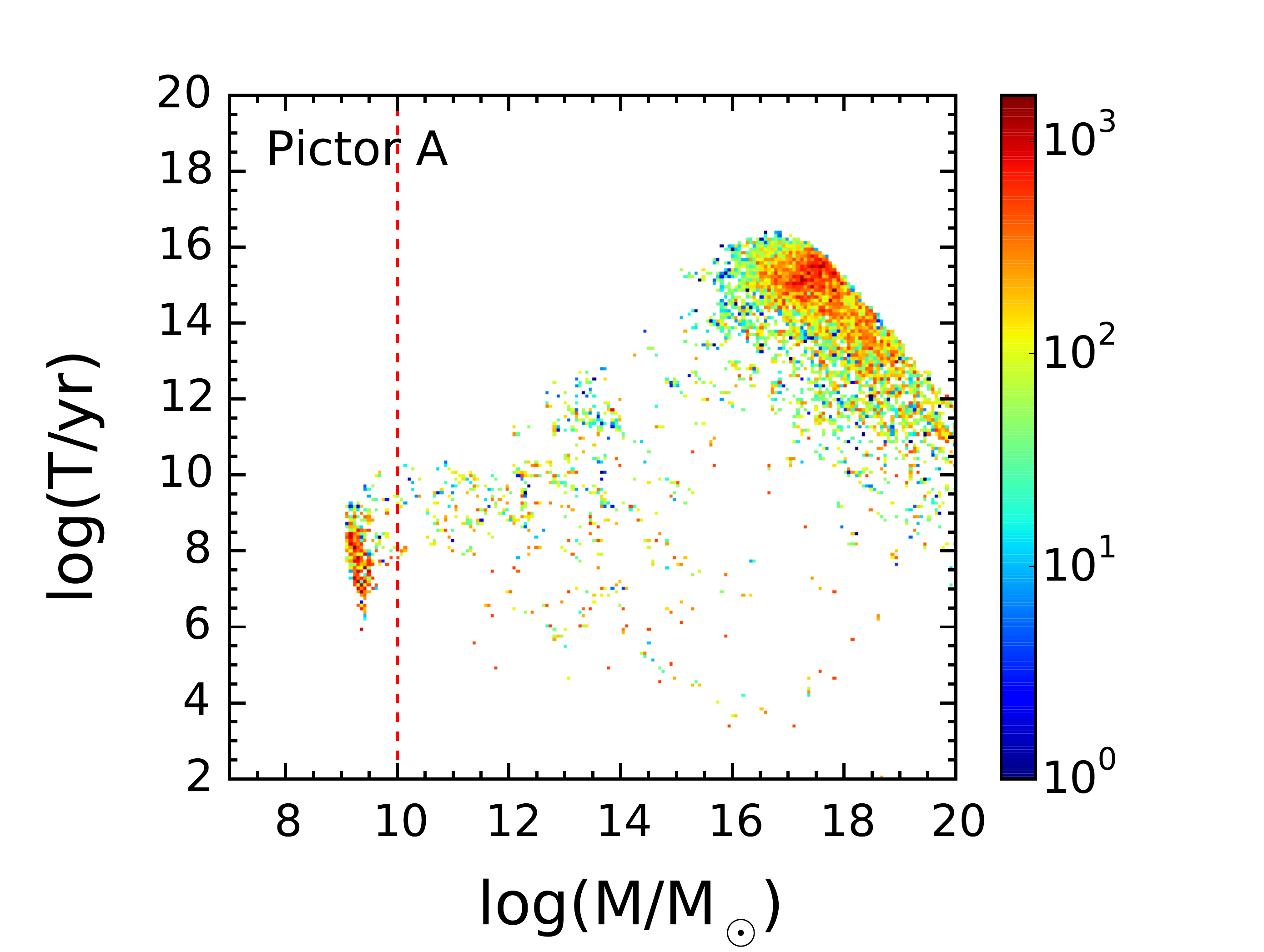}
    \hfill
    \includegraphics[height=7.cm]{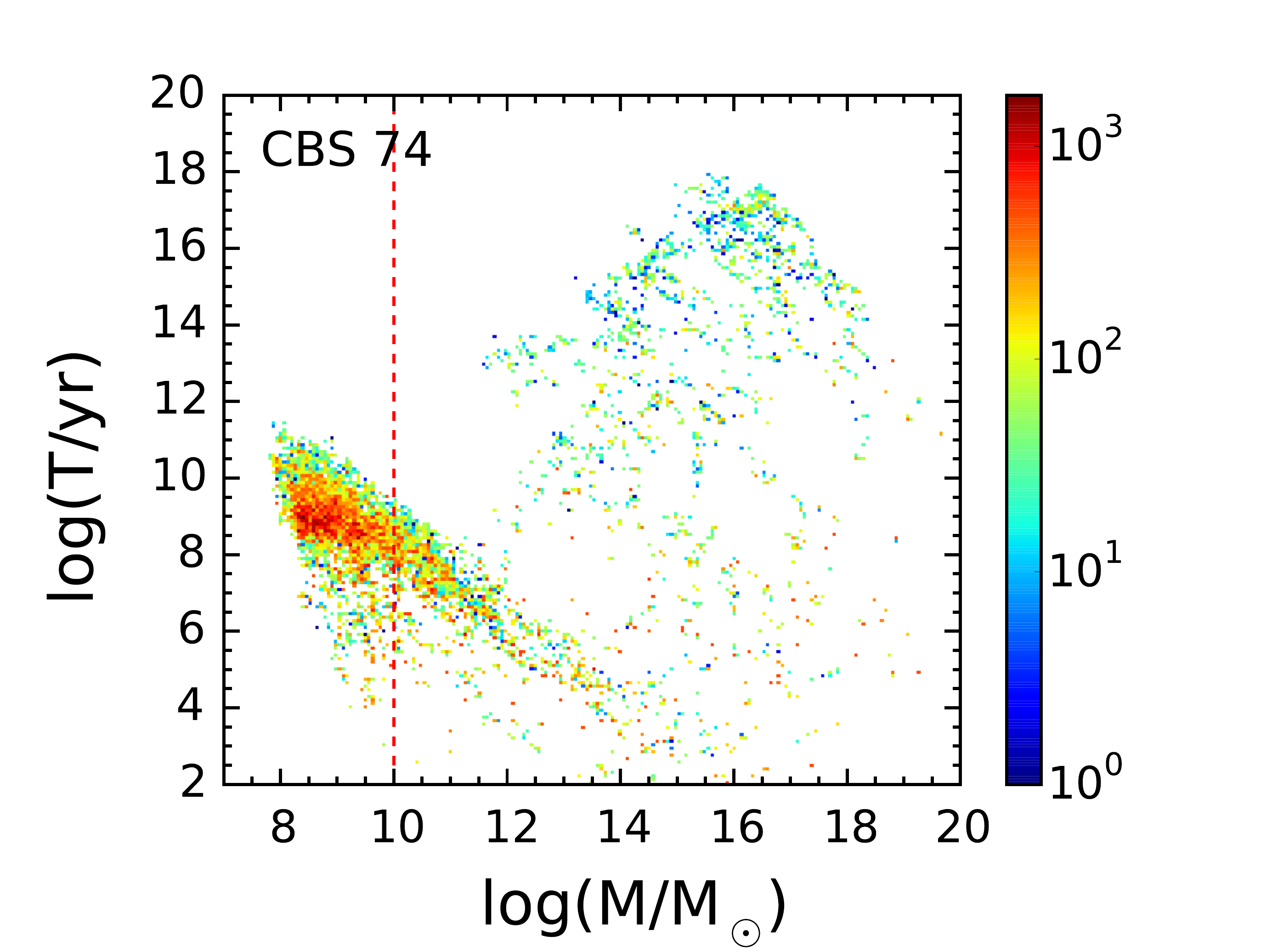}
  }
  \caption{The distribution of solutions BH mass--orbital decay time plane. The mass and orbital decay time were computed from the orbital model parameters as described in Section~\ref{sec:simulation}. The colour indicates the density of solutions within a specific area in the parameter space. Red indicates areas densely populated with solutions, and blue indicates areas sparsely populated. The vertical dashed line indicates a mass of $10^{10} M_{\odot}$. See the discussion in Sections~\ref{sec:results} and \ref{sec:discussion} of the text.}
	\label{fig:massdecay}
\end{figure*}

\begin{figure*}
  \ContinuedFloat
  \hbox{
    \includegraphics[height=7.cm]{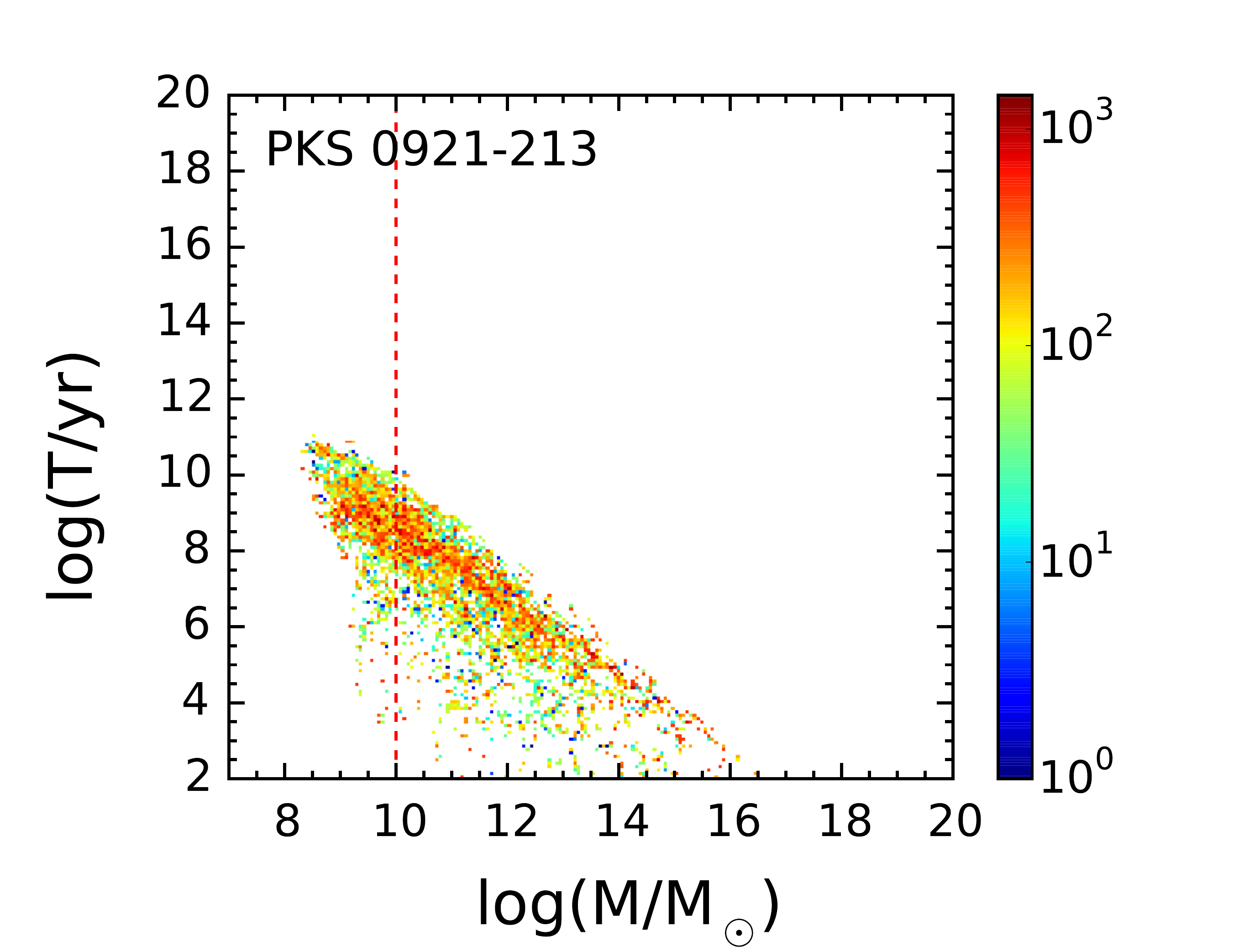}
    \hfill
    \includegraphics[height=7.cm]{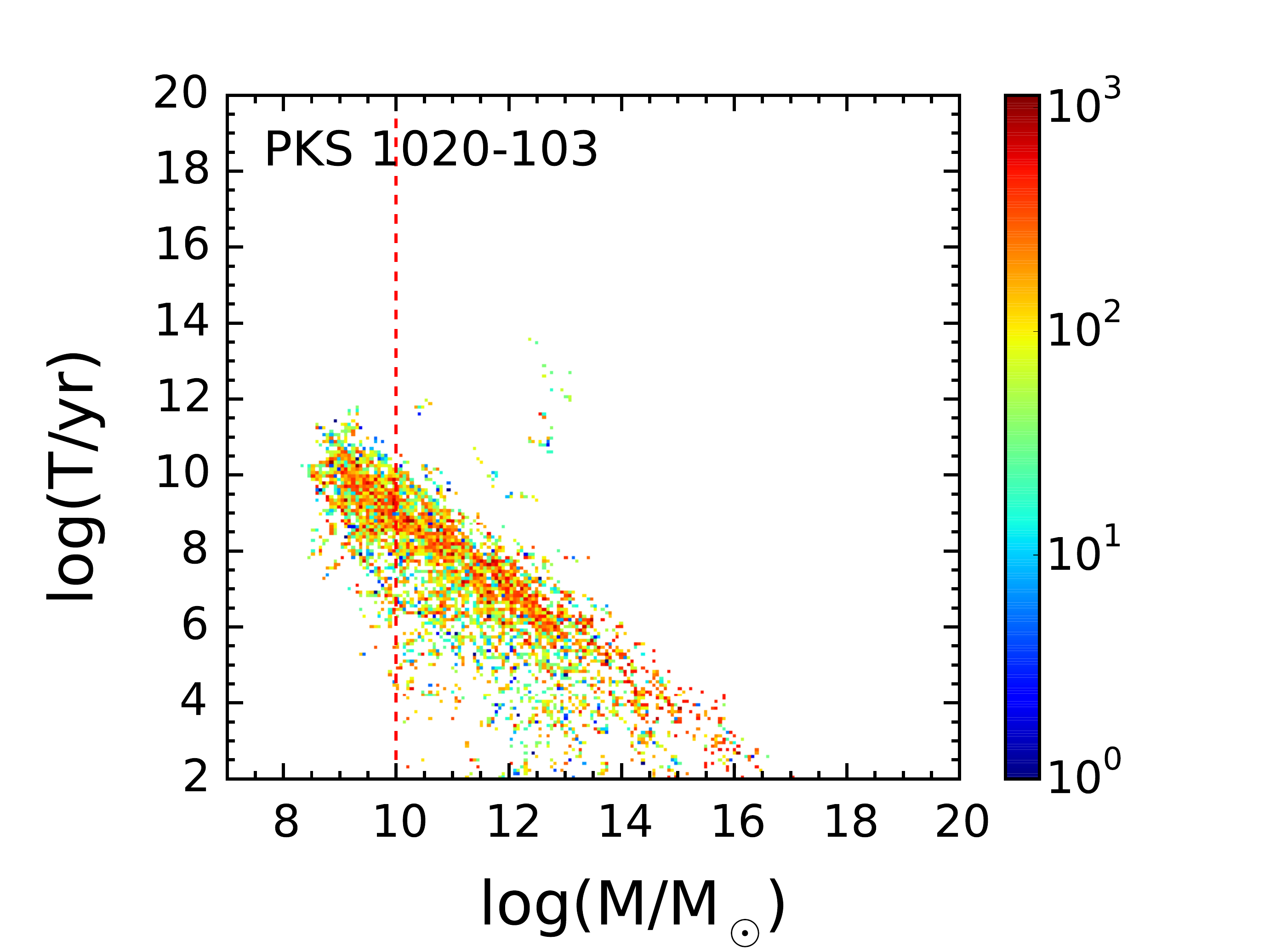}
  }
  \hbox{
    \includegraphics[height=7.cm]{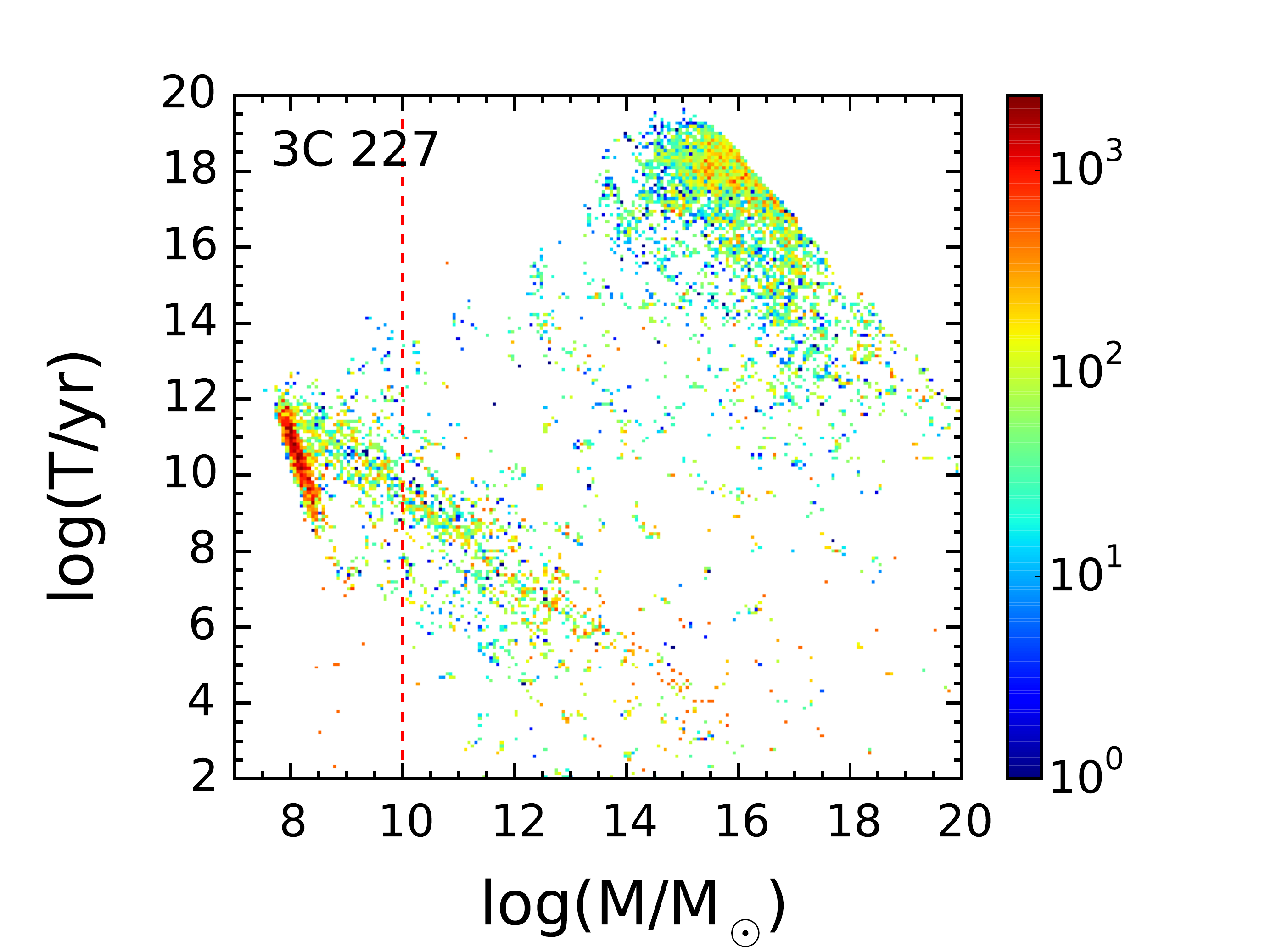}
    \hfill
    \includegraphics[height=7.cm]{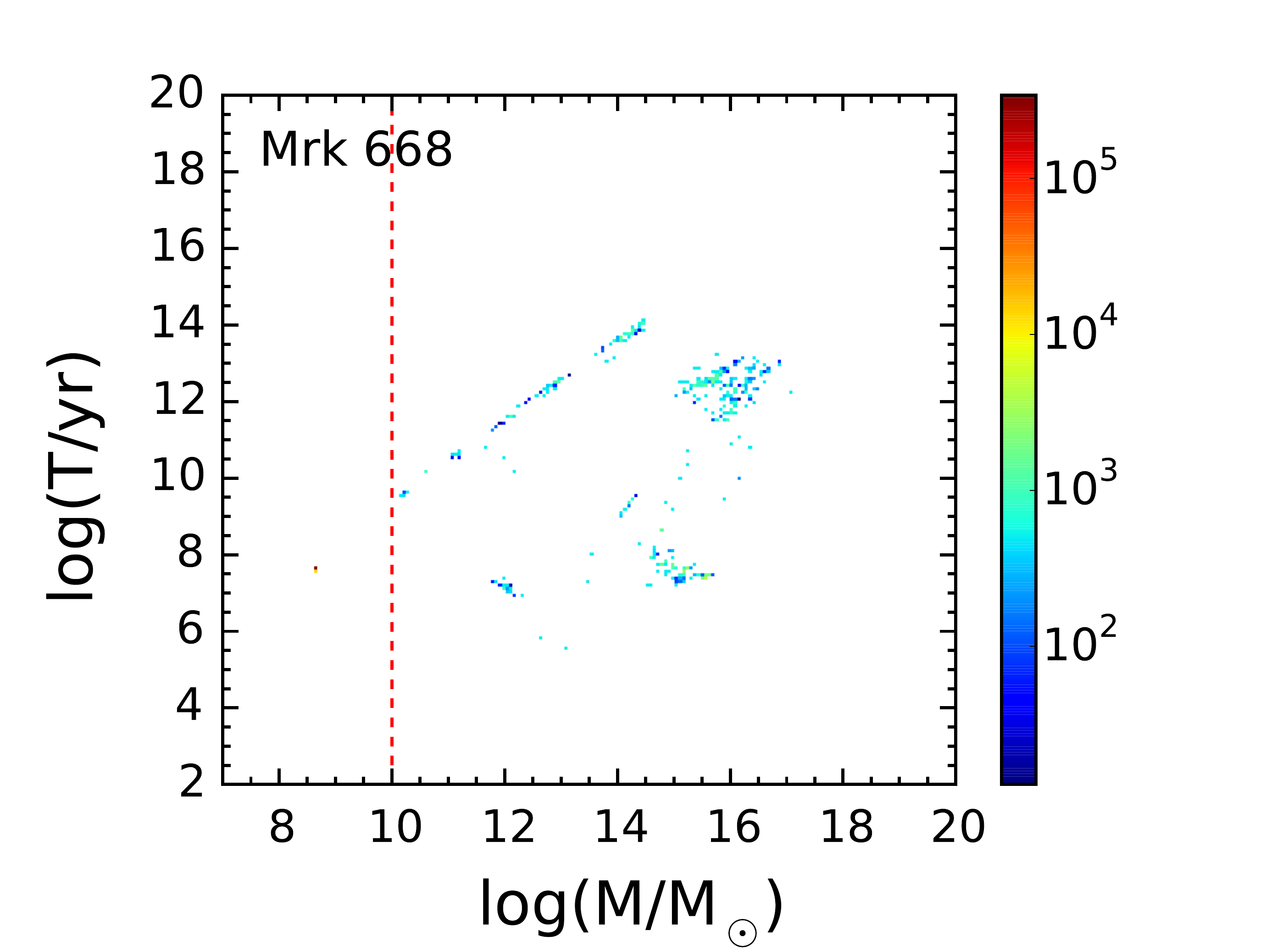}
  }
  \hbox{
    \includegraphics[height=7.cm]{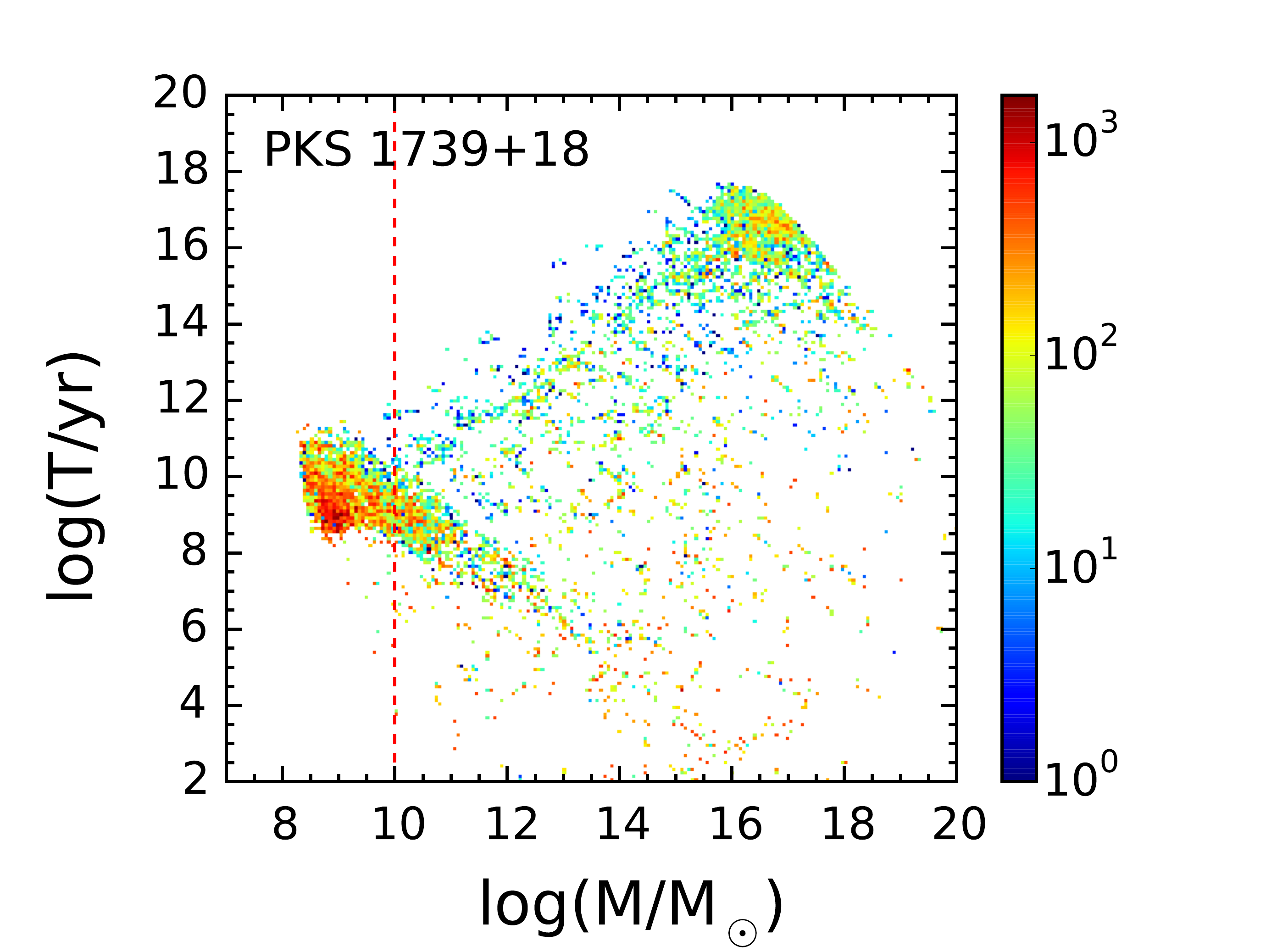}
    \hfill
    \includegraphics[height=7.cm]{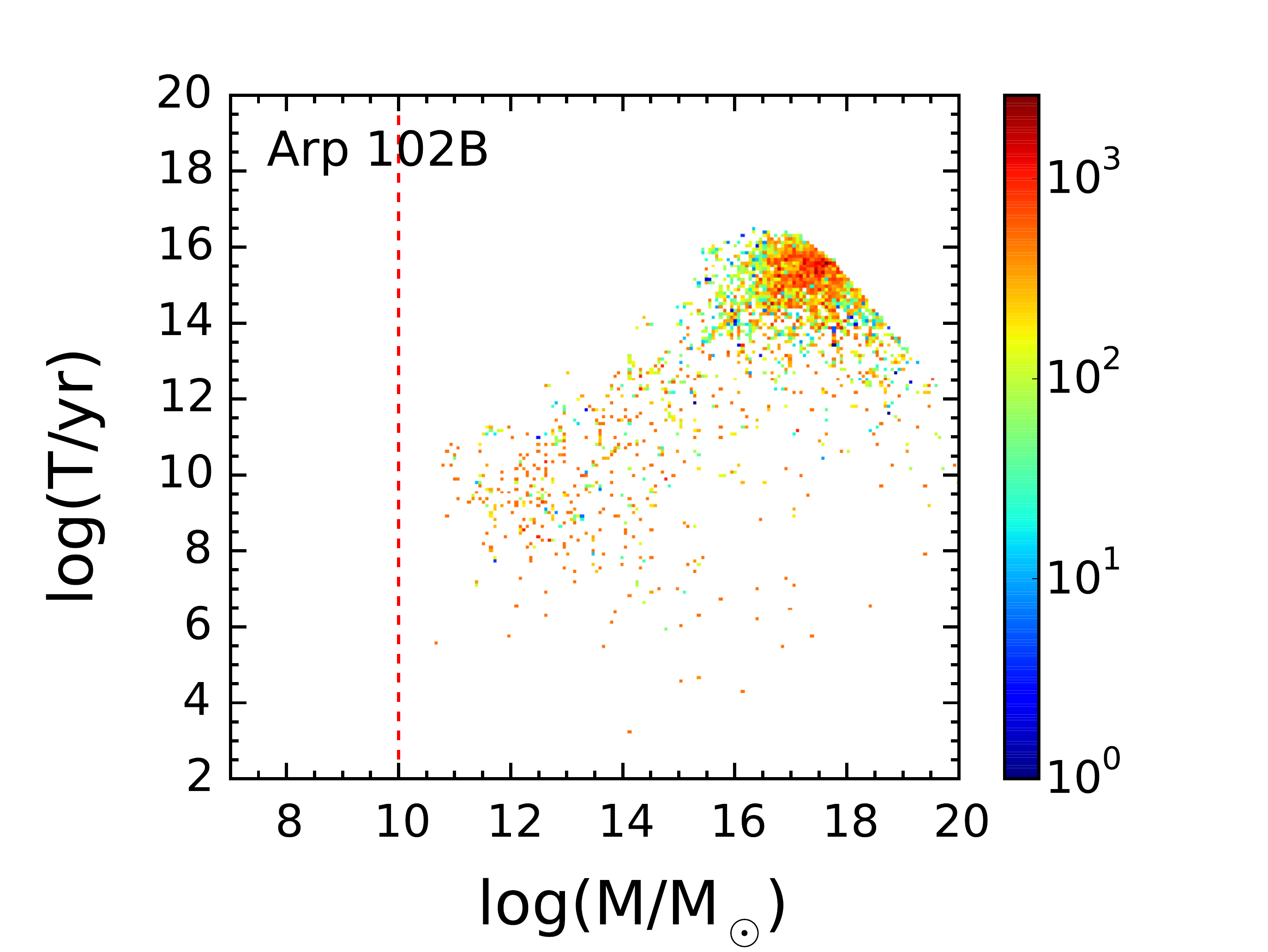}
  }
  \caption{ (continued)}
\end{figure*}

\begin{figure*}
  \ContinuedFloat
  \hbox{
    \includegraphics[height=7.cm]{./PKS1020-103-masslifetime-norm_ver2.pdf}
    \hfill
    \includegraphics[height=7.cm]{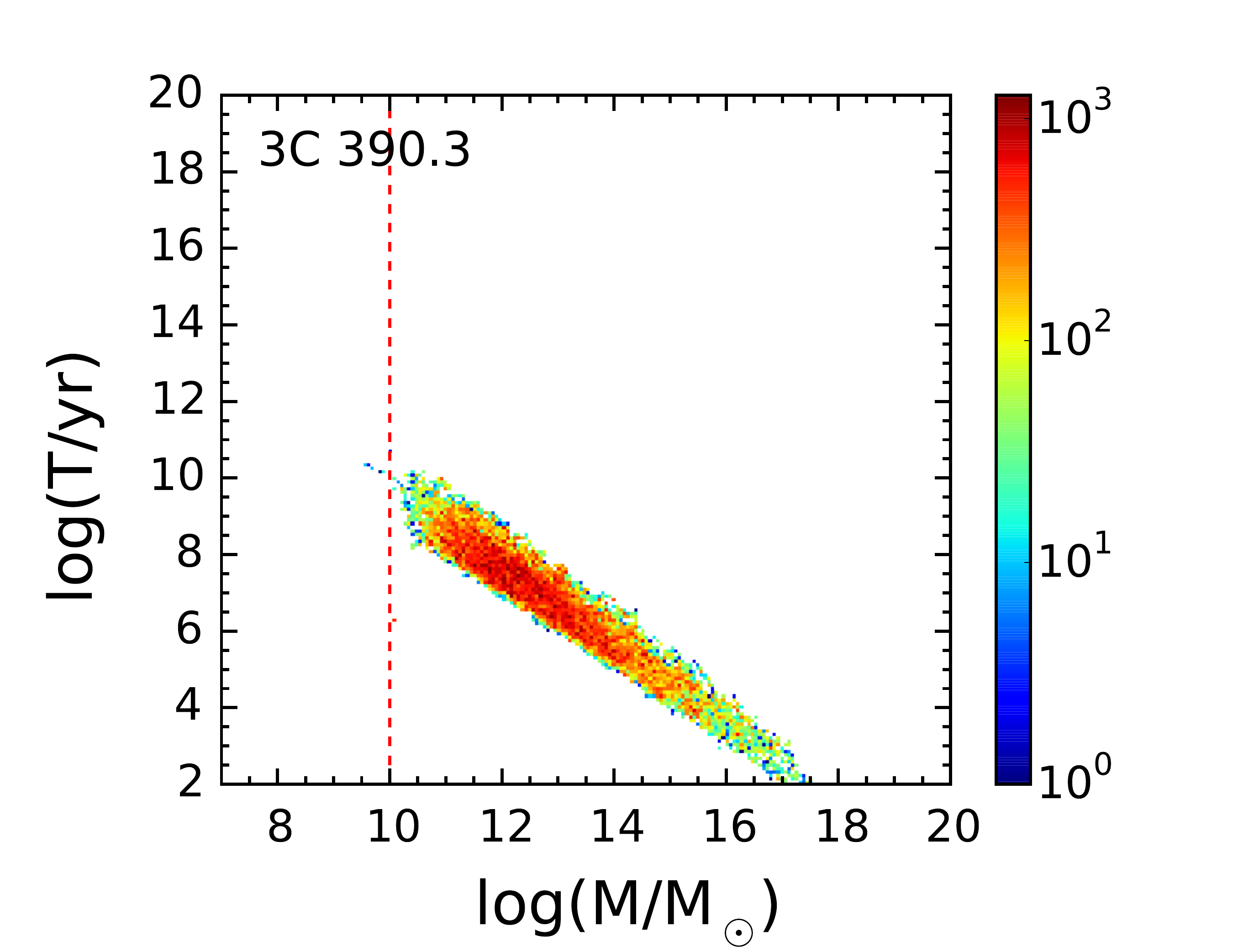}
  }
  \caption{(continued)}
\end{figure*}

%%%%%%%%%%%%%%%%%%%%%%%%%%%%%%%%%%%%%%%%%%%%%%%%%%

% Don't change these lines
\bsp	% typesetting comment
\label{lastpage}
\end{document}